\documentclass[11pt,a4paper]{article}

\usepackage[utf8]{inputenc}

\usepackage[margin=2cm]{geometry}
\usepackage{amsmath,amsthm,mathtools,amssymb}
\usepackage{pdflscape}
\usepackage{thmtools,thm-restate}
\usepackage{mathrsfs}
\usepackage{proba}
\usepackage{booktabs}
\usepackage{threeparttable,subfigure}
\usepackage{enumerate, enumitem}
\usepackage[round]{natbib}
\usepackage{graphicx,setspace,multirow,multicol}
\usepackage{dsfont}
\usepackage{caption}
\usepackage{comment}
\usepackage{xcolor}
\usepackage[normalem]{ulem}
\usepackage{longtable}
\usepackage{blindtext}
\useunder{\uline}{\ul}{}
\usepackage{float}
\usepackage{placeins}

\usepackage{hyperref}

\usepackage{environ}
\NewEnviron{myresizeenv}{\resizebox{\linewidth}{!}{\BODY}}

\usepackage{mathpazo}

\usepackage{ragged2e}

\hypersetup{
    unicode=false,          % non-Latin characters in Acrobat’s bookmarks
    pdftoolbar=true,        % show Acrobat’s toolbar?
    pdfmenubar=true,        % show Acrobat’s menu?
    pdffitwindow=false,     % window fit to page when opened
    pdftitle={},    % title
    pdfauthor={},     % author
    pdfsubject={},   % subject of the document
    pdfcreator={},   % creator of the document
    pdfproducer={},  % producer of the document
    pdfkeywords={}, % list of keywords
    pdfnewwindow=true,      % links in new window
    colorlinks=true,       % false: boxed links; true: colored links
    linkcolor=blue,          % color of internal links
    citecolor=blue,        % color of links to bibliography
    filecolor=blue,      % color of file links
    urlcolor=blue           % color of external links
}
\numberwithin{equation}{section}

\def\wl{\par\vspace{\baselineskip}}

\def\b{\boldsymbol}

\def\argmin{\operatorname*{\mathsf{argmin}}}

\def\diag{\mathsf{diag}\,}

\def\E{\mathbb{E}}
\def\R{\mathbb{R}}

\def\I{\mathcal{I}}

\def\1{\mathds{1}}

\DeclareMathOperator{\var}{\mathrm{Var}}
\DeclareMathOperator{\Cov}{\mathrm{Cov}}

\DeclareMathOperator{\hip}{\mathbb{H}}

\DeclareSymbolFont{sfoperators}{OT1}{cmss}{m}{n}
\DeclareSymbolFontAlphabet{\mathsf}{sfoperators}

\makeatletter
\def\operator@font{\mathgroup\symsfoperators}
\makeatother

\newcommand{\blue}[1]{{\color{blue} #1}}
\definecolor{ao}{rgb}{0.0, 0.0, 1.0}

\definecolor{darkgreen}{rgb}{0.09, 0.45, 0.27}

\definecolor{darkblue}{rgb}{0.0, 0.0, 0.55}

\definecolor{darkred}{rgb}{0.82, 0.1, 0.26}

\definecolor{amber}{rgb}{1.0, 0.75, 0.0}

\definecolor{purple}{rgb}{0.58, 0.0, 0.83}

\definecolor{brown}{rgb}{0.46, 0.21, 0.1}

\title{Forecasting inflation using \\ disaggregates and machine learning}

\author{
{\textbf{Gilberto Boaretto}} \\ [0.15cm]
\small{Dept. of Economics, PUC-Rio, Brazil} \\ [0.1cm]
\small{\url{gilbertoboaretto@hotmail.com} \vspace{-0.4cm}} \\
\and {\textbf{Marcelo C. Medeiros}} \\ [0.15cm] 
\small{Dept. of Economics, UIUC, US} \\ [0.1cm]
\small{\url{marcelom@illinois.edu}}
}

\date{\normalsize This draft: \, \today 
}

\begin{document}

\maketitle

\begin{abstract}
\noindent This paper examines the effectiveness of several forecasting methods for predicting inflation, focusing on aggregating disaggregated forecasts -- also known in the literature as the bottom-up approach. Taking the Brazilian case as an application, we consider different disaggregation levels for inflation and employ a range of traditional time series techniques as well as linear and nonlinear machine learning (ML) models to deal with a larger number of predictors. For many forecast horizons, the aggregation of disaggregated forecasts performs just as well survey-based expectations and models that generate forecasts using the aggregate directly. Overall, ML methods outperform traditional time series models in predictive accuracy, with outstanding performance in forecasting disaggregates. Our results reinforce the benefits of using models in a data-rich environment for inflation forecasting, including aggregating disaggregated forecasts from ML techniques, mainly during volatile periods. Starting from the COVID-19 pandemic, the random forest model based on both aggregate and disaggregated inflation achieves remarkable predictive performance at intermediate and longer horizons.
\wl
\wl
\wl
\noindent
\textbf{Acknowledgements}: Boaretto thanks PUC-Rio, CAPES, CNPq, and FAPERJ for financial support. Medeiros thanks CNPq and CAPES for financial support. We thank Gustavo Araújo, Juliano Assunção, Daniel Coutinho, Marcelo Fernandes, Eduardo Freitas, Cleomar Gomes, Fábio Gomes, João Victor Issler, Guilherme Kira, Márcio Laurini, and Gabriel Vasconcelos for helpful comments as well as all seminar participants of the Federal University of Uberlândia Webinar, PUC-Rio Graduate Workshop, XXII Meeting of the Brazilian Society of Finance (SBFin), Depep/BCB Webinar, Itaú Webinar, University of São Paulo at Ribeirão Preto Seminar, and 44th Meeting of the Brazilian Econometric Society (SBE) for numerous comments and suggestions. 

\end{abstract}

%%%%%%%%%%%%%%%%%%%%%%%%%%%%%%%%%%%%%%%%%%

\newpage

\singlespacing

\tableofcontents

%%%%%%%%%%%%%%%%%%%%%%%%%%%%%%%%%%%%%%%%%%

\newpage

\onehalfspacing

\section{Introduction}

Economists and econometricians aim to provide as accurate inflation forecasts as possible by utilizing the most efficient approaches available. An important question is whether considering disaggregated inflation in different markets or economic classifications can enhance the forecasting performance for aggregate inflation. At first, this approach could capture trend dynamics, seasonality, and short-term changes more effectively \citep{espasa2002}. In other words, using subcomponents would allow the econometric models to capture the heterogeneity underlying the aggregate variable better \citep{bermingham2014}. Given an unknown data-generating process, whether direct or indirect forecasting through aggregating disaggregated forecasts can improve or not forecast accuracy is strictly an empirical question \citep{lutkepohl1984, hendry2011, faust2013}. Nevertheless, a critical challenge that emerges is the increase in estimation uncertainty. To mitigate this problem, we implement a disaggregated analysis using machine learning (ML) methods that can deal with the bias-variance trade-off. Studies such as \citet{inoue2008}, \citet{garcia2017}, and \citet{medeiros2021} point out the benefits of these techniques for inflation forecasting. Our paper employs these techniques in the context of disaggregated analysis, something scarcely explored in the literature.

The broad literature on inflation forecasting documents that the predictive performance of survey-based forecasts is challenging to beat, especially in the short-term horizons -- current and immediate next months \citep{thomas1999, ang2007, croushore2010}. \citet{faust2013} argue that ``purely subjective forecasts are in effect the frontier of our ability to forecast inflation'' because, besides private sectors and central banks having access to econometric models, they add expert judgment to these models. Consequently, ``a useful way of assessing models is by their ability to match survey measures of inflation expectations'' \citep{faust2013}. A potential explanation for this phenomenon is that forecasters are likely to have a richer information set than the econometrician employing a standard set of macroeconomic variables as predictors for inflation \citep{delnegro2011}. Thus, including revealed expectations among the predictors is a way to exploit an information set that is not available. \citet{bacsturk2014}, \citet{altug2016}, \citet{garcia2017}, \citet{fulton2021}, and \citet{banbura2021} find evidence favorable to the incorporation of survey-based forecasts into forecasting econometric models.

This paper examines the effectiveness of various forecasting methods for predicting aggregate inflation, focusing on aggregating disaggregated forecasts -- also known in the literature as the bottom-up approach. Using the Brazilian case as an example, we compare the predictive performance of the bottom-up approach with traditional approaches in the literature, including survey-based forecasts and direct forecasting based exclusively on the aggregate. We explore different levels of disaggregation to assess how forecasts based on disaggregate price levels fare relative to those that rely solely on aggregate. Granularity is a potential advantage of considering disaggregates. Besides the specific effects of the traditional macro-variables related to money, economic activity, government, and external sector, we include lagged and crossed effects between disaggregates. When we compute our forecasts, we also consider \textit{available} survey-based expectations as a predictor to add information not captured by other variables. Finally, we employ a range of traditional time series techniques, as well as linear and nonlinear ML techniques to deal with a larger number of predictors. More specifically, we consider these modeling possibilities:
\begin{enumerate} \itemsep0em
    \item Traditional time series methods: random walk (RW), historical mean, and autoregressive (AR) models;
    \item Shrinkage-based models with or without sparsity, namely, Ridge and adaptive LASSO (adaLASSO);
    \item Factor and target factor augmented models;
    \item FarmPredict, a model that bridges both common factor and sparsity structures. With this method, we can explore remaining sparse idiosyncratic effects after controlling for common factors, and autoregressive, expectation and deterministic components;
    %\item A model called FarmPredict bridges both factor and sparsity.
    \item Complete subset regression (CSR), an ensemble method that combines estimates from all possible linear regression models keeping the number of predictors fixed;
    %\item Random Forest (RF), a non-linear, ``bagged'' tree-based model;
    \item Random forest (RF), a bagged ensemble of non-linear tree-based models;
    \item Model combination via an average of forecasts for each disaggregation.
\end{enumerate}

\vspace{-0.2cm}
The Brazilian case is interesting for several reasons. First, the Broad Consumer Price Index (\textit{Índice de Preços ao Consumidor Amplo} -- IPCA), which serves as the official Brazilian price index, is available monthly and boasts a rich structure to be explored. The index contains several disaggregation levels and all time-varying weights of goods and services in the representative consumption basket are readily available. Second, the Central Bank of Brazil conducts the Focus survey, an extensive daily survey of expectations for multiple forecast horizons for some variables, including inflation. This survey reflects experts' opinions, mainly financial market professionals, and may contain private information that is not available to the econometrician. Beyond its utility as a predictor for generating model-based forecasts, Focus' inflation expectations can be used as a benchmark to assess whether improving survey-based forecasts for a given horizon is possible. Third, due to Brazil's inflationary history, in addition to the official price index, the country has several price indexes that may be used as predictors for inflation. Hence, it is pertinent to examine whether this information is valuable for forecasting Brazilian inflation over future horizons.

%%%%%

\vspace{-0.5cm}
\paragraph{Findings.} Among the main results of this paper are:
\vspace{-0.2cm}
\begin{enumerate} \itemsep0em
    \item[(i)] It is challenging to outperform the survey-based forecast (Focus) before the COVID-19 pandemic; however, it is achievable post-pandemic (including at short horizons);
    \item[(ii)] Taking disaggregated inflation into account tends to generate forecasts as good as survey-based expectations and forecasts based on aggregate inflation directly;
    \item[(iii)] Overall, ML methods tend to outperform traditional time series models in predictive accuracy, with outstanding performance in predicting disaggregates;
    \item[(iv)] There exists high variability in the type of predictors selected by the adaLASSO and FarmPredict;
    \item[(v)] The \textit{available} survey-based inflation expectations and price variables are relevant predictors;
    \item[(vi)] Starting from the pandemic, the RF using aggregate inflation or some disaggregation achieves remarkable predictive performance at intermediate and longer horizons.
\end{enumerate}

%%%%%

\vspace{-0.7cm}
\paragraph{Contributions for the literature.} We can summarize the main contributions of this paper in two fields. First, this paper advances the literature on inflation forecasting via aggregation of disaggregated forecasts by considering many predictors for each disaggregate, as well as several statistical and econometric methods underexplored in this literature. Many papers employ traditional time series models and a limited number of predictors. In this context, some papers find evidence favoring the bottom-up approach for the Euro Area \citep{espasa2002, espasa2007} and various countries \citep{bruneau2007, moser2007, capistran2010, aron2012, carlo2016, fulton2021}. On the other hand, some papers find that aggregating forecasts by components does not necessarily improve aggregate inflation forecasting \citep{benalal2004, hubrich2005, hendry2011}. In turn, \citet{duarte2007}, \citet{ibarra2012}, and \citet{bermingham2014} highlight the benefits of aggregating a large number of disaggregates. By employing a large number of predictors, \citet[Chapter 1]{florido2021} points out the benefits of the disaggregated analysis in inflation nowcasting, while \citet{araujo2023} do not find good results by using a disaggregation in multi-period forecasting. We show that the bottom-up approach can generate multi-period forecasts as accurately as survey-based expectations and direct forecasts.

Second, our analysis extends the literature on machine learning (ML) benefits to forecasting inflation by showing a useful application of these methods considering the aggregation of disaggregated forecasts in a data-rich environment. The employ of ML methods to directly forecast inflation started with factor and principal component models \citep{stockwatson1999, stockwatson2002b, forni2003, bai2008, ibarra2012}, and neural network models \citep{moshiri2000, nakamura2005, choudhary2012}. Several other papers expanded the list of methods to shrinkage-based models (e.g., Ridge and LASSO), Bayesian methods, bagging, boosting, random forest (RF), and complete subset regressions (CSR), but keeping focus on forecast inflation directly from the aggregate \citep{inoue2008, medeiros2016br, garcia2017, zeng2017, baybuza2018, medeiros2021, araujo2023}. \citet[Chapter 1]{florido2021} considers ML techniques, disaggregated inflation, and a broad set of predictors in inflation nowcasting, finding good results. \citet{araujo2023} consider the inflation disaggregated into administered prices, services, industrial goods, and food at home to generate multi-horizon inflation forecasts employing several ML models. However, their results are not favorable to the bottom-up approach. In contrast, our combination between disaggregated analysis, ML, and many predictors yields promising results and opens up new possibilities for further exploration.

We also point out five other minor contributions. First, we corroborate the findings of \citet{bacsturk2014}, \citet{altug2016}, \citet{garcia2017}, \citet{fulton2021}, and \citet{banbura2021} regarding the benefits of incorporating a survey-based expectation as a predictor when econometrician computes their forecasts. The presence of this variable is relevant to improve predictive accuracy even for some disaggregation levels. Second, when estimating a factor-augmented autoregression model using a method that allows predictor selection, we find that the factor that summarizes most of the predictors' variability is not relevant for predicting inflation. Hence, using an estimation method such as the adaLASSO or the approaches of \citet{bai2008, bai2009} instead of least squares may be beneficial. Third, our paper is one of the first to employ the FarmPredict, a model proposed by \citet{farmpredict} that combines factor and sparse linear regressions. We adapt it to allow the simultaneous estimation via adaLASSO of a final model containing lags, common factors, and idiosyncratic components. Fourth, as in \citet{duarte2007}, \citet{ibarra2012}, and \citet{bermingham2014}, we also indicate the potential benefits of considering a high level of disaggregation. However, there are caveats about how to improve the bottom-up approach by considering different models predicting different disaggregates. Finally, our analysis underscores the importance of examining sub-periods and emphasizing the benefits of model-based forecasts in volatile periods, as also pointed out by \citet{altug2016} and \citet{medeiros2021}.

%%%%%

\vspace{-0.5cm}
\paragraph{Outline.} This paper has five more sections in addition to this Introduction. Section \ref{sec:methodology} presents the forecasting methodology, and Section \ref{sec:models} describes the models, estimation, metrics, and test to compute and assess the results. Section \ref{sec:data} displays the data and setup. Section \ref{sec:results} presents the results and provides an economic discussion about them. Finally, Section \ref{sec:conclusion} concludes. Appendixes from \ref{append:groups_subgroups} to \ref{append:selected_models} offer supplementary information and complementary results.

%%%%%%%%%%%%%%%%%%%%%%%%%%%%%%%%%%%%%%%%%%

\vspace{-0.1cm}
\section{Forecasting methodology}
\label{sec:methodology}

\vspace{-0.2cm}
%%%%%
\subsection{``Traditional'' inflation forecasting}
%%%%%
\vspace{-0.1cm}

Let $\pi_t$ be the (aggregate) inflation at period $t$. We compute the inflation from the percentage change in a price index based on a typical consumption basket. For forecasting purposes, assume there are $J$ predictors for inflation. Let $\b{z}_t$ be a $J$-dimensional vector of these explanatory variables \textit{observed} at $t$, that is, the information set available to the econometrician to perform the forecasting. Notice that $\b{z}_t$ can contain both the last available realizations of the predictor variables as well as lags of these variables. Lastly, let $\mathcal{M}_{t,h}$ be a time-varying mapping between explanatory variables (predictors) and inflation $h$ periods ahead. As the estimation is based on moving windows, the mapping is dependent on time, which we indicate by the subscript $t$.

%We estimate this mapping based on past data. 

There are several possibilities to estimate the mapping $\mathcal{M}_{t,h}$. Initially, we choose between linear or non-linear specifications. In a rich-data environment, we can consider dimensionality reduction or shrinkage with or without selecting predictors. Whatever the choices, we must be careful to avoid overfitting. Finally, an $h$-period-ahead forecast is given by
\begin{gather*}
    \widehat{\pi}_{t+h\,|\,t} = \widehat{\mathcal{M}}_{t,h}\big(\b{z}_t\big)
\end{gather*}
where hats indicate estimation.

\vspace{-0.2cm}
%%%%%
\subsection{Aggregation of disaggregated forecasts}
%%%%%
\vspace{-0.1cm}

Besides the general price index and their percentage change, the aggregate inflation $\pi_t$, now consider the availability of $N^d$ disaggregated price indexes (subcomponents of the original price index) indexed by $i = 1,\dots, N^d$. The letter $d$ indicates the disaggregation level. Let $\pi_{it}^d$ be the percentage change of the disaggregate $i$ in disaggregation level $d$ at period $t$. Let $\omega_{it}^d$ be the weight of the disaggregate $i$ at disaggregation level $d$ in the general price index at period $t$. Note that these weights are time-varying since the composition of the representative consumption basket may change over time. The relationship between inflation and price changes in disaggregated indexes is given by
\begin{gather} \label{eq:weights}
    \pi_{t} = \sum_{i=1}^{N^d} \omega_{it}^d \, \pi_{it}^d,
\end{gather}
that is, aggregate inflation is a weighted average of ``disaggregated inflations'' (price changes).

Let $\b{z}_{t}^d$ be a $J^d$-dimensional vector of \textit{all} explanatory variables for price changes \textit{observed} at $t$ by the econometrician at disaggregation level $d$. Note that it is expected that $J^d > J$ since in the disaggregated case we potentially have more information: in addition to all the other explanatory variables available in the aggregated case, we can use the lagged price changes of the other disaggregates as predictors for a specific disaggregate. The question arises as to whether capturing and exploring the crossed dependence between disaggregated prices could enhance inflation forecasting. Let $\mathcal{M}_{i,t,h}^d$ be a time-varying mapping between predictors and $h$-period-ahead price variation of each disaggregate $i = 1, \dots, N^d$ at disaggregation level $d$. Following \eqref{eq:weights}, an $h$-period-ahead forecast for aggregate inflation is given by
\begin{gather*}
    \widehat{\pi}_{t+h\,|\,t} = \sum\limits_{i=1}^{N^d} \widetilde{\omega}_{it}^d \, \widehat{\pi}_{i,t+h} = \sum_{i=1}^{N^d} \widetilde{\omega}_{it}^d \, \widehat{\mathcal{M}}_{i,t,h}^d\big(\b{z}_{t}^d\big),
\end{gather*}
where $\widetilde{\omega}_{it}^d$ is the weight of the disaggregate $i$ at disaggregation level $d$ in aggregate index \textit{observed} at $t$ by the econometrician, that is, the \textit{last available} weight at period $t$ and not the weight \textit{evaluate} for the period $t$ -- which we previously indicate simply by $\omega_{it}^d$.

\vspace{-0.1cm}
%%%%%
\subsection{Direct forecasting approach and expanding window scheme}
\label{subseq:direct_ew}
%%%%%
\vspace{-0.1cm}

\begin{figure}[!ht]
    \centering
    \caption{Direct forecasting approach with expanding window scheme}
    \label{fig:ew}
    \vspace{-0.1cm}
    \includegraphics[width=12cm]{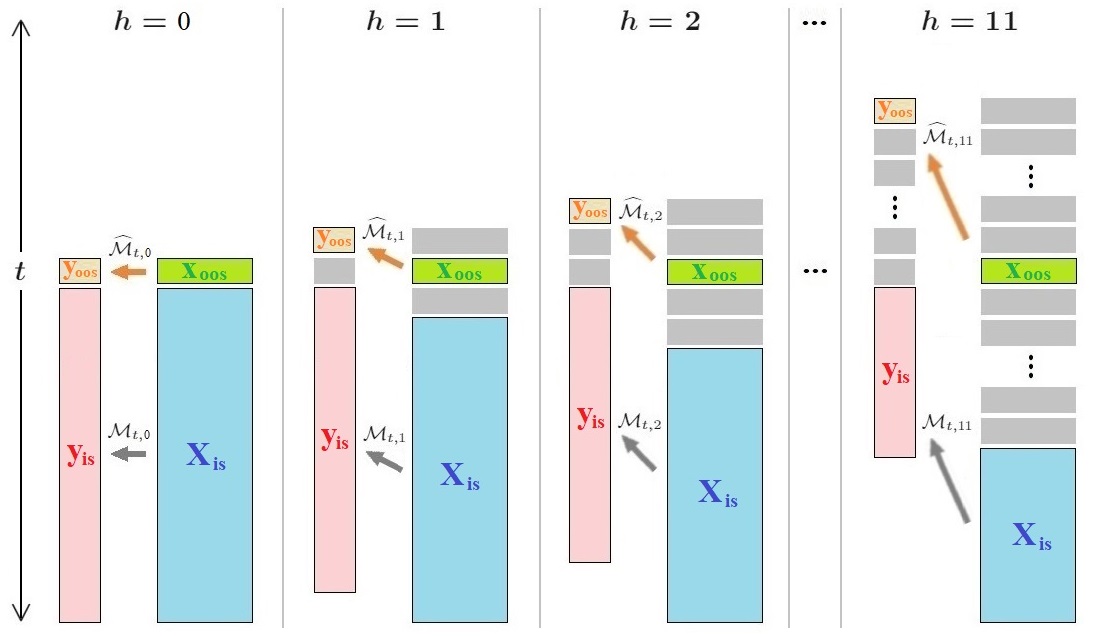}
    \\
    \vspace{0.3cm}
    \footnotesize
    \justifying
    \noindent \textit{Notes:} $y$ indicates the target variable. $\b{X}$ represents predictor variables. Subscripts ``is'' and ``oos'' denote in-sample and out-of-sample, respectively.
\end{figure}

We employ a direct forecast approach considering expanding windows for (monthly) horizons $h \in \{0, 1, \dots, 11\}$. We take the time-adjusted predictors to fit the mapping between them and inflation in this approach. For example, suppose we want to generate a forecast for the current period ($h=0$), which is called \textit{nowcasting}. In that case, we consider the most recently \textit{available} information to estimate the desired mapping. Conversely, when computing a one-month-ahead forecast ($h=1$), we use the information available up until the preceding period in which the forecast is estimated. We continue this way until we calculate the forecast for $h = 11$, utilizing information \textit{available} ten periods prior. Figure \ref{fig:ew} illustrates the exercise. Following the computation of forecasts based on a given period, we advance the time window by one period and repeat the estimation procedure for each forecast horizon, subsequently calculating new forecasts.

%%%%%%%%%%%%%%%%%%%%%%%%%%%%%%%%%%%%%%%%%%

\section{Models and forecast evaluation}
\label{sec:models}

%%%%%
\subsection{Models}
%%%%%

To enhance clarity in presenting the following forecast methods, we omit the superscript $d$ that indicates the level of disaggregation, whenever applicable.

\subsubsection{Benchmarks}

\paragraph{Random walk (RW).} Considering the aggregated case, the forecast of the $h$-period-ahead inflation at period $t$ is given by current inflation, that is, $\widehat{\pi}_{t+h\,|\,t}^{\,\text{RW}} = \pi_t$.

\vspace{-0.4cm}  
\paragraph{Historical mean.} Also for the aggregated case, a prediction for $h$ periods ahead is given by historical average inflation computed at $t$, that is,
\begin{gather*}
    \widehat{\pi}_{t+h\,|\,t}^{\,\text{Hist.\,Mean}} = \overline{\pi}_{t+h\,|\,t} = \frac{1}{S} \sum_{s\,=\,t-S+1}^{t} \pi_{s}
\end{gather*}
where $S$ is the number of previously observed inflation measures (expanding window length).

\vspace{-0.4cm}    
\paragraph{Autoregressive model -- AR($p$).} For both aggregated and disaggregated cases, in the direct forecast approach, for each horizon $h$, we can be written a $p$-order AR model as
\begin{gather*}
    \pi_{t} = \mu + \sum_{l=1}^p \phi_l \, \pi_{t-h-l+1} +  \varepsilon_t
\end{gather*}
where $\varepsilon_t$ is an error term. The order $p$ can be previously fixed or selected via some information criterion (e.g., BIC). Thus, a $h$-period-ahead inflation forecast is given by 
\begin{gather*}
    \widehat{\pi}_{t+h\,|\,t}^{\,\text{AR}} = \widehat{\mu} + \sum_{l=1}^p \widehat{\phi}_l \, \pi_{t-h-l+1}. 
\end{gather*}
where $\widehat{\mu}$ and $\widehat{\phi}$'s are least squares (OLS) estimates.

\vspace{-0.4cm}    
\paragraph{Augmented autoregressive model.} Including seasonal dummies and inflation expectation, we can write the model
\begin{gather} \label{eq:augmented_ar}
    \pi_{t} = \mu + \sum_{l=1}^p \phi_{l} \, \pi_{t-h-l+1} + \eta\;\pi_{t\,|\,t-h}^{e} + \sum_{m=1}^{11} \delta_m \, d_{mt} + \varepsilon_{t}
\end{gather}
where $\pi_{t\,|\,t-h}^{e}$ is the inflation expectation for the period $t$ available at $t-h$, $d_{mt}$ is a seasonal dummy that assumes value 1 for month $m$, and $\delta_{m}$ is a coefficient associated with seasonal dummy $d_{mt}$. In this framework, we estimate the coefficients via OLS, and a $h$-period-ahead forecast is given by
\begin{gather*}
    \widehat{\pi}_{t+h\,|\,t}^{\,\text{Aug.\,AR}} = \widehat{\mu} + \sum_{l=1}^p \widehat{\phi_{l}} \, \pi_{t-l+1} + \widehat{\eta}\;\pi_{t+h\,|\,t}^{e} + \sum_{m=1}^{11} \widehat{\delta}_m \, d_{m,t+h}.
\end{gather*}
    
\vspace{-0.5cm}    
%\newpage
\paragraph{(Empirical) Hybrid New Keynesian Phillips curve (HNKPC).} Following and \textit{adapting} price-setting models such as those presented in \citet{galigertler1999} and \citet{blanchard2007}, we employ a forecasting model for the aggregate inflation based on a hybrid Phillips curve given by
\begin{gather*}
    \pi_{t} = \mu + \sum_{l=1}^p \phi_l \, \pi_{t-h-l+1} + \eta\, \pi_{t\,|\,t-h}^{e} + \psi_1 \, g_{t-h} + \psi_2 \, \Delta s_{t-h} + \varepsilon_t
\end{gather*}
where $g_{t-h}$ is some economic activity measure \textit{observed} at $t-h$, and $\Delta s_{t-h}$ is an exchange rate measure \textit{observed} at $t-h$. We compute the forecast by
\begin{gather*}
    \widehat{\pi}_{t+h\,|\,t}^{\,\text{HNKPC}} = \widehat{\mu} + \sum_{l=1}^p \widehat{\phi}_l \, \pi_{t-l+1} + \widehat{\eta} \, \pi_{t+h\,|\,t}^{e} + \widehat{\psi}_1 \, g_{t} + \widehat{\psi}_2 \, \Delta s_{t} %\qquad h = 1, \dots, 12
\end{gather*}
where $\big(\widehat{\mu},\,\widehat{\b{\phi}},\,\widehat{\eta},\,\widehat{\psi}_1,\,\widehat{\psi}_2\big)$ are OLS estimates.

%%%%%
\subsubsection{Shrinkage-based models}  

\paragraph{Ridge (with incomplete information).} For disaggregated cases, we consider the augmented AR model \eqref{eq:augmented_ar} with the addition of other lagged disaggregates:
\begin{gather*}
    \pi_{it} = \mu + \sum_{i'=1}^{N^d} \sum_{l=1}^p \phi_{i'l} \, \pi_{i',t-h-l+1} + \eta_i\,\pi_{t\,|\,t-h}^{e} + \sum_{m=1}^{11} \delta_{im} \, d_{mt} + \varepsilon_{it} \qquad i = 1, \dots, N^d,
\end{gather*}
where $N^d$ is the number of subcomponents in the disaggregation level indicated by $d$. We consider four disaggregation levels in this paper: aggregate inflation, economic categories defined by the BCB, and groups and subgroups from IPCA (IBGE).

We estimate the coefficients employing the Ridge estimator:
\begin{gather*}
    \left(\widehat{\mu}_i, \, \widehat{\b{\beta}}_{\text{Ridge}, i}(\lambda)\right) = \argmin_{\mu_i,\,\b{\beta}_i} \left\{ \frac{1}{T-h} \sum_{t=1}^{T-h} \left( \pi_{it} - \mu_i - \b{\beta}_i \, \b{z}_{t-h} \right)^2 + \lambda_i \sum_{j=1}^J \,\beta_{ij}^2 \right\}
\end{gather*}
where $\lambda_i$ is a regularization parameter, $\b{z}_{t-h}$ is a vector with all predictors, and $\b{\beta}_i$ is a vector of coefficients. Chosen $\lambda$ via information criteria (e.g., BIC), a prediction for $h$ periods ahead is given by
\begin{gather*}
    \widehat{\pi}_{t+h\,|\,t}^{\,\text{Ridge}} = \sum_{i=1}^{N^d} \omega_{it} \, \widehat{\pi}_{i,t+h\,|\,t}^{\,\text{Ridge}} \qquad \text{with} \qquad \widehat{\pi}_{i, t+h\,|\,t}^{\,\text{Ridge}} = \widehat{\mu}_i + \widehat{\b{\beta}}_{\text{Ridge},i}\,\b{z}_t.
\end{gather*}
%where $\widehat{\b{\beta}}_i$ is the Ridge estimate and $\b{z}_t$ is a vector of more recent observations of the explanatory variables.

\vspace{-0.5cm}    
\paragraph{adaLASSO (with full information).} For all cases, consider the model with full information given by
\begin{gather*}
    \pi_{it} = \mu_i + \sum_{i'=1}^{N^d} \sum_{l=1}^p \phi_{i'l} \, \pi_{i',t-h-l+1} + \eta_i\,\pi_{t\,|\,t-h}^{e} + \sum_{m=1}^{11} \delta_{im} \, d_{mt} + \sum_{j=1}^J \sum_{l=1}^p \theta_{ijl}\,x_{j,t-h-l+1} +
    \varepsilon_{it}
\end{gather*}
where $\b{x}_{t-h} \in \R^{J \cdot p}$ is an expanded vector of potential predictors for $\pi_{it}$. We estimate this model employing the adaptive LASSO (adaLASSO). Introduced by \citet{zou2006}, this method selects predictors and their optimization problem is given by
\begin{gather*}
    \left(\widehat{\mu}_i, \, \widehat{\b{\beta}}_{\text{adaLASSO}}(\lambda, \b{\omega})\right) = \argmin_{\mu_i,\,\b{\beta}_i}
    %\,\in\,\R^{J+1}} 
    \left\{ \frac{1}{T-h} \sum_{t=1}^{T-h} \left( \pi_{it} - \mu_i - \b{\beta}_i \, \b{z}_{t-h} \right)^2 + \xi \sum_{j=1}^V \zeta_{ij} \, |\beta_{ij}| \right\}
\end{gather*}
where $\xi$ is a regularization parameter, $\b{z}_{t-h} \in \R^V$, $V = N^d \cdot p + 12 + J \cdot p$, is a vector of \textit{all} predictors, and $\b{\zeta} = (\zeta_1, \dots, \zeta_V)$ is a vector of weights obtained previously employing LASSO -- a estimator that assumes $\zeta_{ij} = 1$, for all $j$. More precisely, we compute the weights via
\begin{gather*}
    \zeta_{ij} = \left(\left|\widehat{\beta}_{\text{LASSO},ij} \right| + \frac{1}{\sqrt{T}} \right)^{-1},
\end{gather*}
where we add $T^{-1/2}$ to allow a variable that is not selected in the first stage to have a chance of being selected in the second stage.

As before, a $h$-period-ahead forecast is $\widehat{\pi}_{i,t+h\,|\,t}^{\text{adaLASSO}} = \widehat{\mu}_i + \widehat{\b{\beta}}_{\,\text{adaLASSO},i}\,\b{z}_t$.

%%%%%
\subsubsection{Factor models}

\paragraph{(Augmented) factor model.} Consider that all regressors are normalized for both aggregate and disaggregate cases. Thus, for $i = 1, \dots, N^d$, a factor-augmented autoregression model is described by
\begin{gather}
    \label{eq:factor_pca}
    \b{x}_{t} = \sum_{k=1}^K \b{\lambda}_{k} \, f_{kt} + \b{u}_{t}
    \\
    \nonumber
    \pi_{it} = \mu_i + \sum_{i'=1}^D \sum_{l=1}^p \phi_{i'l} \, \pi_{i',t-h-l+1} + \eta_i \, \pi_{t\,|\,t-h}^{e}
    + \sum_{m=1}^{11} \delta_{im} \, d_{mt} + \sum_{k=1}^K \beta_{ik} \, \widehat{f}_{k,t-h} +  \varepsilon_{it}
\end{gather}
from which we compute common factors $\widehat{\b{f}}_{t} = \left(\widehat{f}_{1t},\dots,\widehat{f}_{Kt}\right)$ and factor loadings $\b{\lambda}_k = \left(\lambda_{1k},\dots,\lambda_{Jk}\right)$ by combining principal component analysis (PCA) and OLS. Finally, we compute $\big(\widehat{\mu}_i, \, \widehat{\b{\phi}}, \, \widehat{\eta}_i, \, \widehat{\b{\beta}}\big)$ via adaLASSO.

For identification purposes, we assume that 
\begin{gather*}
    \E(\b{f}_t\,|\,\b{u}_t) = 0, \quad \Cov(\b{u}_t, \varepsilon_t) = 0, \quad \var(\b{f}_t) = \b{I}_K, \quad \var(\b{u}_t) = \b{\Omega} = \diag(\sigma_1^2, \dots, \sigma_p^2).
\end{gather*}
The number of factors $K$ is selected via information criterion $\text{IC}_{p2}$ of \citet{bai2002}, and the forecast $h$ periods ahead is given by
\begin{gather*}
    \widehat{\pi}_{i,\,t+h\,|\,t}^{\,\text{Factor}} = \widehat{\mu}_i + \sum_{i'=1}^D \sum_{l=1}^p \widehat{\phi}_{i'l}\,\pi_{i',t-l+1} + \widehat{\eta}_i \, \pi_{t+h\,|\,t}^{e} + \sum_{m=1}^{11} \widehat{\delta}_{im} \, d_{m,t+h} + \sum_{k=1}^K \widehat{\beta}_{ik} \, \widehat{f}_{kt}
\end{gather*}
where $\widehat{f}_{kt}$ is the $k$-th factor evaluated at $t$.

\vspace{-0.5cm}
%\newpage
\paragraph{Target factor model.} Proposed by \citet{bai2008}, in this ``hard thresholding'' version, this approach controls for the participation of normalized explanatory variables in the factor construction. In a previous stage, for each predictor indexed by $j = 1, \dots, J$, and disaggregate indexed by $i = 1, \dots, N^d$, we estimate
\begin{gather*}
    \pi_{it} = \mu_i + \sum_{i'=1}^{N^d} \sum_{l=1}^p \phi_{i'l} \, \pi_{i', \,t-l} + \eta_i\,\pi_{t\,|\,t-h}^{e} + \sum_{m=1}^{11} \delta_{im} \, d_{mt} + \theta_{ij} \, x_{j,t-h} + \nu_{it}
\end{gather*}
and run the hypothesis test \, $\theta_{ij} = 0 \, \times \, \theta_{ij} \neq 0$ \, for some significance level $\alpha$. If $\theta_{ij}$ is statistically different from zero, we employ $x_j$ in the factor estimation. 
Let $\b{x}_t(\alpha, i)$ be the set of selected variables for $i$-th disaggregation. Finally, we proceed as in the traditional factor-augmented autoregressive model: we perform
\begin{gather*}
    \b{x}_{t}(\alpha, i) = \sum_{k=1}^K \b{\lambda}_k \, f_{kt} + \b{u}_{t}
\end{gather*}
which $\widehat{\b{f}}_{t}$ and $\widehat{\b{\lambda}}_k$ are computed via PCA and OLS. Then we estimate the augmented (target) factor model via adaLASSO and compute the forecast as before.

%%%%%
\subsubsection{FarmPredict}

Some idiosyncratic errors of the factor model, that is, some $\b{u}_t$ entries in Equation \eqref{eq:factor_pca}, can impact the price variation, which the common factor structure does not capture. Defining $\widehat{\b{u}}_{t} = \b{x}_t - \sum_{k=1}^{K} \widehat{\b{\lambda}}_{k} \, \widehat{f}_{k,t}$, a $J$-dimensional vector, we can introduce lags of $\b{u}_t$ on the factor model: 
\begin{align} \nonumber
    \pi_{it} = \mu_i \, + \, \sum_{i'=1}^{N^d} \sum_{l=1}^p \phi_{i'l}\,&\pi_{i',t-h-l+1} +  \gamma_i\,\pi_{t\,|\,t-h}^{e} + \sum_{m=1}^{11} \delta_{im} \, d_{mt} \\
    \label{eq:farmpredict}
    &+ \sum_{k=1}^K \sum_{l=1}^p \beta_{ikl} \, \widehat{f}_{k,t-h-l+1} 
    + \sum_{j=1}^J \sum_{l=1}^p \theta_{ijl}\,\widehat{u}_{j,t-h-l+1} + \varepsilon_{it}.
\end{align}

This model is a specific form of a general model called FarmPredict proposed by \citet{farmpredict}. Here, we estimate the ``final equation'' \eqref{eq:farmpredict} with all regressors \textit{simultaneously} employing the adaLASSO. Next, we compute the forecast.

%%%%%
\subsubsection{Complete subset regression (CSR)}

Introduced by \citet{elliott2013, elliott2015}, this ensemble method combines estimates from all (or several) possible linear regression models, keeping the number of predictors fixed. Let $p$ be the total available predictors and $k \leqslant p$ be the number of ``selected'' predictors (complete subsets). The CSR involves the estimation of $\frac{\textstyle k!}{\textstyle (k-p)!k!}$ linear models. Variables when ``non-selected'' has their coefficients set to zero. The final CSR estimate is the average of all estimates. Thus, subset regression has a shrinkage interpretation since when averaging parameters that sometimes assume zero value, this average generates shrunken estimates of the coefficients, which can contribute to more accurate forecasts. Due to the high computational cost arising from a large number of predictors, we (pre-)select $\widetilde{p} \leqslant p$ predictors based on a ranking of t-statistics in absolute value, as in \citet{garcia2017} and \citet{medeiros2021}. This procedure is similar to that used in the target factor model. So, instead of considering all available $p$ predictors, we run the CSR considering $\widetilde{p}$ pre-selected predictors.

%%%%%
\subsubsection{Random forest (RF)}

\begin{figure}[!h]
    \caption{A regression tree with two explanatory variables $(X_1, X_2)$}
    \label{fig:rf}
    \centering
    \vspace{-0.35cm}
    \includegraphics[width=10cm]{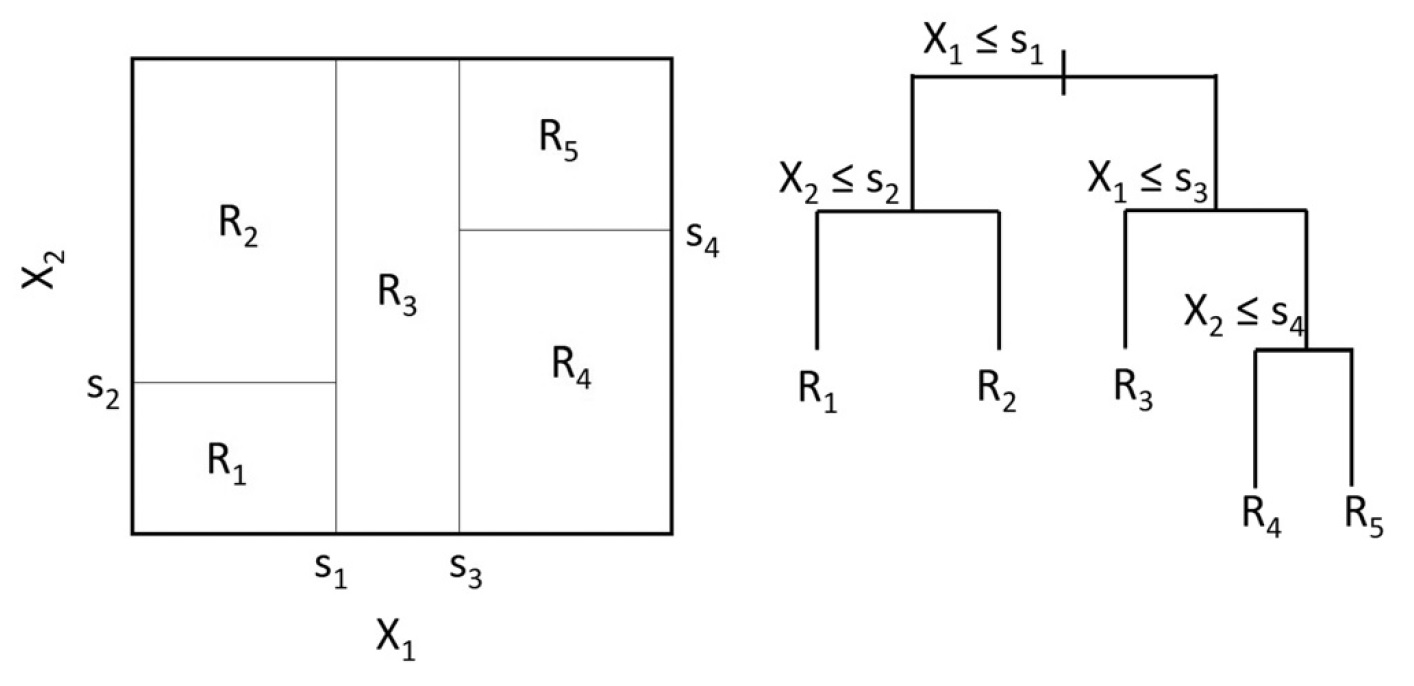}
    \\
    %\begin{center}
        \footnotesize
        \justifying
        \vspace{-0.1cm}
        \noindent \textit{Notes:} $s_i, \; i = 1, \dots, 4$, indicate splits, while $R_k, \; k=1,\dots,5$, denote regions. Example extracted from \citet{medeiros2021}.
    %\end{center}
\end{figure}

\citet{breiman2001} introduces the random forest (RF), a model that combines several based-tree regressions using bagging. A regression tree is a nonparametric model that approximates an unknown nonlinear function with local predictions via recursive partitioning, as illustrated in Figure \ref{fig:rf}.

Formally, a regression tree model can be written as follows:
\begin{gather*}
    \pi_{it} = \sum_{k=1}^K c_k\,\I_k(\b{x}_{t-h} \in R_k)
\end{gather*}
where $\I_k(\b{x}_{t-h} \in R_k)$ is an indicator function that assumes the value 1 when $\b{x}_{t-h}$ belongs to the $k$-th region $R_k$, and $c_k$ is the average of $\pi_t$ in this region. We have to set the minimum number of observations per region. Then, we obtain $B$ trees by implementing a double draw: we draw on the observation dimension using block bootstrap, and we draw variables to incorporate in the estimation of the tree. The idea is that this double draw will ensure the variability of the trees. Let $K_b$ be the number of regions of the $b$-th tree, $b = 1, \dots, B$. Lastly, the final forecast is given by the average of the forecasts obtained by each tree evaluated in the original data, that is,
\begin{gather*}
    \widehat{\pi}_{i,\,t+h\,|\,t}^{\,\text{RF}} = \frac{1}{B}\sum_{b=1}^B \sum_{k=1}^{K_b} \widehat{c}_{k,b} \,\I_{k,b}(\b{x}_{t-h} \in R_{k,b})
\end{gather*}
where $R_{k,b}$ is the $k$-th region of the $b$-th tree.

\vspace{-0.1cm}
%%%%%
\subsection{Model combinations via average of forecasts}
%%%%%
\vspace{-0.1cm}

Methods may perform differently for distinct disaggregates or even over time for the same disaggregate. To mitigate instabilities associated with some method for some disaggregate or at some point in time, for each disaggregation, we will compute a combined forecast given by the average of forecasts generated by all methods applied to this disaggregation and Focus expectations \textit{available} when the econometrician computes their forecasts, that is,
\begin{gather*} %\label{eq:comb}
    \widehat{\pi}_{t+h\,|\,t}^{\,\text{Comb},\,d} = \frac{1}{M^d+1} \left(\sum_{m=1}^{M^d} \sum_{i=1}^{N^d} \widetilde{\omega}_{it}^d \, \widehat{\pi}_{i,t+h\,|\,t}^{\,m,\,d} + \pi_{t+h\,|\,t}^{e} \right),
\end{gather*}
where $d$ indicates one of four possible disaggregations levels addressed in this paper (aggregate inflation, BCB categories, IBGE groups, and IBGE subgroups), $m$ indicates a method, $M^d$ is the number of methods employed to forecast the inflation for the disaggregation $d$, $N^d$ is the number of disaggregates in the disaggregation $d$, and $\widetilde{\omega}_{it}^d$ is the weight of disaggregate $i$ of the disaggregation level $d$ in the aggregate index \textit{observed} at $t$ by the econometrician. The idea is to investigate whether this simple combination leads to improvements in forecast performance.

%%%%%
\subsection{Evaluation: metrics and test}
%%%%%

\paragraph{Metrics.} We use out-of-sample root mean squared error (RMSE) as the main metric to evaluate the forecast performance. For each horizon $h$, this metric is described by
\begin{gather*}
    \text{RMSE}_{h}^{\,m,\,d} = \left[\frac{1}{T} \sum_{t=1}^T\big(\pi_{t+h} - \widehat{\pi}_{t+h\,|\,t}^{\,m,\,d}\big)^2\right]^{1/2}
    \end{gather*}
where $\widehat{\pi}_{t+h\,|\,t}^{\,m,\,d}$ indicates a forecast generated by the model $m$ considering the disaggregation level $d$. The smaller the $\text{RMSE}_{h,\,m}$, the better the model's predictive performance. For the Diebold-Mariano test, we consider the mean squared error (MSE) defined by
\begin{gather*}
    \text{MSE}_{h}^{\,m,\,d} = \frac{1}{T} \sum_{t=1}^T\big(\pi_{t+h} - \widehat{\pi}_{t+h\,|\,t}^{\,m,\,d}\big)^2.
\end{gather*}

%We use out-of-sample R-squared ($R^2_{\text{oos}}$) and mean squared error (MSE) as main metric to evaluate forecast performance. For each forecast horizon $h$, this metric is described by
%\begin{gather*}
%    R^2_{\text{oos},\,h} = 1 - \frac{\sum_{t=1}^T \big(\pi_{t+h} - \widehat{\pi}_{t+h\,|\,t}\big)^2}{\sum_{t=1}^T \big(\pi_{t+h} - \overline{\pi}_t\big)^2}
%\end{gather*}
%where $\overline{\pi}_t = \frac{1}{S} \sum_{s\,=\,t-S}^{t-1} \pi_{s}$ is an average computed from the inflation observed in the rolling window ended at $t$. The higher the $R^2_{\text{oos}}$, the better the model's predictive performance. Notice that $R^2_{\text{oos}}$ compares a model and the historical average. If $R^2_{\text{oos}} > 0$, then the forecast performance of the model is superior to the historical average. By construction, the $R^2_{\text{oos}}$ of the historical average is zero. 

%The lower the MSE, the better the model's predictive performance. Notice that the MSE is contained in the second term of $R^2_{\text{oos}}$ (multiplied by $T$). As this second term of $R^2_{\text{oos}}$ is negative, the ordering of best models in the MSE is the inverse of the ordering in $R^2_{\text{oos}}$.

%\noindent \textit{Out-of-sample Mean Absolute Error (MAE).} \, Finally, the third metric is given by
%\begin{gather*}
%    \text{MAE}_{h} = \frac{1}{T} \sum_{t=1}^T\big{|}\pi_{t+h} - \widehat{\pi}_{t+h\,|\,t}\big{|}.
%\end{gather*}

\vspace{-0.5cm}
\paragraph{Test.} To assess the results, we consider the widely employed test developed by \citet{dmtest}. Let $\widehat{v}_{t+h\,|\,t}^{\,m} = \pi_{t+h} - \widehat{\pi}_{t+h\,|\,t}^{\,m}$ be a forecast error of the model $m$. Here, we omit the disaggregation level $d$. Let $g(\cdot)$ be a metric to be applied to $\widehat{v}_{t+h\,|\,t,\,m}$ (e.g., MSE). The Diebold-Mariano (DM) test statistic is given by
\begin{gather*}
    d_{m,m'} = \frac{1}{T} \sum_{t=1}^T \left(g\big(\widehat{v}_{t+h\,|\,t}^{\,m}\big) - g\big(\widehat{v}_{t+h\,|\,t}^{\,m'}\big)\right)
\end{gather*}
where $m'$ indicates another model, a competitor (i.e., a benchmark model or specific forecast, for example). We will consider that the normality of DM statistics is likely a trustworthy approximation, including for model-based forecasts. %Diebold and Mariano show the asymptotic and finite-sample convergences of this statistic to proceed with the test.

%\red{Model Confidence Set (MCS).}

%%%%%%%%%%%%%%%%%%%%%%%%%%%%%%%%%%%%%%%%%%

\section{Data and setup}
\label{sec:data}

\paragraph{Data.} We analyze the period from January 2004 to June 2022, totalizing 18,5 years of monthly data. For aggregate inflation, we employ the IPCA, the official Brazilian price index computed by the Brazilian Institute of Geography and Statistics (\textit{Instituto Brasileiro de Geografia e Estatística} -- IBGE). For disaggregations, we consider all groups and subgroups of the IPCA. There are nine groups and 19 subgroups throughout the period analyzed. Subgroups are subdivisions and, in some cases, the group itself. For definition of groups and subgroups, and their respective average weights in the IPCA, see Table \ref{tab:groups_subgroups} in Appendix \ref{append:groups_subgroups}. In addition, we use a disaggregation defined by the Central Bank of Brazil (BCB) based on IBGE data. The BCB disaggregation consists of administrated, non-tradables, and tradables items. The use of this last disaggregation is interesting because, in principle, it presents more economic intuition, which can contribute to better forecast performance.

%Namely, the nine groups are foods and beverages, housing, household goods, apparel, transportation, medical and personal care, personal expenses, education, and communication. 

We consider inflation expectations of the Central Bank of Brazil's Focus survey and lags of the predicted variables among the admissible predictors. To forecast a disaggregate, we consider lags of other disaggregates in the same disaggregation, which allows capturing potential lagged ``cross-effects''. The Focus survey has a daily frequency and contains inflation expectations formed by many economic agents (experts) for several horizons (months) ahead. Reflecting the opinion of experts, the Focus may contain private information that is not available to the econometrician -- hence the importance of considering this variable in our information set. We consider the latest available inflation expectation for the horizon of interest when we generate our forecast. Moreover, there are eighty-nine other predictor variables (and their lags) divided into ten categories: prices and money (17), commodities prices (4), economic activity (19), employment (5), electricity (4), confidence (3), finance (12), credit (4), government (12), and exchange and international transactions (9). In Appendix \ref{append:variables}, Table \ref{tab:variables} presents a description of these variables, the delay for each to become \textit{available} and transformations implemented to guarantee the stationarity. We structure our dataset to closely approximate real-time data for the Brazilian case. The main challenge lies in our lack of access to the first releases of some economic activity variables, including the IBC-Br (a Brazilian Economic Activity Index) and industrial production (and their subcomponents). %Figure \ref{fig:variables_corr} contains an illustration of correlations between \textit{contemporaneously available} variables.

\vspace{-0.5cm}
\paragraph{Setup.} The reference day to compute our forecasts is the last business day of each month. For the results shown in the following section, we consider three lags for all predictive variables, including variables mentioned above, factors in factor models, idiosyncratic components in FarmPredict, and lags of aggregate and all disaggregates. The only exception is the factors in the target factor model for which we employ only one (target) factor. As mentioned in Subsection \ref{subseq:direct_ew}, the main results are generated based on expanding windows. In this setup, we generate 114 forecasts for each horizon. The regularization parameters ($\lambda$'s) of the Ridge, LASSO, and adaLASSO are obtained via Bayesian Information Criterion (BIC). We restrict the number of possible selected variables by the ceiling of $\sqrt{T}$ to enforce discipline. The number $K$ of latent factors in factor models is selected via \citet{bai2002} information criterion $IC_{p2}$. For CSR, we set $\widetilde{p} = 20$ (number of pre-selected predictors) and $p = 4$ (number of selected variables by CSR). For pre-selecting of both target factor and CSR models, we adopt the 5\% significance level ($\alpha = 0.05$). In its turn, for the RF models, we allow the trees to grow until five observations by leaf. We set the proportion of selected variables in each split to 1/3 and the number of bootstrap samples to 500 ($B = 500$). All settings are similar to those adopted by \citet{garcia2017} and \citet{medeiros2021}. Finally, to estimate the empirical Phillips curve, we use the Central Bank of Brazil's economic activity index (IBC-Br) and BIS' real effective exchange rate (REER) as a proxy for economic activity and exchange rate, respectively.

%The decrease in forecasts by horizon is due to the increase in lags used as predictors and the lack of actual values to verify the predictive performance. \red{We also did robustness exercises in which we consider the 30th day as the reference day instead of the 15th, three lags instead of two lags, and rolling windows of 9 years instead of 10 years.}

%%%%%%%%%%%%%%%%%%%%%%%%%%%%%%%%%%%%%%%%%%

\section{Results}
\label{sec:results}

%%%%%
\subsection{Entire period: forecasts from January 2014 to June 2022}
%%%%%

Table \ref{tab:rmse_allsample} exhibits the results of forecast performance in terms of root mean squared errors (RMSE) for different models and horizons ranging from nowcasting ($h=0$) to eleven months ahead ($h=11$), as well as for 12-month accumulated inflation. We normalize every RMSE to relative terms by computing their ratio to the RMSE of the Focus consensus -- the median expectation of the available Focus survey. Thus, a value lower than one indicates that a model numerically outperforms the Focus consensus, while a value greater than one suggests underperformance compared to the same benchmark. At this first moment, the results consider the entire period for which we compute predictions, from January 2014 to June 2022. Each panel of Table \ref{tab:rmse_allsample} considers a group of competitors. In panel A, we have the available and \textit{ex-post} Focus, the latter released by the Central Bank in the following week reflecting the experts' opinions on the same day we compute our forecasts. We note virtually no difference between the available and ex-post Focus for longer horizons. However, in the short term ($h \leqslant 3$), there is evidence that ex-post Focus statistically outperforms available Focus. Despite being only a few days apart, the informational gain is considerable for shorter horizons, which does not occur for more distant periods since it is unlikely that very relevant information about them will emerge within a few days.

% latex table generated in R 4.1.3 by xtable 1.8-4 package
% version adapted by Gilberto Boaretto
% Fri Mar 17 19:53:31 2023
\begin{table}[!ht]
\centering
\caption{Out-of-sample RMSE with respect to the available Focus: Jan/2014 to Jun/2022} 
\label{tab:rmse_allsample}
\resizebox{1\linewidth}{!}{
\begin{tabular}{llllllllllllll}
  \toprule
\multicolumn{1}{c}{Estimator/Model} & $h = 0\quad$ & $h = 1\quad$ & $h = 2\quad$ & $h = 3\quad$ & $h = 4\quad$ & $h = 5\quad$ & $h = 6\quad$ & $h = 7\quad$ & $h = 8\quad$ & $h = 9\quad$ & $h = 10\;\,$ & $h = 11\;\,$ & $\Sigma \, 12m$ \\ 
  \cmidrule(lr){1-14}
\vspace{-0.3cm} \\ \multicolumn{14}{c}{\textbf{\underline{A. Survey}}} \\ \vspace{-0.3cm} &  &  &  &  &  &  &  &  &  &  &  &  &  \\ 
  Focus (available) & 1.000 & 1.000 & 1.000 & 1.000 & 1.000 & 1.000 & 1.000 & 1.000 & 1.000 & 1.000 & 1.000 & 1.000 & 1.000 \\ 
  Focus (\textit{ex-post}) & $\textbf{\blue{0.933}}^{\,***}$ & $0.972^{\,***}$ & $0.993^{\,***}$ & 1.001 & 1.000 & 1.000 & 1.000 & 0.999 & 0.999 & 1.000 & 1.001 & 1.001 & 0.999 \\ 
  \vspace{-0.3cm} \\ \multicolumn{14}{c}{\textbf{\underline{B. Aggregate inflation}}} \\ \vspace{-0.3cm} &  &  &  &  &  &  &  &  &  &  &  &  &  \\ 
  RW & 2.782 & 1.471 & 1.212 & 1.251 & 1.279 & 1.271 & 1.234 & 1.180 & 1.087 & 1.063 & 1.102 & 1.138 & 1.485 \\ 
  Historical Mean & 3.117 & 1.342 & 1.084 & 1.053 & 1.034 & 1.018 & 0.998 & 0.990 & 0.987 & 0.983 & 0.982 & 0.984 & 1.041 \\ 
  AR & 2.534 & 1.260 & 1.053 & 1.077 & 1.090 & 1.064 & 1.008 & 0.961 & $0.939^{\,*}$ & $0.941^{\,*}$ & \textit{\blue{0.964}} & 0.988 & 0.987 \\ 
  HNKPC & $\textit{\blue{0.946}}^{\,**}$ & $\textit{\blue{0.957}}^{\,**}$ & $0.972^{\,*}$ & \textit{\blue{0.981}} & 1.036 & 1.030 & 0.980 & 0.968 & $0.963^{\,*}$ & 0.974 & 0.998 & 1.021 & 0.984 \\ 
  Augmented AR & 0.970 & 0.981 & 0.989 & 1.012 & 1.023 & 0.999 & 0.963 & $\textit{\blue{0.943}}^{\,*}$ & $\textit{\blue{0.936}}^{\,**}$ & 0.961 & 0.993 & 1.016 & $0.954^{\,*}$ \\ 
  adaLASSO & $\textit{\blue{0.944}}^{\,**}$ & $\textit{\blue{0.948}}^{\,***}$ & $0.978^{\,*}$ & 0.993 & $\textit{\blue{0.977}}^{\,**}$ & 0.989 & 0.976 & 0.976 & 0.989 & 1.004 & 1.043 & 0.992 & $0.961^{\,*}$ \\ 
  Factor & $\textit{\blue{0.943}}^{\,**}$ & $\textit{\blue{0.931}}^{\,***}$ & $\textit{\blue{0.966}}^{\,**}$ & $\textbf{\blue{0.964}}^{\,**}$ & \textit{\blue{0.974}} & 0.980 & 0.989 & 0.986 & 1.016 & 1.016 & 1.052 & 1.045 & 0.989 \\ 
  FarmPredict & $\textit{\blue{0.944}}^{\,**}$ & $\textit{\blue{0.942}}^{\,***}$ & $\textit{\blue{0.961}}^{\,**}$ & 0.987 & 0.999 & 1.001 & 1.004 & 1.020 & 1.020 & 1.032 & 1.027 & 1.051 & 1.001 \\ 
  Target Factor & 1.280 & 1.129 & 1.045 & 1.155 & 1.122 & 1.069 & $\textbf{\blue{0.937}}^{\,*}$ & 0.989 & 0.990 & 0.980 & 1.019 & 1.102 & 0.970 \\ 
  CSR & 0.961 & $\textbf{\blue{0.927}}^{\,**}$ & $\textit{\blue{0.962}}^{\,*}$ & 0.983 & 1.022 & 0.996 & 0.969 & $\textit{\blue{0.940}}^{\,**}$ & $0.940^{\,**}$ & 0.975 & 1.042 & 1.071 & $0.935^{\,**}$ \\ 
  Random Forest & 1.665 & 1.110 & 1.025 & 1.012 & 0.996 & 0.987 & 0.959 & 0.949 & $\textbf{\blue{0.920}}^{\,**}$ & $\textit{\blue{0.918}}^{\,**}$ & $\textit{\blue{0.908}}^{\,**}$ & $\textbf{\blue{0.909}}^{\,**}$ & $\textit{\blue{0.918}}^{\,*}$ \\ 
  \vspace{-0.3cm} \\ \multicolumn{14}{c}{\textbf{\underline{C. Disaggregation: tradable, nontradable and monitored prices (BCB)}}} \\ \vspace{-0.3cm} &  &  &  &  &  &  &  &  &  &  &  &  &  \\ 
  AR & 2.579 & 1.241 & 1.039 & 1.057 & 1.089 & 1.085 & 1.015 & 1.037 & 0.990 & 0.989 & 1.026 & 1.036 & 1.050 \\ 
  Augmented AR & 0.978 & 0.978 & 1.012 & 1.061 & 1.094 & 1.072 & 1.010 & 1.010 & 0.982 & 1.003 & 1.052 & 1.073 & 1.048 \\ 
  Ridge & 1.018 & 1.015 & 1.007 & 1.019 & 1.045 & 1.083 & 0.999 & 0.978 & 0.969 & $0.960^{\,*}$ & 1.013 & 0.981 & 1.032 \\ 
  adaLASSO & 1.099 & 0.998 & \textit{\blue{0.968}} & \textit{\blue{0.969}} & \textit{\blue{0.977}} & $\textit{\blue{0.948}}^{\,**}$ & $0.962^{\,*}$ & $0.957^{\,*}$ & 0.979 & 0.964 & 0.972 & $\textit{\blue{0.957}}^{\,*}$ & $\textit{\blue{0.903}}^{\,**}$ \\ 
  Factor & 1.023 & 0.988 & 0.991 & 0.990 & \textit{\blue{0.977}} & $\textbf{\blue{0.934}}^{\,**}$ & $\textit{\blue{0.940}}^{\,**}$ & 0.977 & 1.000 & 1.007 & 1.009 & 1.009 & $0.943^{\,*}$ \\ 
  FarmPredict & 1.129 & 1.024 & 1.003 & \textit{\blue{0.981}} & \textbf{\blue{0.969}} & $\textit{\blue{0.944}}^{\,**}$ & $0.956^{\,*}$ & 0.984 & 1.003 & 1.009 & 1.008 & 0.991 & 0.963 \\ 
  Target Factor & 1.331 & 1.052 & 0.976 & 1.090 & 1.131 & 1.079 & \textit{\blue{0.943}} & 0.962 & 0.964 & 0.991 & 1.016 & 1.077 & 0.936 \\ 
  CSR & 2.176 & 1.178 & 1.016 & 1.015 & 1.018 & 0.983 & 0.957 & 0.977 & 0.959 & 0.967 & 1.022 & 1.052 & 0.952 \\ 
  Random Forest & 1.934 & 1.177 & 1.020 & 1.051 & 1.014 & 1.001 & 0.976 & 0.963 & $\textit{\blue{0.930}}^{\,*}$ & $\textbf{\blue{0.912}}^{\,**}$ & $\textbf{\blue{0.902}}^{\,**}$ & $\textit{\blue{0.915}}^{\,**}$ & \textit{\blue{0.923}} \\ 
  \vspace{-0.3cm} \\ \multicolumn{14}{c}{\textbf{\underline{D. Disaggregation: groups (IBGE)}}} \\ \vspace{-0.3cm} &  &  &  &  &  &  &  &  &  &  &  &  &  \\ 
  AR & 2.731 & 1.296 & 1.048 & 1.063 & 1.054 & 1.032 & 1.067 & 1.087 & 1.118 & 1.145 & 1.105 & 1.095 & 1.067 \\ 
  Augmented AR & 1.004 & 1.046 & 1.042 & 1.056 & 1.088 & 1.046 & 1.067 & 1.065 & 1.069 & 1.105 & 1.134 & 1.180 & 1.028 \\ 
  Ridge & 1.656 & 1.288 & 1.044 & 1.032 & 1.016 & 1.010 & 0.986 & 0.989 & 0.972 & 0.983 & 0.989 & 0.999 & 1.005 \\ 
  adaLASSO & 1.223 & 1.090 & 1.026 & 1.013 & 1.003 & 0.994 & 1.019 & 0.989 & 0.972 & 1.004 & 0.988 & 1.000 & 0.954 \\ 
  Factor & 1.219 & 1.075 & 1.034 & 1.031 & 1.035 & 1.021 & 1.024 & 0.967 & 0.974 & 1.021 & 1.000 & 0.990 & 1.005 \\ 
  FarmPredict & 1.334 & 1.089 & 1.025 & 1.049 & 1.040 & 1.016 & 1.023 & 0.991 & 0.971 & 1.029 & 0.997 & 0.998 & 1.020 \\ 
  Target Factor & 1.268 & 1.082 & 1.068 & 1.142 & 1.047 & 0.997 & 0.976 & 0.961 & 1.003 & 0.970 & 1.026 & 1.092 & 0.977 \\ 
  CSR & 2.146 & 1.144 & 1.048 & 1.047 & 1.002 & 0.980 & $\textit{\blue{0.940}}^{\,*}$ & $\textbf{\blue{0.926}}^{\,*}$ & 0.975 & 0.981 & 0.981 & 0.985 & $\textbf{\blue{0.900}}^{\,*}$ \\ 
  Random Forest & 2.096 & 1.217 & 1.026 & 1.047 & 1.009 & 0.998 & 0.971 & 0.953 & 0.944 & $\textit{\blue{0.931}}^{\,*}$ & $\textit{\blue{0.920}}^{\,**}$ & $\textit{\blue{0.921}}^{\,*}$ & 0.955 \\ 
  \vspace{-0.3cm} \\ \multicolumn{14}{c}{\textbf{\underline{E. Disaggregation: subgroups (IBGE)}}} \\ \vspace{-0.3cm} &  &  &  &  &  &  &  &  &  &  &  &  &  \\ 
  AR & 3.111 & 1.502 & 1.186 & 1.212 & 1.244 & 1.184 & 1.212 & 1.253 & 1.264 & 1.316 & 1.275 & 1.362 & 1.318 \\ 
  Augmented AR & 1.197 & 1.260 & 1.290 & 1.288 & 1.290 & 1.191 & 1.164 & 1.212 & 1.239 & 1.310 & 1.314 & 1.440 & 1.208 \\ 
  Ridge & 3.100 & 1.308 & 1.058 & 1.045 & 1.023 & 1.026 & 1.008 & 1.002 & 1.000 & 0.993 & 0.995 & 1.016 & 1.068 \\ 
  adaLASSO & 1.380 & 1.122 & 1.096 & 1.056 & 1.030 & 1.010 & 0.990 & 0.990 & 0.983 & 1.011 & 1.049 & 1.039 & 1.005 \\ 
  Factor & 1.364 & 1.147 & 1.068 & 1.050 & 1.039 & 1.029 & 1.009 & 0.985 & 0.996 & 1.027 & 1.040 & 1.005 & 1.034 \\ 
  FarmPredict & 1.384 & 1.149 & 1.047 & 1.040 & 1.028 & 1.008 & 1.010 & 0.980 & 0.982 & 1.033 & 1.060 & 1.042 & 1.036 \\ 
  Target Factor & 1.230 & 1.059 & 1.145 & 1.126 & 1.117 & 1.059 & 0.976 & \textit{\blue{0.937}} & 1.028 & 0.979 & 1.051 & 1.116 & 1.021 \\ 
  CSR & 2.222 & 1.210 & 1.076 & 1.067 & 1.027 & 1.028 & 0.977 & 0.955 & \textit{\blue{0.939}} & 0.987 & 1.034 & 1.030 & 0.974 \\ 
  Random Forest & 2.106 & 1.232 & 1.032 & 1.063 & 1.021 & 1.022 & 0.990 & 0.962 & 0.949 & $\textit{\blue{0.925}}^{\,*}$ & $\textit{\blue{0.917}}^{\,**}$ & $\textit{\blue{0.921}}^{\,*}$ & 0.974 \\ 
  \vspace{-0.3cm} \\ \multicolumn{14}{c}{\textbf{\underline{F. Model combinations for disaggregates}}} \\ \vspace{-0.3cm} &  &  &  &  &  &  &  &  &  &  &  &  &  \\ 
  Aggreg. Comb. & 1.201 & 0.989 & $\textbf{\blue{0.952}}^{\,**}$ & $\textit{\blue{0.972}}^{\,*}$ & 0.982 & \textit{\blue{0.969}} & $\textit{\blue{0.942}}^{\,**}$ & $\textit{\blue{0.934}}^{\,**}$ & $\textit{\blue{0.935}}^{\,**}$ & $\textit{\blue{0.939}}^{\,**}$ & 0.971 & 0.986 & $\textit{\blue{0.935}}^{\,**}$ \\ 
  BCB Comb. & 1.243 & 1.014 & 0.969 & 0.993 & 0.997 & 0.977 & $0.945^{\,*}$ & $0.954^{\,*}$ & $0.950^{\,*}$ & $0.952^{\,*}$ & 0.975 & 0.986 & 0.950 \\ 
  Groups Comb. & 1.401 & 1.068 & 0.976 & 1.001 & 0.988 & \textit{\blue{0.969}} & 0.970 & 0.954 & 0.959 & 0.982 & 0.978 & 0.995 & 0.961 \\ 
  Subgroups Comb. \; & 1.583 & 1.129 & 1.035 & 1.045 & 1.031 & 1.007 & 0.981 & 0.966 & 0.985 & 1.004 & 1.026 & 1.046 & 1.029 \\ 
   \bottomrule
\end{tabular}
}
\\
\footnotesize
\justifying
\singlespacing
\onehalfspacing
\noindent \textit{Notes:} $^{***}$, $^{**}$, and $^{*}$ indicate that for a specific forecast horizon, a model $m$ performed statistically better than the median of the available Focus at 1, 5, and 10\% significance levels in a one-tailed Diebold-Mariano test with $\hip_0: \text{MSE}\big(\widehat{\pi}_{t+h\, |\,t}^{m}\big) = \text{MSE}\big(\pi_{t+h\,|\,t}^{\text{Focus}}\big) \; \textit{ versus } \hip_1: \text{MSE}\big(\widehat{\pi}_{t+h\,|\,t}^{m}\big) < \text{MSE}\big(\pi_{t+h\,|\,t}^{\text{Focus}}\big)$. The value highlighted in bold blue indicates the best model for each horizon in terms of RMSE ratio with respect to \textit{ex-post} Focus, and the values in blue italics indicate the second and third best models.
\end{table}

Panels B to E of Table \ref{tab:rmse_allsample} show the results for each model considering different levels of disaggregation: aggregate inflation, disaggregations from the Central Bank of Brazil (BCB), and disaggregations into IBGE groups and subgroups, respectively. Perhaps not surprisingly, it is hard to outperform the Focus survey in nowcasting. Exceptions are due to models that forecast aggregate inflation directly. However, such models perform better only than the available Focus. Considering the whole period, no alternative beats the ex-post  Focus, which delivers almost 7\% RMSE reduction compared to the available survey. However, it is appreciable that some models are competitive with the ex-post Focus. Specialists who report their expectations to the BCB often have access to information unavailable to econometricians, such as private data. Since models do not have this additional information and other advantages, such as including personal judgments, as pointed out by \cite{faust2013}, their ability to outperform available expectations and get closer to \textit{ex-post} survey-based expectations is a great result. We note that models forecasting disaggregates do not deliver good performance for nowcasting. Lastly, the combinations of models in each level of disaggregation, whose results are shown in Panel F, also do not generate forecasts better.

For other horizons, the contribution of the models becomes more effective. Despite the challenge of surpassing survey-based expectations for short-term horizons such as $h=1$ and $h=2$, several models for aggregate inflation (Panel B) achieve good results for these horizons. Like occurred for $h=0$, the hybrid Phillips curve, adaLASSO, factor model, FarmPredict, and, additionally, the complete subset regression (CSR), deliver the best forecast performances for one and two months ahead. All are statistically superior to the available Focus according to the Diebold-Mariano (DM) test considering at least the more slack significance level (i.e., 10\%). Furthermore, these models also numerically outperform the ex-post Focus. On the other hand, the adaLASSO using BCB disaggregation (Panel C) is the only model employing any disaggregation among the best models. However, this model is not statistically superior to the available Focus by the DM test. Finally, regarding these shorter horizons, it is worth highlighting the performance of the average forecast of the models for aggregate inflation, which achieves the highest accuracy for $h = 2$ by presenting a statistically significant reduction of almost 5\% in RMSE.

Models considering some disaggregation for the inflation yield better results starting from the 4-month horizon. The adaLASSO, factor model, and FarmPredict, all using the BCB disaggregation, perform well for forecast horizons ranging from fourth to seventh months. These models are statistically superior to available or ex-post Focus at various periods. Regarding the use of disaggregated inflation data in groups from the IBGE, it is worth mentioning the good performance of the CSR, which achieves the best result among all the options for $h = 7$. Another highlight is the combination of forecasts generated by models that directly forecast the aggregate inflation, which achieves a statistically significant reduction of 6\% in RMSE from 6 to 9 months ahead. For $h \geqslant 8$, there is broad dominance of the random forest (RF), whether using aggregate inflation or some disaggregation. Frequently, for these more distant horizons, the RF registers a statistically significant reduction in RMSE ranging from 7\% to 10\% in comparison to the survey-based expectations. This result highlights that the RF, employing IBGE group disaggregation, achieves the best performance among all competitors for inflation accumulated over 12 months (see last column of Table \ref{tab:rmse_allsample}), closely followed by the adaLASSO using BCB disaggregation, which achieves a similar RMSE reduction.

\vspace{-0.5cm}
\paragraph{Remarks.} Considering the forecast performance of various models from January 2014 to June 2022, we observe that different approaches are more effective at different times. In the short term, machine learning models that deal directly with aggregate inflation perform better, whereas for intermediate horizons of 4 to 7 months, considering the BCB disaggregation lead to significant benefits. For the period between 6 and 9 months ahead, the average of forecasts obtained from models that used only aggregate inflation also perform well. Finally, for longer horizons of 8 months or more, regardless of the approach, the RF delivers the best forecast performances. While \citet{garcia2017} points to the superiority of the CSR in several horizons, we only verify the prevalence of the CSR for $h = 1$ and other isolated good performances. The RF's performance in predicting inflation had already been pointed out by \citet{medeiros2021} when analyzing the case of the United States and highlighting the benefits of this method for dealing with non-linearities. The advent of the COVID-19 pandemic in Brazil changes the price dynamics considerably from 2020 onwards. Because of that, in what follows, we divide our analysis into two sub-periods: (i) before the pandemic, from January 2014 to February 2020, and (ii) after the pandemic, from March 2020 onwards.

\vspace{-0.2cm}
%%%%%
\subsection{Forecasts before and after of COVID-19 pandemic}
%%%%%
\vspace{-0.2cm}

\begin{figure}[!ht]
    \centering
    \caption{Forecasts by each horizon and 12-month cumulative period}
    \label{fig:forecasts}
    \vspace{-0.2cm}
    \includegraphics[width=0.98\linewidth]{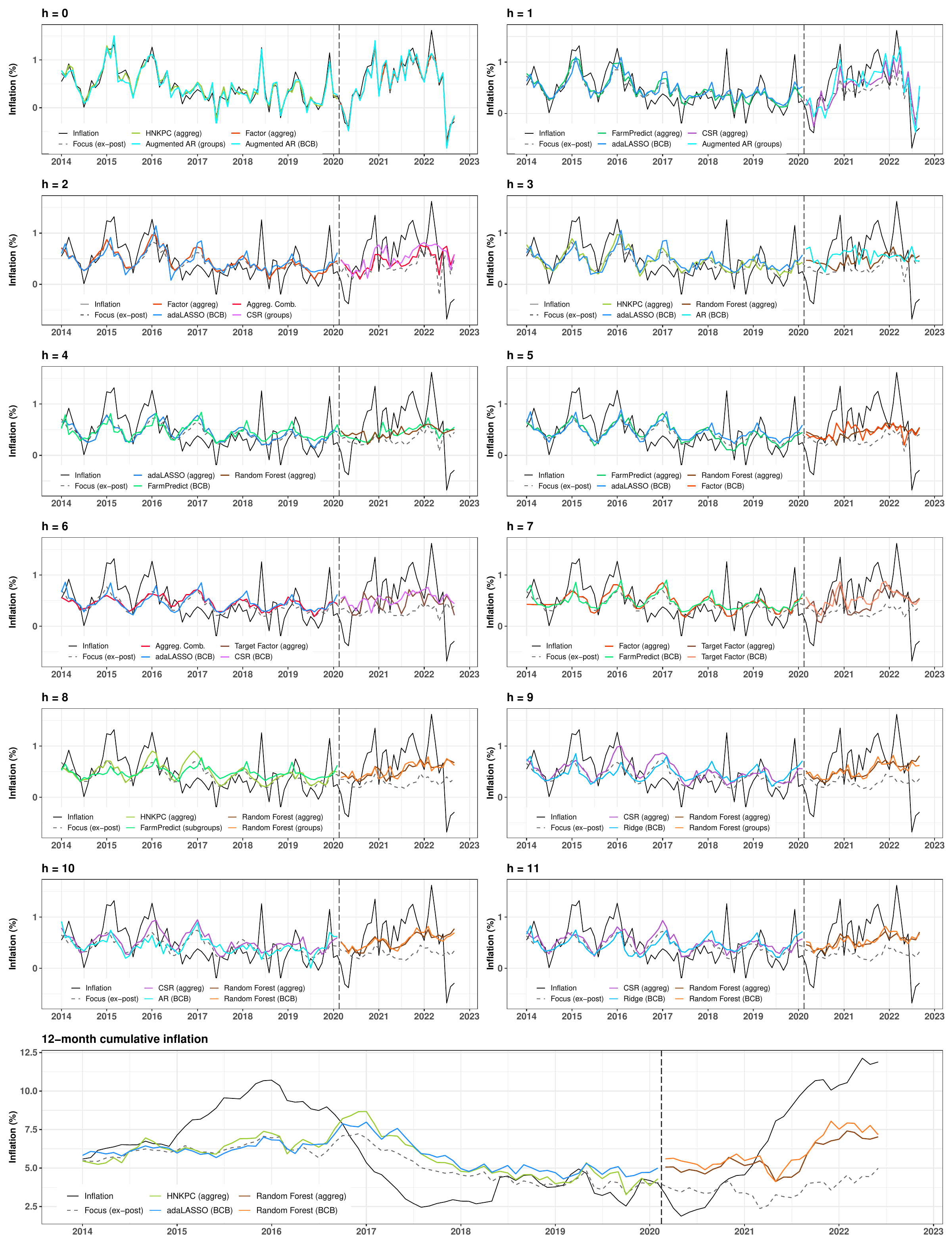}
    \\
    \vspace{-0.3cm}
    \footnotesize
    \justifying
    \singlespacing
    \onehalfspacing
    \noindent \textit{Notes:} Black solid lines indicate the actual inflation. Gray dashed lines indicate the median of the ex-post inflation expectations from the Focus survey on the last business day of each month. Solid-colored lines indicate forecasts generated by different models or a combination of models. The vertical dashed lines separate the period before and after the COVID-19 pandemic.
\end{figure}

Table \ref{tab:rmse_sample1} shows that the sub-period between January 2014 and February 2020 is quite challenging for model-based forecasts. Almost no RMSE ratios are below 1, with exceptions mainly in the short term. Nevertheless, no model is able to beat the ex-post Focus or be statistically superior to the available Focus for nowcasting. For $h = 1$, only the factor model and FarmPredict are statistically superior to the available Focus at the 10\% significance level, with these two tying with the predictive performance of the ex-post Focus. For the 3-month forecast, the hybrid Phillips curve for aggregate inflation is subtly superior to the Focus in numerical terms but without statistical significance. For the other horizons, no model performs better than the survey-based expectations. There are some potential explanations for this poor performance of the models. First, since we are analyzing the first sub-period, a small sample may have affected the estimates, contributing to the models' poor forecast performance. Second, the instabilities in the Brazilian economy in 2014 and 2015 that resulted in a sharp increase in inflation in 2015, as well as the rapid disinflation that occurred from the second half of 2016, are challenging events to anticipate, especially without extensive historical data. Conversely, from 2017 until the begging of the COVID-19 pandemic, Brazilian inflation remained reasonably controlled and close to the inflation target, leaving limited opportunities for models to enhance survey-based expectations. These dynamics of the Brazilian inflation can be observed from Figure \ref{fig:forecasts} later on.

% latex table generated in R 4.1.3 by xtable 1.8-4 package
% version adapted by Gilberto Boaretto
% Fri Mar 17 19:59:28 2023
\begin{table}[!ht]
\centering
\caption{Out-of-sample RMSE with respect to the available Focus: Jan/2014 to Feb/2020} 
\label{tab:rmse_sample1}
\resizebox{1\linewidth}{!}{
\begin{tabular}{llllllllllllll}
  \toprule
\multicolumn{1}{c}{Estimator/Model} & $h = 0\quad$ & $h = 1\quad$ & $h = 2\quad$ & $h = 3\quad$ & $h = 4\quad$ & $h = 5\quad$ & $h = 6\quad$ & $h = 7\quad$ & $h = 8\quad$ & $h = 9\quad$ & $h = 10\;\,$ & $h = 11\;\,$ & $\Sigma \, 12m$ \\ 
  \cmidrule(lr){1-14}
\vspace{-0.3cm} \\ \multicolumn{14}{c}{\textbf{\underline{A. Survey}}} \\ \vspace{-0.3cm} &  &  &  &  &  &  &  &  &  &  &  &  &  \\ 
  Focus (available) & 1.000 & \textit{\blue{1.000}} & \textit{\blue{1.000}} & \textit{\blue{1.000}} & \textit{\blue{1.000}} & \textbf{\blue{1.000}} & \textit{\blue{1.000}} & \textit{\blue{1.000}} & \textit{\blue{1.000}} & \textit{\blue{1.000}} & \textit{\blue{1.000}} & \textbf{\blue{1.000}} & \textit{\blue{1.000}} \\ 
  Focus (\textit{ex-post}) & $\textbf{\blue{0.932}}^{\,***}$ & $\textbf{\blue{0.969}}^{\,***}$ & $\textbf{\blue{0.995}}^{\,**}$ & \textit{\blue{1.001}} & \textbf{\blue{0.999}} & \textit{\blue{1.001}} & \textbf{\blue{0.999}} & \textbf{\blue{1.000}} & \textbf{\blue{1.000}} & \textbf{\blue{1.000}} & \textbf{\blue{0.999}} & \textit{\blue{1.002}} & $\textbf{\blue{0.996}}^{\,***}$ \\ 
  \vspace{-0.3cm} \\ \multicolumn{14}{c}{\textbf{\underline{B. Aggregate inflation}}} \\ \vspace{-0.3cm} &  &  &  &  &  &  &  &  &  &  &  &  &  \\ 
  RW & 3.088 & 1.752 & 1.428 & 1.474 & 1.574 & 1.547 & 1.516 & 1.427 & 1.260 & 1.240 & 1.307 & 1.382 & 2.114 \\ 
  Historical Mean & 3.309 & 1.492 & 1.246 & 1.208 & 1.184 & 1.173 & 1.155 & 1.146 & 1.133 & 1.125 & 1.127 & 1.134 & 1.422 \\ 
  AR & 2.781 & 1.466 & 1.225 & 1.240 & 1.266 & 1.244 & 1.172 & 1.122 & 1.093 & 1.098 & 1.121 & 1.155 & 1.339 \\ 
  HNKPC & \textit{\blue{0.966}} & 1.008 & 1.022 & \textbf{\blue{0.994}} & 1.083 & 1.087 & \textit{\blue{1.061}} & \textit{\blue{1.043}} & \textit{\blue{1.046}} & \textit{\blue{1.050}} & \textit{\blue{1.072}} & 1.079 & \textit{\blue{1.089}} \\ 
  Augmented AR & 1.015 & 1.067 & 1.081 & 1.090 & 1.114 & 1.101 & 1.082 & 1.076 & 1.068 & 1.086 & 1.117 & 1.126 & 1.161 \\ 
  adaLASSO & \textit{\blue{0.990}} & \textit{\blue{0.979}} & \textit{\blue{1.009}} & 1.052 & \textit{\blue{1.031}} & 1.074 & 1.086 & 1.084 & 1.134 & 1.136 & 1.176 & 1.097 & 1.136 \\ 
  Factor & \textit{\blue{0.990}} & $\textit{\blue{0.970}}^{\,*}$ & \textit{\blue{1.001}} & \textit{\blue{1.016}} & \textit{\blue{1.056}} & \textit{\blue{1.055}} & \textit{\blue{1.056}} & \textit{\blue{1.043}} & 1.106 & 1.121 & 1.113 & 1.115 & \textit{\blue{1.109}} \\ 
  FarmPredict & \textit{\blue{0.990}} & $\textit{\blue{0.970}}^{\,*}$ & \textit{\blue{1.003}} & \textit{\blue{1.024}} & \textit{\blue{1.065}} & \textit{\blue{1.040}} & 1.062 & 1.094 & 1.113 & 1.129 & 1.105 & 1.163 & \textit{\blue{1.102}} \\ 
  Target Factor & 1.499 & 1.225 & 1.122 & 1.338 & 1.334 & 1.215 & 1.110 & 1.223 & 1.212 & 1.231 & 1.171 & 1.118 & 1.333 \\ 
  CSR & \textit{\blue{0.968}} & 1.011 & 1.035 & 1.081 & 1.139 & 1.104 & 1.086 & 1.063 & \textit{\blue{1.056}} & \textit{\blue{1.047}} & \textit{\blue{1.043}} & \textit{\blue{1.054}} & 1.113 \\ 
  Random Forest & 1.918 & 1.276 & 1.156 & 1.151 & 1.137 & 1.144 & 1.130 & 1.130 & 1.101 & 1.113 & 1.110 & 1.127 & 1.319 \\ 
  \vspace{-0.3cm} \\ \multicolumn{14}{c}{\textbf{\underline{C. Disaggregation: tradable, nontradable and monitored prices (BCB)}}} \\ \vspace{-0.3cm} &  &  &  &  &  &  &  &  &  &  &  &  &  \\ 
  AR & 2.943 & 1.535 & 1.268 & 1.275 & 1.304 & 1.318 & 1.241 & 1.261 & 1.150 & 1.086 & 1.079 & 1.077 & 1.519 \\ 
  Augmented AR & 1.049 & 1.120 & 1.131 & 1.150 & 1.182 & 1.189 & 1.164 & 1.186 & 1.137 & 1.163 & 1.189 & 1.153 & 1.335 \\ 
  Ridge & 1.100 & 1.190 & 1.134 & 1.129 & 1.167 & 1.242 & 1.125 & 1.111 & 1.088 & 1.069 & 1.144 & \textit{\blue{1.060}} & 1.363 \\ 
  adaLASSO & 1.225 & 1.098 & 1.060 & 1.092 & 1.112 & \textit{\blue{1.049}} & 1.064 & 1.069 & 1.110 & 1.095 & 1.106 & \textit{\blue{1.062}} & 1.168 \\ 
  Factor & 1.094 & 1.098 & 1.120 & 1.142 & 1.142 & 1.082 & 1.075 & 1.074 & 1.089 & 1.084 & 1.098 & 1.101 & 1.178 \\ 
  FarmPredict & 1.226 & 1.118 & 1.095 & 1.116 & 1.110 & 1.072 & 1.072 & 1.050 & 1.092 & 1.107 & 1.101 & 1.085 & 1.180 \\ 
  Target Factor & 1.569 & 1.183 & 1.078 & 1.221 & 1.372 & 1.260 & 1.127 & 1.215 & 1.123 & 1.239 & 1.175 & 1.118 & 1.309 \\ 
  CSR & 2.364 & 1.401 & 1.221 & 1.188 & 1.196 & 1.162 & 1.168 & 1.184 & 1.133 & 1.123 & 1.153 & 1.105 & 1.362 \\ 
  Random Forest & 2.223 & 1.322 & 1.161 & 1.167 & 1.163 & 1.182 & 1.164 & 1.161 & 1.131 & 1.107 & 1.112 & 1.134 & 1.380 \\ 
  \vspace{-0.3cm} \\ \multicolumn{14}{c}{\textbf{\underline{D. Disaggregation: groups (IBGE)}}} \\ \vspace{-0.3cm} &  &  &  &  &  &  &  &  &  &  &  &  &  \\ 
  AR & 3.236 & 1.659 & 1.301 & 1.319 & 1.272 & 1.199 & 1.176 & 1.221 & 1.236 & 1.268 & 1.237 & 1.314 & 1.465 \\ 
  Augmented AR & 1.084 & 1.305 & 1.270 & 1.263 & 1.276 & 1.238 & 1.199 & 1.195 & 1.177 & 1.261 & 1.269 & 1.346 & 1.380 \\ 
  Ridge & 2.043 & 1.443 & 1.205 & 1.171 & 1.149 & 1.136 & 1.111 & 1.107 & 1.084 & 1.098 & 1.106 & 1.116 & 1.329 \\ 
  adaLASSO & 1.361 & 1.273 & 1.159 & 1.146 & 1.128 & 1.110 & 1.154 & 1.155 & 1.163 & 1.186 & 1.148 & 1.146 & 1.302 \\ 
  Factor & 1.411 & 1.270 & 1.164 & 1.163 & 1.157 & 1.157 & 1.127 & 1.105 & 1.105 & 1.114 & 1.109 & 1.119 & 1.289 \\ 
  FarmPredict & 1.510 & 1.226 & 1.148 & 1.174 & 1.181 & 1.138 & 1.108 & 1.113 & 1.089 & 1.139 & 1.128 & 1.149 & 1.322 \\ 
  Target Factor & 1.553 & 1.323 & 1.227 & 1.315 & 1.215 & 1.172 & 1.166 & 1.181 & 1.244 & 1.167 & 1.210 & 1.185 & 1.358 \\ 
  CSR & 2.410 & 1.434 & 1.318 & 1.285 & 1.231 & 1.159 & 1.083 & 1.074 & 1.171 & 1.148 & 1.138 & 1.111 & 1.248 \\ 
  Random Forest & 2.282 & 1.342 & 1.174 & 1.216 & 1.185 & 1.170 & 1.154 & 1.170 & 1.167 & 1.146 & 1.120 & 1.127 & 1.409 \\ 
  \vspace{-0.3cm} \\ \multicolumn{14}{c}{\textbf{\underline{E. Disaggregation: subgroups (IBGE)}}} \\ \vspace{-0.3cm} &  &  &  &  &  &  &  &  &  &  &  &  &  \\ 
  AR & 3.467 & 1.745 & 1.461 & 1.503 & 1.503 & 1.464 & 1.448 & 1.538 & 1.510 & 1.501 & 1.453 & 1.665 & 1.788 \\ 
  Augmented AR & 1.345 & 1.529 & 1.547 & 1.547 & 1.545 & 1.496 & 1.418 & 1.462 & 1.498 & 1.494 & 1.476 & 1.745 & 1.634 \\ 
  Ridge & 3.208 & 1.448 & 1.214 & 1.185 & 1.151 & 1.142 & 1.124 & 1.107 & 1.108 & 1.111 & 1.110 & 1.118 & 1.390 \\ 
  adaLASSO & 1.533 & 1.261 & 1.264 & 1.174 & 1.141 & 1.117 & 1.105 & 1.166 & 1.149 & 1.150 & 1.200 & 1.166 & 1.308 \\ 
  Factor & 1.521 & 1.266 & 1.180 & 1.149 & 1.156 & 1.133 & 1.108 & 1.106 & 1.104 & 1.110 & 1.137 & 1.114 & 1.280 \\ 
  FarmPredict & 1.541 & 1.273 & 1.136 & 1.148 & 1.146 & 1.121 & 1.117 & 1.098 & 1.081 & 1.143 & 1.176 & 1.173 & 1.302 \\ 
  Target Factor & 1.450 & 1.296 & 1.337 & 1.278 & 1.280 & 1.184 & 1.141 & 1.153 & 1.255 & 1.171 & 1.218 & 1.244 & 1.407 \\ 
  CSR & 2.457 & 1.394 & 1.319 & 1.244 & 1.201 & 1.155 & 1.096 & 1.107 & 1.104 & 1.138 & 1.177 & 1.119 & 1.253 \\ 
  Random Forest & 2.325 & 1.360 & 1.180 & 1.224 & 1.194 & 1.199 & 1.182 & 1.168 & 1.150 & 1.117 & 1.106 & 1.117 & 1.407 \\ 
  \vspace{-0.3cm} \\ \multicolumn{14}{c}{\textbf{\underline{F. Model combinations for disaggregates}}} \\ \vspace{-0.3cm} &  &  &  &  &  &  &  &  &  &  &  &  &  \\ 
  Aggreg. Comb. & 1.273 & 1.077 & 1.024 & 1.042 & 1.076 & 1.059 & \textit{\blue{1.048}} & \textit{\blue{1.046}} & \textit{\blue{1.050}} & \textit{\blue{1.058}} & \textit{\blue{1.075}} & 1.083 & 1.142 \\ 
  BCB Comb. & 1.393 & 1.168 & 1.097 & 1.119 & 1.138 & 1.122 & 1.096 & 1.107 & 1.084 & 1.086 & 1.093 & 1.066 & 1.275 \\ 
  Groups Comb. & 1.610 & 1.271 & 1.142 & 1.158 & 1.133 & 1.100 & 1.083 & 1.090 & 1.104 & 1.117 & 1.107 & 1.126 & 1.290 \\ 
  Subgroups Comb. \; & 1.739 & 1.290 & 1.187 & 1.180 & 1.163 & 1.138 & 1.105 & 1.118 & 1.133 & 1.130 & 1.143 & 1.172 & 1.341 \\ 
   \bottomrule
\end{tabular}
}
\\
\footnotesize
\justifying
\singlespacing
\noindent \textit{Notes:} see Table \ref{tab:rmse_allsample}.
\end{table}

Since the COVID-19 pandemic, most models perform statistically better than the Focus survey, as seen in Table \ref{tab:rmse_sample2}. Models that look directly at aggregate inflation or use some inflation disaggregation tend to perform very well at all horizons, including nowcasting. Specifically, the adaLASSO, factor model, and FarmPredict working directly with the aggregate inflation achieve RMSE 10\% lower than the RMSE of the available Focus and smaller RMSE than the ex-post Focus, something challenging to imagine before the pandemic. Meanwhile, augmented AR and ridge, both employing BCB disaggregation, also perform well. One month ahead, even some models employing the highest level of disaggregation (i.e., subgroups) deliver good results. From the results in this second sub-period, we understand the good performance of the RF for the entire period. Considering aggregate inflation, the method already registers the best performances from $h = 3$, with similar results when using any disaggregation. Frequently, the RF can reduce the RMSE of the available and ex-post Focus by up to 30\% for longer horizons, regardless of whether considering aggregate inflation or some disaggregation. For 12-month accumulated inflation, the RF obtains the best performance when using the BCB disaggregation: a 35\% reduction compared to the RMSE of the main benchmarks. Finally, regarding model combinations, each is statistically superior to Focus for all $h \geqslant 1$, often at a 1\% significance level. However, the combinations do not beat some individual models with good predictive performance in the sub-period.

% latex table generated in R 4.1.3 by xtable 1.8-4 package
% version adapted by Gilberto Boaretto
% Fri Mar 17 20:13:27 2023
\begin{table}[!ht]
\centering
\caption{Out-of-sample RMSE with respect to the available Focus: Mar/2020 to Jun/2022} 
\label{tab:rmse_sample2}
\resizebox{1\linewidth}{!}{
\begin{tabular}{llllllllllllll}
  \toprule
\multicolumn{1}{c}{Estimator/Model} & $h = 0\quad$ & $h = 1\quad$ & $h = 2\quad$ & $h = 3\quad$ & $h = 4\quad$ & $h = 5\quad$ & $h = 6\quad$ & $h = 7\quad$ & $h = 8\quad$ & $h = 9\quad$ & $h = 10\;\,$ & $h = 11\;\,$ & $\Sigma \, 12m$ \\ 
  \cmidrule(lr){1-14}
\vspace{-0.3cm} \\ \multicolumn{14}{c}{\textbf{\underline{A. Survey}}} \\ \vspace{-0.3cm} &  &  &  &  &  &  &  &  &  &  &  &  &  \\ 
  Focus (available) & 1.000 & 1.000 & 1.000 & 1.000 & 1.000 & 1.000 & 1.000 & 1.000 & 1.000 & 1.000 & 1.000 & 1.000 & 1.000 \\ 
  Focus (\textit{ex-post}) & $0.934^{\,**}$ & $0.975^{\,***}$ & $0.991^{\,***}$ & 1.001 & 1.000 & 0.999 & 1.001 & 0.999 & 0.999 & 1.000 & 1.002 & 1.001 & 1.000 \\ 
  \vspace{-0.3cm} \\ \multicolumn{14}{c}{\textbf{\underline{B. Aggregate inflation}}} \\ \vspace{-0.3cm} &  &  &  &  &  &  &  &  &  &  &  &  &  \\ 
  RW & 2.443 & 1.189 & 1.023 & 1.056 & 1.003 & 1.024 & 0.983 & 0.965 & 0.939 & 0.908 & 0.920 & 0.920 & 1.128 \\ 
  Historical Mean & 2.915 & 1.203 & 0.948 & $0.921^{\,**}$ & $0.906^{\,**}$ & $0.888^{\,***}$ & $0.868^{\,***}$ & $0.862^{\,***}$ & $0.863^{\,***}$ & $0.862^{\,***}$ & $0.860^{\,***}$ & $0.857^{\,***}$ & $0.835^{\,***}$ \\ 
  AR & 2.264 & 1.060 & $0.904^{\,*}$ & $0.938^{\,*}$ & $0.937^{\,*}$ & $0.910^{\,**}$ & $0.871^{\,***}$ & $0.827^{\,***}$ & $0.807^{\,***}$ & $0.803^{\,***}$ & $\textit{\blue{0.828}}^{\,***}$ & $\textit{\blue{0.846}}^{\,***}$ & $0.799^{\,***}$ \\ 
  HNKPC & $0.927^{\,*}$ & $0.913^{\,***}$ & $0.933^{\,***}$ & 0.971 & 0.999 & 0.986 & $0.917^{\,***}$ & $0.911^{\,**}$ & $0.896^{\,***}$ & $0.913^{\,***}$ & $0.940^{\,***}$ & $0.976^{\,*}$ & $0.938^{\,***}$ \\ 
  Augmented AR & $0.924^{\,*}$ & $0.904^{\,**}$ & $0.915^{\,**}$ & 0.951 & $0.949^{\,*}$ & $0.916^{\,**}$ & $0.868^{\,***}$ & $0.834^{\,***}$ & $0.825^{\,***}$ & $0.856^{\,***}$ & $0.889^{\,***}$ & $0.927^{\,**}$ & $0.854^{\,***}$ \\ 
  adaLASSO & $\textit{\blue{0.896}}^{\,**}$ & $0.922^{\,**}$ & $0.954^{\,***}$ & $0.948^{\,***}$ & $0.935^{\,***}$ & $0.921^{\,***}$ & $0.889^{\,***}$ & $0.890^{\,***}$ & $0.868^{\,***}$ & $0.893^{\,***}$ & $0.933^{\,***}$ & $0.908^{\,***}$ & $0.879^{\,***}$ \\ 
  Factor & $\textbf{\blue{0.894}}^{\,**}$ & $0.896^{\,***}$ & $0.939^{\,***}$ & $0.923^{\,***}$ & $0.907^{\,***}$ & $0.921^{\,***}$ & $0.938^{\,***}$ & $0.943^{\,***}$ & $0.944^{\,***}$ & $0.931^{\,***}$ & 1.005 & 0.990 & $0.935^{\,***}$ \\ 
  FarmPredict & $\textit{\blue{0.896}}^{\,**}$ & $0.918^{\,***}$ & $0.929^{\,***}$ & $0.958^{\,**}$ & $0.947^{\,**}$ & 0.971 & $0.959^{\,**}$ & $0.964^{\,**}$ & $0.947^{\,**}$ & $0.953^{\,**}$ & $0.965^{\,*}$ & $0.962^{\,*}$ & $0.956^{\,**}$ \\ 
  Target Factor & 1.018 & 1.043 & 0.986 & 0.998 & 0.934 & 0.948 & $\textit{\blue{0.789}}^{\,***}$ & $\textit{\blue{0.778}}^{\,***}$ & $0.788^{\,***}$ & $\textit{\blue{0.740}}^{\,**}$ & $0.889^{\,*}$ & 1.089 & $0.772^{\,***}$ \\ 
  CSR & 0.954 & $0.851^{\,***}$ & $0.904^{\,***}$ & $0.903^{\,**}$ & $0.925^{\,**}$ & $0.909^{\,**}$ & $0.875^{\,***}$ & $0.840^{\,***}$ & $0.845^{\,***}$ & $0.917^{\,**}$ & 1.041 & 1.083 & $0.851^{\,***}$ \\ 
  Random Forest & 1.370 & 0.951 & $0.916^{\,*}$ & $0.896^{\,***}$ & $0.877^{\,***}$ & $0.855^{\,***}$ & $0.815^{\,***}$ & $0.794^{\,***}$ & $\textit{\blue{0.761}}^{\,***}$ & $\textit{\blue{0.740}}^{\,***}$ & $\textit{\blue{0.723}}^{\,***}$ & $\textbf{\blue{0.710}}^{\,***}$ & $\textit{\blue{0.689}}^{\,***}$ \\ 
  \vspace{-0.3cm} \\ \multicolumn{14}{c}{\textbf{\underline{C. Disaggregation: tradable, nontradable and monitored prices (BCB)}}} \\ \vspace{-0.3cm} &  &  &  &  &  &  &  &  &  &  &  &  &  \\ 
  AR & 2.161 & 0.931 & $\textit{\blue{0.831}}^{\,**}$ & $\textit{\blue{0.861}}^{\,**}$ & $0.895^{\,*}$ & $0.876^{\,**}$ & $0.815^{\,***}$ & $0.841^{\,***}$ & $0.853^{\,***}$ & $0.910^{\,*}$ & 0.985 & 1.006 & $0.779^{\,***}$ \\ 
  Augmented AR & $\textit{\blue{0.903}}^{\,*}$ & $\textit{\blue{0.841}}^{\,**}$ & $0.914^{\,**}$ & 0.991 & 1.024 & 0.977 & $0.883^{\,***}$ & $0.863^{\,***}$ & $0.849^{\,***}$ & $0.865^{\,***}$ & 0.938 & 1.011 & $0.904^{\,***}$ \\ 
  Ridge & $0.930^{\,*}$ & $0.842^{\,**}$ & $0.903^{\,**}$ & $0.929^{\,**}$ & $0.944^{\,**}$ & $0.950^{\,*}$ & $0.898^{\,***}$ & $0.870^{\,***}$ & $0.871^{\,***}$ & $0.870^{\,***}$ & $0.904^{\,***}$ & $0.919^{\,***}$ & $0.859^{\,***}$ \\ 
  adaLASSO & 0.959 & $0.907^{\,*}$ & $0.895^{\,***}$ & $0.866^{\,***}$ & $0.864^{\,***}$ & $0.867^{\,***}$ & $0.881^{\,***}$ & $0.868^{\,***}$ & $0.869^{\,***}$ & $0.854^{\,***}$ & $0.859^{\,***}$ & $0.873^{\,***}$ & $0.769^{\,***}$ \\ 
  Factor & 0.948 & $0.886^{\,***}$ & $0.884^{\,***}$ & $\textit{\blue{0.861}}^{\,***}$ & $\textit{\blue{0.832}}^{\,***}$ & $\textbf{\blue{0.808}}^{\,***}$ & $0.829^{\,***}$ & $0.901^{\,***}$ & $0.928^{\,***}$ & $0.946^{\,***}$ & $0.937^{\,***}$ & $0.938^{\,***}$ & $0.827^{\,***}$ \\ 
  FarmPredict & 1.024 & $0.940^{\,**}$ & $0.928^{\,***}$ & $0.867^{\,***}$ & $\textit{\blue{0.849}}^{\,***}$ & $\textit{\blue{0.838}}^{\,***}$ & $0.863^{\,***}$ & $0.934^{\,***}$ & $0.932^{\,***}$ & $0.929^{\,***}$ & $0.933^{\,***}$ & $0.916^{\,***}$ & $0.858^{\,***}$ \\ 
  Target Factor & 1.046 & 0.931 & $0.893^{\,**}$ & 0.982 & 0.910 & 0.924 & $\textit{\blue{0.784}}^{\,***}$ & $\textbf{\blue{0.727}}^{\,***}$ & $0.827^{\,***}$ & $0.754^{\,***}$ & $0.881^{\,*}$ & 1.046 & $0.729^{\,***}$ \\ 
  CSR & 1.973 & 0.954 & $\textit{\blue{0.834}}^{\,***}$ & $\textit{\blue{0.864}}^{\,**}$ & $0.860^{\,**}$ & $\textit{\blue{0.827}}^{\,***}$ & $\textbf{\blue{0.769}}^{\,***}$ & $0.796^{\,***}$ & $0.808^{\,***}$ & $0.831^{\,***}$ & $0.912^{\,**}$ & 1.011 & $\textit{\blue{0.719}}^{\,***}$ \\ 
  Random Forest & 1.600 & 1.042 & $0.902^{\,*}$ & 0.955 & $0.886^{\,***}$ & $\textit{\blue{0.845}}^{\,***}$ & $\textit{\blue{0.814}}^{\,***}$ & $0.790^{\,***}$ & $\textit{\blue{0.748}}^{\,***}$ & $\textit{\blue{0.733}}^{\,***}$ & $\textbf{\blue{0.707}}^{\,***}$ & $\textit{\blue{0.716}}^{\,***}$ & $\textbf{\blue{0.648}}^{\,***}$ \\ 
  \vspace{-0.3cm} \\ \multicolumn{14}{c}{\textbf{\underline{D. Disaggregation: groups (IBGE)}}} \\ \vspace{-0.3cm} &  &  &  &  &  &  &  &  &  &  &  &  &  \\ 
  AR & 2.118 & 0.888 & $\textit{\blue{0.814}}^{\,*}$ & $\textbf{\blue{0.825}}^{\,**}$ & $\textit{\blue{0.856}}^{\,**}$ & $0.890^{\,**}$ & 0.981 & 0.979 & 1.022 & 1.043 & 0.995 & $0.903^{\,**}$ & $0.850^{\,**}$ \\ 
  Augmented AR & 0.918 & $\textbf{\blue{0.767}}^{\,**}$ & $0.836^{\,**}$ & $0.871^{\,*}$ & $0.924^{\,*}$ & $0.878^{\,**}$ & 0.960 & 0.961 & 0.983 & 0.973 & 1.024 & 1.043 & $0.841^{\,***}$ \\ 
  Ridge & 1.153 & 1.143 & $0.907^{\,**}$ & $0.916^{\,**}$ & $0.904^{\,**}$ & $0.907^{\,***}$ & $0.884^{\,***}$ & $0.894^{\,***}$ & $0.881^{\,***}$ & $0.887^{\,***}$ & $0.893^{\,***}$ & $0.905^{\,***}$ & $0.835^{\,***}$ \\ 
  adaLASSO & 1.070 & $0.910^{\,*}$ & $0.915^{\,**}$ & $0.901^{\,**}$ & $0.898^{\,***}$ & $0.900^{\,***}$ & $0.909^{\,**}$ & $0.849^{\,***}$ & $0.803^{\,***}$ & $0.844^{\,***}$ & $0.851^{\,***}$ & $0.879^{\,***}$ & $0.767^{\,***}$ \\ 
  Factor & 0.993 & $0.880^{\,**}$ & $0.926^{\,**}$ & $0.922^{\,**}$ & $0.934^{\,**}$ & $0.909^{\,***}$ & $0.942^{\,**}$ & $0.854^{\,***}$ & $0.866^{\,***}$ & $0.946^{\,**}$ & $0.911^{\,***}$ & $0.885^{\,***}$ & $0.861^{\,***}$ \\ 
  FarmPredict & 1.134 & 0.960 & $0.923^{\,**}$ & $0.946^{\,*}$ & $0.922^{\,**}$ & $0.916^{\,**}$ & $0.957^{\,*}$ & $0.893^{\,***}$ & $0.874^{\,***}$ & $0.938^{\,**}$ & $0.888^{\,***}$ & $0.871^{\,***}$ & $0.865^{\,***}$ \\ 
  Target Factor & \textit{\blue{0.904}} & $\textit{\blue{0.831}}^{\,**}$ & 0.934 & 0.995 & $0.901^{\,*}$ & $0.846^{\,**}$ & $\textit{\blue{0.812}}^{\,***}$ & $\textit{\blue{0.765}}^{\,***}$ & $\textit{\blue{0.779}}^{\,***}$ & $0.792^{\,***}$ & $0.865^{\,**}$ & 1.019 & $0.768^{\,***}$ \\ 
  CSR & 1.849 & $\textit{\blue{0.831}}^{\,*}$ & $\textbf{\blue{0.792}}^{\,***}$ & $\textit{\blue{0.828}}^{\,***}$ & $\textbf{\blue{0.789}}^{\,***}$ & $\textit{\blue{0.825}}^{\,***}$ & $0.823^{\,***}$ & $0.802^{\,***}$ & $0.799^{\,***}$ & $0.834^{\,***}$ & $0.846^{\,***}$ & $0.882^{\,***}$ & $\textit{\blue{0.710}}^{\,***}$ \\ 
  Random Forest & 1.895 & 1.102 & $0.902^{\,*}$ & $0.902^{\,**}$ & $\textit{\blue{0.855}}^{\,**}$ & $0.851^{\,***}$ & $0.815^{\,***}$ & $\textit{\blue{0.760}}^{\,***}$ & $\textbf{\blue{0.738}}^{\,***}$ & $\textbf{\blue{0.731}}^{\,***}$ & $\textit{\blue{0.737}}^{\,***}$ & $\textit{\blue{0.735}}^{\,***}$ & $\textit{\blue{0.685}}^{\,***}$ \\ 
  \vspace{-0.3cm} \\ \multicolumn{14}{c}{\textbf{\underline{E. Disaggregation: subgroups (IBGE)}}} \\ \vspace{-0.3cm} &  &  &  &  &  &  &  &  &  &  &  &  &  \\ 
  AR & 2.715 & 1.264 & 0.931 & 0.939 & 1.008 & 0.927 & 1.009 & 0.999 & 1.047 & 1.158 & 1.126 & 1.089 & 1.067 \\ 
  Augmented AR & 1.030 & 0.982 & 1.060 & 1.056 & 1.061 & 0.905 & 0.939 & 0.995 & 1.006 & 1.153 & 1.180 & 1.168 & 0.980 \\ 
  Ridge & 2.990 & 1.179 & $0.927^{\,*}$ & $0.927^{\,**}$ & $0.915^{\,**}$ & $0.931^{\,**}$ & $0.916^{\,***}$ & $0.920^{\,**}$ & $0.912^{\,**}$ & $0.896^{\,***}$ & $0.901^{\,***}$ & $0.934^{\,**}$ & $0.903^{\,***}$ \\ 
  adaLASSO & 1.210 & 0.992 & 0.954 & 0.960 & $0.938^{\,**}$ & $0.925^{\,**}$ & $0.898^{\,***}$ & $0.841^{\,***}$ & $0.841^{\,***}$ & $0.893^{\,***}$ & $0.922^{\,***}$ & $0.936^{\,***}$ & $0.849^{\,***}$ \\ 
  Factor & 1.189 & 1.037 & 0.977 & 0.971 & $0.943^{\,**}$ & $0.946^{\,**}$ & $0.932^{\,***}$ & $0.887^{\,***}$ & $0.909^{\,***}$ & $0.961^{\,*}$ & $0.962^{\,**}$ & $0.918^{\,***}$ & $0.914^{\,***}$ \\ 
  FarmPredict & 1.209 & 1.034 & 0.975 & 0.951 & $0.930^{\,**}$ & $0.916^{\,**}$ & $0.925^{\,***}$ & $0.886^{\,***}$ & $0.903^{\,***}$ & $0.942^{\,**}$ & $0.965^{\,**}$ & $0.935^{\,***}$ & $0.904^{\,***}$ \\ 
  Target Factor & 0.964 & $\textit{\blue{0.810}}^{\,*}$ & 0.979 & 0.998 & 0.977 & 0.959 & $0.838^{\,**}$ & $\textit{\blue{0.744}}^{\,***}$ & $0.823^{\,***}$ & $0.807^{\,**}$ & 0.908 & 1.012 & $0.810^{\,***}$ \\ 
  CSR & 1.963 & 1.033 & $0.853^{\,***}$ & $0.914^{\,**}$ & $0.875^{\,***}$ & $0.923^{\,***}$ & $0.882^{\,***}$ & $0.829^{\,***}$ & $0.795^{\,***}$ & $0.857^{\,***}$ & $0.913^{\,***}$ & 0.961 & $0.832^{\,***}$ \\ 
  Random Forest & 1.865 & 1.114 & $0.908^{\,*}$ & $0.926^{\,*}$ & $0.869^{\,**}$ & $0.870^{\,**}$ & $0.825^{\,***}$ & $0.781^{\,***}$ & $\textit{\blue{0.768}}^{\,***}$ & $\textit{\blue{0.750}}^{\,***}$ & $\textit{\blue{0.746}}^{\,***}$ & $\textit{\blue{0.746}}^{\,***}$ & $0.725^{\,***}$ \\ 
  \vspace{-0.3cm} \\ \multicolumn{14}{c}{\textbf{\underline{F. Model combinations for disaggregates}}} \\ \vspace{-0.3cm} &  &  &  &  &  &  &  &  &  &  &  &  &  \\ 
  Aggreg. Comb. & 1.127 & $0.909^{\,***}$ & $0.895^{\,***}$ & $0.918^{\,***}$ & $0.906^{\,***}$ & $0.898^{\,***}$ & $0.858^{\,***}$ & $0.845^{\,***}$ & $0.840^{\,***}$ & $0.840^{\,***}$ & $0.886^{\,***}$ & $0.910^{\,***}$ & $0.835^{\,***}$ \\ 
  BCB Comb. & 1.075 & $0.866^{\,***}$ & $0.862^{\,***}$ & $0.888^{\,***}$ & $0.877^{\,***}$ & $0.856^{\,***}$ & $0.820^{\,***}$ & $0.827^{\,***}$ & $0.838^{\,***}$ & $0.839^{\,***}$ & $0.878^{\,***}$ & $0.923^{\,***}$ & $0.777^{\,***}$ \\ 
  Groups Comb. & 1.159 & $0.864^{\,**}$ & $\textit{\blue{0.833}}^{\,***}$ & $0.866^{\,***}$ & $0.864^{\,***}$ & $0.861^{\,***}$ & $0.880^{\,***}$ & $0.842^{\,***}$ & $0.836^{\,***}$ & $0.868^{\,***}$ & $0.869^{\,***}$ & $0.888^{\,***}$ & $0.786^{\,***}$ \\ 
  Subgroups Comb. \; & 1.411 & 0.975 & $0.906^{\,**}$ & $0.932^{\,**}$ & $0.921^{\,***}$ & $0.899^{\,***}$ & $0.882^{\,***}$ & $0.840^{\,***}$ & $0.860^{\,***}$ & $0.898^{\,**}$ & $0.929^{\,**}$ & $0.944^{\,**}$ & $0.869^{\,***}$ \\ 
   \bottomrule
\end{tabular}
}
\\
\footnotesize
\justifying
\singlespacing
\noindent \textit{Notes:} see Table \ref{tab:rmse_allsample}.
\end{table}

Figure \ref{fig:forecasts} presents the temporal evolution of actual inflation, Focus survey expectations, and the best aggregate- and disaggregates-based models for each horizon. Looking at the projections for $h = 0$, we partially understand why it is not easy to outperform the survey in the very-short term. The Focus consensus is very close to the actual values. Furthermore, as we will see in Subsection \ref{subsec:selection}, available inflation expectations are the primary predictor for model-based nowcasting. The survey contains much relevant information unavailable to the econometrician, so we already expect this result. When analyzing the other horizons ($h \geqslant 1$), we note that it is challenging for the survey and models to predict peaks and valleys of inflation. Already at $h = 1$, we observe outstanding forecasting errors. One consequence of COVID-19, which start is highlighted by vertical dashed lines on each plot, is that the Focus survey initially overestimated inflation and afterward systematically underestimated one. Despite the challenges of generating accurate forecasts in such an uncertain period, several models perform better than expert forecasts for all horizons. A punctual example is the adaLASSO that, using both aggregates and BCB disaggregations, as well as other models, achieves a great result in forecasting the peak observed in December 2020 at $h = 1$, a point at which the available inflation expectation is far from the actual value. Thus, other variables besides the available inflation expectations are fundamental for the performance of model-based forecasts.

\vspace{-0.5cm}
%\newpage
\paragraph{Remarks.} The good performance of the RF is mainly due to its ability to capture the higher level of future inflation from the second half of 2020. The model generates forecasts closer to the actual inflation than the Focus expectations, which systematically underestimate inflation in that period. Since the pandemic, models for disaggregated inflation tend to provide more accurate forecasts than models for aggregate inflation, except for nowcasting. For each $h \geqslant 2$, we note that adaLASSO, factor model, FarmPredict, and CSR using any disaggregation of inflation deliver forecasts with lower RMSE than the respective models using aggregate inflation, with a few exceptions for the CSR using groups and subgroups that do not outperform the CSR using aggregate inflation. In turn, the RF performs well regardless of the target variable. These findings underscore the use of models in inflation forecasting, including the junction between disaggregated analysis and machine learning techniques, particularly during periods of higher economic instability, such as a pandemic. Previous studies such as \citet{altug2016} and \citet{medeiros2021} have also shown that models perform well during more volatile periods. Next, we will analyze the forecasts for disaggregates and identify the predictors selected by adaLASSO and FarmPredict, methods that allow variable selection.

%%%%%
\vspace{-0.1cm}
\subsection{Forecast of disaggregates and variable selection}
\label{subsec:selection}
%%%%%
\vspace{-0.1cm}

%%%%%
\subsubsection{Disaggregation into BCB categories}
\vspace{-0.1cm}

\paragraph{Predictive performance.} Now we consider the predictive performance of the models using different disaggregations, starting with the disaggregation from the BCB. Since we lack long time series for survey-based expectations for disaggregated inflation, we use the AR model as a benchmark. Of all the BCB disaggregations, monitored prices are the most challenging to forecast since they are subject to many unexpected changes resulting from government decisions that often do not freely follow supply and demand movements. According to the results displayed in Table \ref{tab:rmse_disaggreg_bcb}, other models manage to beat the AR only in short or more distant horizons for this disaggregate (Panel A). In nowcasting, other methods perform significantly better than the AR, with augmented AR and Ridge standing out by obtaining more than a 30\% reduction in RMSE. For $h = 1$, augmented AR achieves a 10\% reduction in RMSE, the only statistically significant at any level. Some models present minor RMSE for intermediate horizons, but the results are not statistically significant according to the DM test. For ten and eleven months ahead, Ridge delivers reductions of 4\% and 6\% in RMSE, respectively, compared to the AR, and, specifically for $h = 11$, adaLASSO, factor model, and FarmPredict also statistically outperform the AR, with RMSE reductions ranging from 3\% and 4\%. Putting all horizons together, Ridge, adaLASSO, FarmPredict, and factor model generate more accurate forecasts than the AR by delivering RMSEs 2\% to 4\% lower than the AR model, all statistically significant at the 1\% level. The good result of the augmented AR model is restricted to the short term.

% latex table generated in R 4.1.3 by xtable 1.8-4 package
% version adapted by Gilberto Boaretto
% Tue Mar 14 17:31:37 2023
\begin{table}[!t]
\centering
\caption{\centering Out-of-sample RMSE for BCB disaggregate (in terms of RMSE of the AR model): Jan/2014 to Jun/2022, by disaggregate and horizon} 
\label{tab:rmse_disaggreg_bcb}
\resizebox{1\linewidth}{!}{
\begin{tabular}{llllllllllllll}
  \toprule
\multicolumn{1}{c}{Estimator/Model \;\;\;} & $h = 0\quad\,$ & $h = 1\quad\,$ & $h = 2\quad\,$ & $h = 3\quad\,$ & $h = 4\quad\,$ & $h = 5\quad\,$ & $h = 6\quad\,$ & $h = 7\quad\,$ & $h = 8\quad\,$ & $h = 9\quad\,$ & $h = 10\;\;$ & $h = 11\;\;$ & all $h$\;\;\;\; \\ 
  \cmidrule(lr){1-14}
\vspace{-0.3cm} \\ \multicolumn{13}{c}{\textbf{\underline{A. Monitored Prices}}} \\ \vspace{-0.3cm} &  &  &  &  &  &  &  &  &  &  &  &  &  \\ 
  AR & 1.000 & 1.000 & 1.000 & 1.000 & 1.000 & 1.000 & \textit{\blue{1.000}} & 1.000 & 1.000 & 1.000 & 1.000 & 1.000 & 1.000 \\ 
  Augmented AR & $\textbf{\blue{0.694}}^{\,***}$ & $\textbf{\blue{0.903}}^{\,***}$ & 1.015 & 1.022 & 1.025 & 1.023 & 1.026 & 1.035 & 1.038 & 1.037 & 1.021 & 1.033 & 0.996 \\ 
  Ridge & $\textit{\blue{0.709}}^{\,***}$ & \textit{\blue{0.954}} & \textit{\blue{0.987}} & \textbf{\blue{0.984}} & \textbf{\blue{0.984}} & \textit{\blue{0.988}} & 1.001 & \textit{\blue{0.985}} & \textbf{\blue{0.988}} & \textbf{\blue{0.986}} & $\textbf{\blue{0.957}}^{\,**}$ & $\textbf{\blue{0.943}}^{\,**}$ & $\textbf{\blue{0.960}}^{\,***}$ \\ 
  adaLASSO & $0.756^{\,***}$ & 0.960 & \textbf{\blue{0.986}} & \textit{\blue{0.988}} & \textit{\blue{0.987}} & \textbf{\blue{0.987}} & 1.004 & 0.989 & \textit{\blue{0.996}} & 1.004 & 0.971 & $0.962^{\,*}$ & $\textit{\blue{0.969}}^{\,***}$ \\ 
  Factor & $0.734^{\,***}$ & 0.970 & 1.008 & 1.003 & 0.999 & 0.999 & 1.016 & 1.009 & 1.006 & \textit{\blue{0.996}} & 0.976 & $0.961^{\,*}$ & $0.977^{\,***}$ \\ 
  FarmPredict & $0.774^{\,***}$ & 0.980 & 1.002 & 0.998 & 0.997 & 1.009 & \textbf{\blue{0.998}} & \textbf{\blue{0.982}} & 0.996 & 0.998 & 0.974 & $0.965^{\,*}$ & $0.975^{\,***}$ \\ 
  Target Factor & $0.750^{\,***}$ & 1.033 & 1.045 & 1.082 & 1.063 & 1.079 & 1.086 & 1.149 & 1.143 & 1.097 & 1.073 & 1.080 & 1.063 \\ 
  CSR & 0.974 & 1.002 & 1.022 & 1.034 & 1.061 & 1.053 & 1.034 & 1.046 & 1.066 & 1.057 & 1.021 & 1.027 & 1.034 \\ 
  Random Forest & $0.872^{\,***}$ & 1.020 & 1.039 & 1.060 & 1.058 & 1.076 & 1.098 & 1.068 & 1.061 & 1.033 & \textit{\blue{0.967}} & \textit{\blue{0.956}} & 1.027 \\ 
  \vspace{-0.3cm} \\ \multicolumn{13}{c}{\textbf{\underline{B. Non-Tradables}}} \\ \vspace{-0.3cm} &  &  &  &  &  &  &  &  &  &  &  &  &  \\ 
  AR & 1.000 & 1.000 & 1.000 & 1.000 & 1.000 & 1.000 & 1.000 & 1.000 & 1.000 & 1.000 & 1.000 & 1.000 & 1.000 \\ 
  Augmented AR & $\textit{\blue{0.843}}^{\,**}$ & $0.831^{\,***}$ & $0.844^{\,***}$ & $0.820^{\,***}$ & $0.810^{\,***}$ & $0.787^{\,***}$ & $0.829^{\,***}$ & $0.853^{\,**}$ & $0.907^{\,**}$ & 0.963 & 0.985 & 0.975 & $0.869^{\,***}$ \\ 
  Ridge & $\textbf{\blue{0.824}}^{\,***}$ & $0.817^{\,***}$ & $0.830^{\,***}$ & $0.808^{\,***}$ & $0.796^{\,***}$ & $0.780^{\,***}$ & $0.823^{\,***}$ & $\textit{\blue{0.847}}^{\,***}$ & $\textit{\blue{0.900}}^{\,**}$ & \textit{\blue{0.944}} & 0.957 & 0.950 & $0.854^{\,***}$ \\ 
  adaLASSO & $0.896^{\,*}$ & $\textbf{\blue{0.784}}^{\,***}$ & $\textbf{\blue{0.778}}^{\,***}$ & $\textit{\blue{0.777}}^{\,***}$ & $\textit{\blue{0.748}}^{\,***}$ & $\textit{\blue{0.733}}^{\,***}$ & $\textit{\blue{0.812}}^{\,***}$ & $0.850^{\,**}$ & 0.944 & 0.977 & 0.955 & 0.962 & $\textit{\blue{0.847}}^{\,***}$ \\ 
  Factor & $0.889^{\,**}$ & $0.809^{\,***}$ & $0.806^{\,***}$ & $0.804^{\,***}$ & $0.765^{\,***}$ & $0.774^{\,***}$ & $0.823^{\,***}$ & $0.866^{\,**}$ & 0.951 & 1.006 & 0.968 & 0.950 & $0.864^{\,***}$ \\ 
  FarmPredict & $0.895^{\,*}$ & $\textit{\blue{0.798}}^{\,***}$ & $\textit{\blue{0.782}}^{\,***}$ & $\textbf{\blue{0.768}}^{\,***}$ & $\textbf{\blue{0.744}}^{\,***}$ & $0.739^{\,***}$ & $0.842^{\,***}$ & $0.855^{\,**}$ & 0.955 & 1.000 & 0.968 & \textit{\blue{0.938}} & $0.853^{\,***}$ \\ 
  Target Factor & 1.040 & $0.846^{\,**}$ & 0.896 & 1.022 & 0.935 & $0.799^{\,***}$ & $0.885^{\,*}$ & 1.030 & 1.026 & 1.015 & \textit{\blue{0.944}} & 1.045 & $0.954^{\,**}$ \\ 
  CSR & $0.892^{\,***}$ & $0.864^{\,***}$ & $0.887^{\,***}$ & $0.887^{\,***}$ & $0.867^{\,***}$ & $0.816^{\,***}$ & $0.856^{\,***}$ & $0.875^{\,***}$ & $0.931^{\,*}$ & 0.956 & 0.968 & 1.013 & $0.899^{\,***}$ \\ 
  Random Forest & $0.868^{\,***}$ & $0.855^{\,***}$ & $0.816^{\,***}$ & $0.803^{\,***}$ & $0.778^{\,***}$ & $\textbf{\blue{0.729}}^{\,***}$ & $\textbf{\blue{0.791}}^{\,***}$ & $\textbf{\blue{0.805}}^{\,***}$ & $\textbf{\blue{0.854}}^{\,***}$ & $\textbf{\blue{0.886}}^{\,**}$ & $\textbf{\blue{0.893}}^{\,**}$ & $\textbf{\blue{0.918}}^{\,*}$ & $\textbf{\blue{0.829}}^{\,***}$ \\ 
  \vspace{-0.3cm} \\ \multicolumn{13}{c}{\textbf{\underline{C. Tradables}}} \\ \vspace{-0.3cm} &  &  &  &  &  &  &  &  &  &  &  &  &  \\ 
  AR & 1.000 & 1.000 & 1.000 & 1.000 & 1.000 & 1.000 & 1.000 & 1.000 & 1.000 & 1.000 & 1.000 & 1.000 & 1.000 \\ 
  Augmented AR & $0.832^{\,**}$ & $0.911^{\,**}$ & 0.995 & 1.070 & 1.082 & 1.051 & 0.998 & $0.948^{\,**}$ & $0.951^{\,*}$ & 0.989 & 1.044 & 1.055 & 1.003 \\ 
  Ridge & $0.810^{\,***}$ & $0.920^{\,**}$ & $0.954^{\,**}$ & 0.982 & $0.960^{\,**}$ & 1.042 & $0.899^{\,***}$ & $0.874^{\,***}$ & $0.917^{\,**}$ & $0.935^{\,**}$ & 1.030 & $0.889^{\,***}$ & $0.940^{\,***}$ \\ 
  adaLASSO & $\textbf{\blue{0.763}}^{\,***}$ & $0.941^{\,*}$ & $0.943^{\,**}$ & $\textit{\blue{0.940}}^{\,***}$ & $\textit{\blue{0.909}}^{\,***}$ & $\textit{\blue{0.844}}^{\,***}$ & $0.848^{\,***}$ & $0.861^{\,***}$ & $0.918^{\,**}$ & $0.886^{\,***}$ & $\textit{\blue{0.912}}^{\,**}$ & $\textit{\blue{0.834}}^{\,***}$ & $\textit{\blue{0.886}}^{\,***}$ \\ 
  Factor & $0.814^{\,***}$ & $0.928^{\,**}$ & 0.972 & $0.956^{\,*}$ & $0.927^{\,**}$ & $0.860^{\,***}$ & $0.845^{\,***}$ & $0.882^{\,***}$ & $0.932^{\,**}$ & 0.966 & $0.937^{\,*}$ & $0.906^{\,***}$ & $0.913^{\,***}$ \\ 
  FarmPredict & $0.811^{\,***}$ & 0.967 & 0.980 & 0.967 & $0.929^{\,**}$ & $0.872^{\,***}$ & $0.858^{\,***}$ & $0.897^{\,***}$ & $0.940^{\,*}$ & $0.956^{\,*}$ & $0.937^{\,*}$ & $0.899^{\,***}$ & $0.920^{\,***}$ \\ 
  Target Factor & $\textit{\blue{0.792}}^{\,**}$ & $\textbf{\blue{0.886}}^{\,**}$ & $\textit{\blue{0.907}}^{\,**}$ & 0.979 & 1.000 & 0.935 & $\textit{\blue{0.821}}^{\,***}$ & $\textbf{\blue{0.769}}^{\,***}$ & $\textbf{\blue{0.831}}^{\,***}$ & $0.896^{\,**}$ & 0.990 & $0.926^{\,*}$ & $0.900^{\,***}$ \\ 
  CSR & $0.884^{\,**}$ & 1.045 & 1.079 & 1.003 & $0.917^{\,*}$ & $0.886^{\,***}$ & $0.909^{\,***}$ & $0.884^{\,***}$ & $0.866^{\,***}$ & $\textit{\blue{0.873}}^{\,***}$ & $0.925^{\,**}$ & $0.932^{\,***}$ & $0.933^{\,***}$ \\ 
  Random Forest & $0.820^{\,***}$ & $\textit{\blue{0.903}}^{\,***}$ & $\textbf{\blue{0.906}}^{\,***}$ & $\textbf{\blue{0.902}}^{\,***}$ & $\textbf{\blue{0.853}}^{\,***}$ & $\textbf{\blue{0.824}}^{\,***}$ & $\textbf{\blue{0.819}}^{\,***}$ & $\textit{\blue{0.810}}^{\,***}$ & $\textit{\blue{0.836}}^{\,***}$ & $\textbf{\blue{0.848}}^{\,***}$ & $\textbf{\blue{0.841}}^{\,***}$ & $\textbf{\blue{0.810}}^{\,***}$ & $\textbf{\blue{0.847}}^{\,***}$ \\ 
   \bottomrule
\end{tabular}
}
\\
\footnotesize
\justifying
\singlespacing
\onehalfspacing
\noindent \textit{Notes:} $^{***}$, $^{**}$, and $^{*}$ indicate that for a specific disaggregate and forecast horizon, a model $m$ performed statistically better than an AR model at 1, 5, and 10\% significance levels in a one-tailed Diebold-Mariano test with $\hip_0: \text{MSE}\big(\widehat{\pi}_{i,\,t+h\, |\,t}^{m}\big) = \text{MSE}\big(\pi_{i,\,t+h\,|\,t}^{\text{AR}}\big) \; \textit{ versus } \hip_1: \text{MSE}\big(\widehat{\pi}_{i,\,t+h\,|\,t}^{m}\big) < \text{MSE}\big(\pi_{i,\,t+h\,|\,t}^{\text{AR}}\big)$. The value highlighted in bold blue indicates the best model for each horizon in terms of RMSE ratio with respect to the AR model, and the values in blue italics indicate the second and third best models. The average weights of each disaggregate in IPCA are: monitored prices (25\%); non-tradables (41.5\%); and tradables ( 33.5\%).
\end{table}

Looking at the other BCB disaggregates, namely non-tradable and tradable items, we notice that machine learning models deliver better results than the traditional AR model (Panels B and C of Table \ref{tab:rmse_disaggreg_bcb}). For non-tradables, once again augmented AR and Ridge stand out in nowcasting with RMSE reductions of 18\% and 16\%, while adaLASSO and FarmPredict achieve the best performances between one and five months ahead, with RMSE reductions oscillating between 20\% and 27\%. For all $h \geqslant 5$, the random forest dominates by delivering the lowest RMSE. Aggregating all horizons, all ML models perform statistically better than AR for non-tradable items. In turn, the results for tradables are similar, with the ML methods yielding subtly smaller improvements. Also for tradables, there is a predominance of the RF: this method obtains the best or second-best performance for all $h \geqslant 1$, with RMSE reductions ranging from 10\% to 19\%. Other methods that stand out are adaLASSO, which obtains the best result in nowcasting (almost 24\% reduction in RMSE), and target factor, which registers the best performance one, seven, and eight months ahead. By gathering the forecasts for all periods, the RF obtains an average reduction of 15\% in RMSE compared to the AR model, with adaLASSO coming close behind, with a reduction of 11\% in RMSE. All models, except the augmented AR, outperform the AR at the 1\% significance level. These findings suggest that models that include more predictors, impose restrictions on parameters, or assume other functional forms can be more advantageous in inflation forecasting than traditional time-series models such as the AR model. Lastly, we note that the improvements due to ML methods in a data-rich environment are not just observed for the pandemic period, as seen in Tables \ref{tab:rmse_disaggreg_bcb_sample1} and \ref{tab:rmse_disaggreg_bcb_sample2} in Appendix \ref{append:disaggreg_by_horizon}.

\begin{landscape}
    \begin{figure}[!ht]
        \centering
        \caption{adaLASSO selection: aggregate inflation and BCB disaggregates (\% of extending windows)}
        \label{fig:sel_adalasso}
        \vspace{-0.35cm}
        \includegraphics[width=0.88\linewidth]{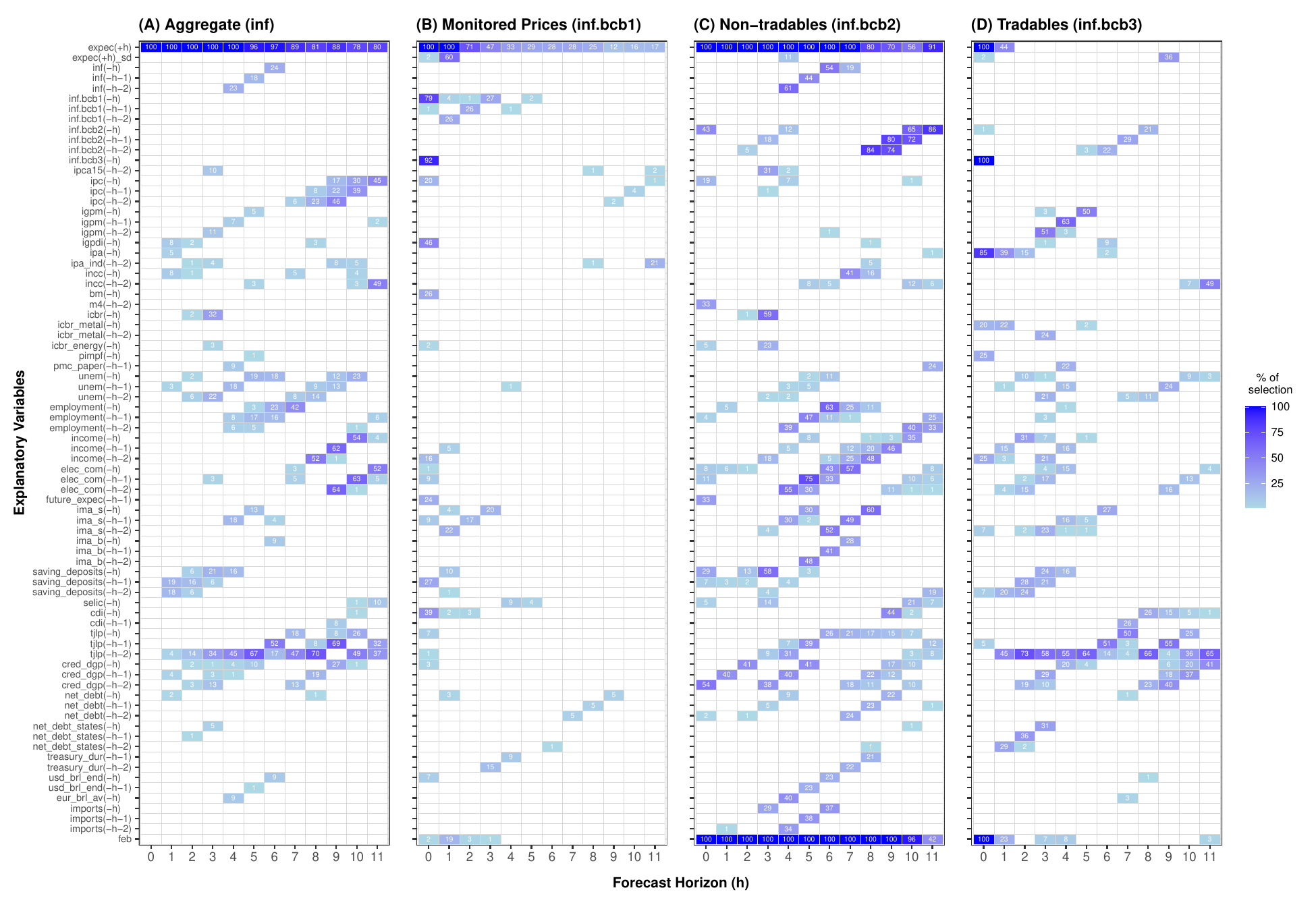}
        \\
        \vspace{-0.35cm}
        \footnotesize
        \centering
        \onehalfspacing
        \noindent \textit{Notes:} We cut out variables that do not exceed 20\% of selection at least one forecast horizon. The definitions of the variables are in Table \ref{tab:variables}.
    \end{figure}
\end{landscape}

\begin{landscape}
    \begin{figure}[!ht]
        \centering
        \caption{FarmPredict selection: aggregate inflation and BCB disaggregates (\% of extending windows)}
        \label{fig:sel_farmpredict}
        \vspace{-0.35cm}
        \includegraphics[width=0.88\linewidth]{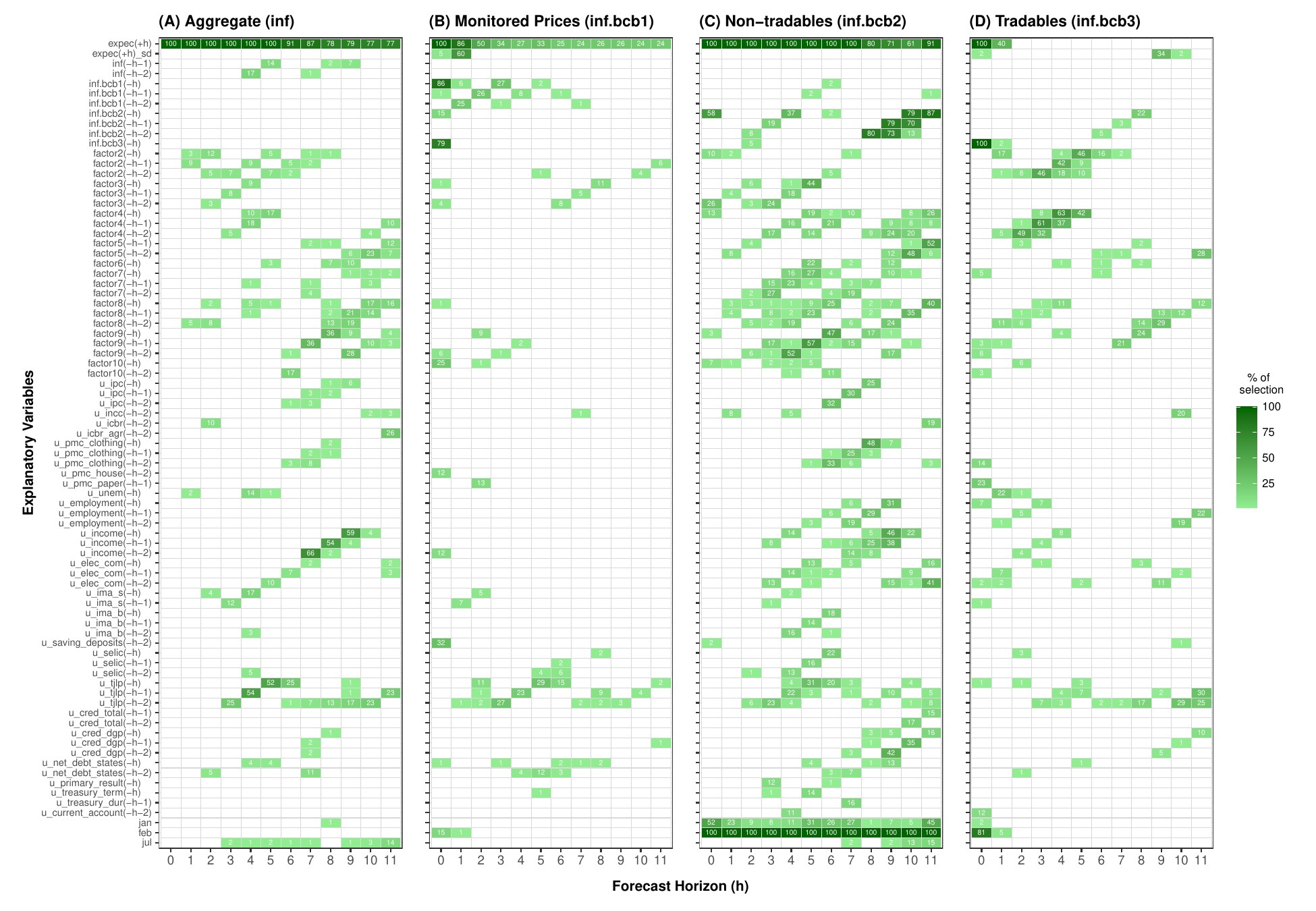}
        \\
        \vspace{-0.35cm}
        \footnotesize
        \centering
        \onehalfspacing
        \noindent \textit{Notes:} We cut out variables that do not exceed 11.5\% of selection at least one forecast horizon. The definitions of the variables are in Table \ref{tab:variables}.
    \end{figure}
\end{landscape}

\vspace{-0.5cm}
\paragraph{Variable selection.} To explore potential economic intuitions for the results from aggregate inflation and BCB disaggregation, we compare what is behind the two approaches in terms of variable selections by adaLASSO and FarmPredict, as shown in Figures \ref{fig:sel_adalasso} and \ref{fig:sel_farmpredict}, respectively. In both Figures, panel A brings the predictors that each method selected to forecast the aggregate inflation directly. Meanwhile, panels B, C, and D display the predictors selected to predict price variation of administrated, non-tradable, and tradable items. To make the presentation viable, we restricted ourselves to the variables chosen at least 20\% and 11.5\% of the time in at least one forecast horizon, respectively. Variables definitions are shown in Table \ref{tab:variables} in Appendix \ref{append:variables}. The prefix ``u\_'' in some variables in Figure \ref{fig:sel_farmpredict} indicates that the variable had the common factors ``discounted'' and, therefore, only its idiosyncratic component is left. These variables are indicated by $\widehat{u}_j$ in Equation \eqref{eq:farmpredict}.

We can summarize the results of the variable selections in the following topics:

\vspace{-0.3cm}
\begin{enumerate} \itemsep0em
    \item \textit{High variability in the type of selected predictors}. The adaLASSO selects all ten classes of predictors (see Table \ref{tab:variables} in Appendix \ref{append:variables}). Even controlling for common factors (FarmPredict), each class of predictors still appears. This result shows the importance of considering a broad set of information.
    
    \item \textit{Low participation of economic activity variables}. Variables related to economic activity appear little when we control for common factors via FarmPredict and almost never for adaLASSO. There exists some correlation between activity variables and monetary base, M1 and M2 money supplies, variables occasionally select, for example. Beyond that, the information linked to economic activity variables may be contained in other relevant variables (e.g., inflation expectations), including the possibility of a non-linear relationship between these variables. Thus, we must be careful not to conclude that economic activity variables are irrelevant for forecasting inflation. %A more profound answer to these questions is beyond the scope of this work.

    %Consequently, do other variables contain information on economic activity? When we control for them, do the activity variables cease to be relevant, or do they not contribute to forecasting inflation?

    \item[3.] \textit{Available inflation expectations (survey) are frequently picked}. adaLASSO and FarmPredict for aggregate and non-tradable items frequently select the inflation expectation (\texttt{expec}) at all horizons. For administrated and tradable items, the selection of the expectation decreases as the horizon increases, appearing only in the very short term for tradables. We note that the expectation is about aggregate inflation, so it is reasonable that it is not relevant to explain some specific disaggregate. Furthermore, due to their greater share, non-tradables present a more remarkable similarity with aggregate inflation than the other breakdowns.
    
    \item[4.] \textit{Prices (including commodities) are often chosen}. According to the variable selections, Brazilian inflation indexes and commodities price variations are relevant to forecast official Brazilian inflation. Various indexes and price variations carry relevant information to forecast Brazilian inflation. In addition, due to past inflationary history, Brazil has several monthly indexes calculated by different organizations that cover different periods (e.g., days 1 to 30, 11 to 10, and 21 to 30). Thus, we must consider this information when forecasting Brazilian inflation.
    
    \item[5.] \textit{Factors that explain most of the variability of predictors are not always more relevant to forecasting inflation.} Interestingly, the common factor that explains most of the variability of the predictors (i.e., \texttt{factor1}) is rarely select to forecast aggregate inflation or any disaggregate. Instead, the following factors up to the tenth are chosen for various horizons, mainly to predict non-tradables, lesser extent for aggregate and tradables, and very little for administrated ones.
    
    %\item \textit{Brazilian long-term interest rate has great explanatory power for both aggregate and tradables inflation in the mid- and long-term}.
    
    \item[6.] \textit{Non-tradables record the richer structure of predictors; monitored items, the poorer.} The good predictability of non-tradables is potentially related to the larger number of predictors. Note that non-tradables have more predictors than aggregate inflation itself. This finding underscores the importance of looking at disaggregates. For example, the February dummy's relevance for forecasting inflation only appears when we consider the forecast of non-tradable inflation. This dummy is crucial because it captures the variation in education prices, a sector whose contracts are usually updated in January. In turn, the price variation of monitored items has few predictors, contributing to this disaggregate being the most challenging for forecasting.
\end{enumerate}

%%%%%
%\vspace{-0.1cm}
\subsubsection{Disaggregation into IBGE groups}
%\vspace{-0.1cm}

Now we address the predictive performance for each IBGE group. The identification of each group, as well as their respective participation in the IPCA, are available in Table \ref{tab:groups_subgroups} in Appendix \ref{append:groups_subgroups}. From Table \ref{tab:rmse_disaggreg_groups}, we note that ML methods perform statistically better than the AR model and numerically better than the augmented AR for all components by stacking the horizons. An exception is the target factor, which does not perform well for some groups. Furthermore, from Table \ref{tab:rmse_disaggreg_groups_append} in Appendix \ref{append:disaggreg_by_horizon}, which presents detailed results of each disaggregate by forecast horizon, we notice that there are infrequent horizons that do not have an ML model performing statistically better than the benchmark.

% latex table generated in R 4.1.3 by xtable 1.8-4 package
% version adapted by Gilberto Boaretto
% Tue Mar 14 10:57:23 2023
\begin{table}[!ht]
\centering
\caption{\centering Out-of-sample RMSE for IBGE groups (in terms of RMSE of the AR model): Jan/2014 to Jun/2022, by disaggregate, joining all horizons} 
\label{tab:rmse_disaggreg_groups}
\resizebox{1\linewidth}{!}{
\begin{tabular}{llllllllll}
  \toprule
\multicolumn{1}{c}{Estimator/Model \;\;\;} & inf.g1$\qquad$ & inf.g2$\qquad$ & inf.g3$\qquad$ & inf.g4$\qquad$ & inf.g5$\qquad$ & inf.g6$\qquad$ & inf.g7$\qquad$ & inf.g8$\qquad$ & inf.g9$\qquad$ \\ 
  \cmidrule(lr){1-10}
AR & 1.000 & 1.000 & 1.000 & 1.000 & 1.000 & 1.000 & 1.000 & 1.000 & 1.000 \\ 
  Augmented AR & 1.007 & $0.977^{\,**}$ & 1.062 & $0.874^{\,***}$ & 1.001 & $\textbf{\blue{0.881}}^{\,***}$ & 1.006 & $0.451^{\,***}$ & 1.066 \\ 
  Ridge & $\textit{\blue{0.886}}^{\,***}$ & $\textbf{\blue{0.908}}^{\,***}$ & $0.956^{\,***}$ & $\textbf{\blue{0.836}}^{\,***}$ & $0.966^{\,**}$ & $\textit{\blue{0.889}}^{\,***}$ & $0.917^{\,***}$ & $\textbf{\blue{0.446}}^{\,***}$ & $0.805^{\,***}$ \\ 
  adaLASSO & $0.900^{\,***}$ & $\textit{\blue{0.913}}^{\,***}$ & $0.905^{\,***}$ & $\textit{\blue{0.844}}^{\,***}$ & $0.961^{\,***}$ & 0.995 & $0.886^{\,***}$ & $0.457^{\,***}$ & $\textbf{\blue{0.790}}^{\,***}$ \\ 
  Factor & $0.911^{\,***}$ & $0.915^{\,***}$ & $0.900^{\,***}$ & $0.863^{\,***}$ & $\textit{\blue{0.959}}^{\,***}$ & 0.994 & $\textbf{\blue{0.876}}^{\,***}$ & $\textit{\blue{0.450}}^{\,***}$ & $\textit{\blue{0.795}}^{\,***}$ \\ 
  FarmPredict & $0.909^{\,***}$ & $0.919^{\,***}$ & $0.912^{\,***}$ & $0.920^{\,***}$ & $0.966^{\,***}$ & 1.015 & $\textit{\blue{0.882}}^{\,***}$ & $0.457^{\,***}$ & $0.798^{\,***}$ \\ 
  Target Factor & $0.932^{\,***}$ & 0.992 & $0.905^{\,***}$ & $0.884^{\,***}$ & 1.026 & $0.909^{\,***}$ & 1.008 & $0.487^{\,***}$ & 1.164 \\ 
  CSR & $0.954^{\,**}$ & $0.949^{\,***}$ & $\textit{\blue{0.849}}^{\,***}$ & $0.898^{\,***}$ & $\textbf{\blue{0.951}}^{\,***}$ & $0.974^{\,*}$ & $0.927^{\,***}$ & $0.698^{\,***}$ & $0.863^{\,***}$ \\ 
  Random Forest & $\textbf{\blue{0.844}}^{\,***}$ & $0.961^{\,***}$ & $\textbf{\blue{0.845}}^{\,***}$ & $0.866^{\,***}$ & $0.968^{\,***}$ & $0.945^{\,***}$ & $0.931^{\,***}$ & $0.573^{\,***}$ & $0.838^{\,***}$ \\ 
   \bottomrule
\end{tabular}
}
\\
\footnotesize
\justifying
\singlespacing
\noindent \textit{Notes:} see Table \ref{tab:rmse_disaggreg_bcb}. For a definition of the groups and their respective weights in the IPCA, see Table \ref{tab:groups_subgroups} in Appendix \ref{append:groups_subgroups}.
\end{table}

Different models perform better for different disaggregates. Education (\texttt{inf.g8}) is the group most benefited from using other techniques. However, given the good performance of the augmented AR, we infer that predictive improvement is mainly due to the inclusion of the February dummy. Except for augmented AR and target factor, the models also achieve predictive improvement for communication (\texttt{inf.g9}), including all horizons individually. Transportation is the group for which the models beat the AR model by stacking the horizons with the smallest margin (\texttt{inf.g5}). In addition, the ML methods are not statistically superior to AR in half of the forecast horizons for this disaggregate. The transportation group comprises public transport fares and expenses with own vehicle and fuel, mostly items whose prices are administered by the government, which are difficult to forecast. However, except once again for augmented AR and target factor, the models deliver a statistically significant average reduction of at least 3\% in RMSE compared to the AR model, with Ridge's predictive gain around 10\% between 6- and 9-month-ahead. Lastly, it is worth highlighting the good performance of the RF to forecast the price variation of foods and beverages (\texttt{inf.g1}) at all horizons.

%%%%%
\subsubsection{Disaggregation into IBGE subgroups}

Lastly, we examine the predictive performance for each IBGE subgroup. The results stacking all horizons are shown in Table \ref{tab:rmse_disaggreg_subgroups}, and results for each horizon are in Table \ref{tab:rmse_disaggreg_subgroups_append} in Appendix \ref{append:disaggreg_by_horizon}. Descriptions of subgroups are in Table \ref{tab:groups_subgroups} in Appendix \ref{append:groups_subgroups}. Similar to what happens with the disaggregation into groups, ML models achieve more accurate forecasts in comparison to AR models for each subgroup, but with no single model emerging as a dominant predictor for different disaggregates. The reduction of RMSE reaches 71\% in the case of courses, reading, and stationery (\texttt{inf.sg18}), 40\% for communication (\texttt{inf.sg19}), and 35\% for household operations (\texttt{inf.sg7}) and personal services (\texttt{inf.sg16}). RF stout out for delivering the best predictive performances considering all forecast horizons for food at home (\texttt{inf.sg1}), appliances (\texttt{inf.sg6}), household operations (\texttt{inf.sg7}), and fabrics (\texttt{inf.sg10}). The Ridge performs well at all horizons for domestic fuels and energy (\texttt{inf.sg4}) and jewelry (\texttt{inf.sg10}), and the adaLASSO performs well for communication (\texttt{inf.sg19}) at all horizons as well. Furthermore, we show outstanding performances for specific horizons: adaLASSO performs well in the long-term for food away from home (\texttt{inf.sg2}), target factor in short- and intermediate-term for pharmaceutical and optical products (\texttt{inf.sg13}), while the RF performs well at short term for that one and at more distant horizons for the latter.

% latex table generated in R 4.1.3 by xtable 1.8-4 package
% version adapted by Gilberto Boaretto
% Mon Mar 27 17:09:12 2023
\begin{table}[ht]
\centering
\caption{\centering Out-of-sample RMSE for IBGE subgroups (in terms of RMSE of the AR model): Jan/2014 to Jun/2022, by disaggregate, joining all horizons} 
\label{tab:rmse_disaggreg_subgroups}
\resizebox{1\linewidth}{!}{
\begin{tabular}{lllllllllll}
  \toprule
 \multicolumn{1}{c}{Estimator/Model \;\;\;} & inf.sg1$\quad\,$ & inf.sg2$\quad\,$ & inf.sg3$\quad\,$ & inf.sg4$\quad\,$ & inf.sg5$\quad\,$ & inf.sg6$\quad\,$ & inf.sg7$\quad\,$ & inf.sg8$\quad\,$ & inf.sg9$\quad\,$ & inf.sg10$\;\;$ \\ 
   \cmidrule(lr){1-11}
AR & 1.000 & 1.000 & 1.000 & 1.000 & 1.000 & 1.000 & 1.000 & 1.000 & 1.000 & 1.000 \\ 
  Augmented AR & 1.062 & 1.053 & 1.080 & 1.000 & 1.063 & 1.085 & 1.050 & $0.902^{\,***}$ & 1.040 & 1.106 \\ 
  Ridge & $\textit{\blue{0.742}}^{\,***}$ & $0.798^{\,***}$ & $0.804^{\,***}$ & $\textbf{\blue{0.794}}^{\,***}$ & $0.854^{\,***}$ & $0.827^{\,***}$ & $0.681^{\,***}$ & $0.886^{\,***}$ & $0.832^{\,***}$ & $\textbf{\blue{0.693}}^{\,***}$ \\ 
  adaLASSO & $0.766^{\,***}$ & $\textbf{\blue{0.759}}^{\,***}$ & $\textit{\blue{0.781}}^{\,***}$ & $\textit{\blue{0.803}}^{\,***}$ & $\textit{\blue{0.823}}^{\,***}$ & $0.826^{\,***}$ & $\textit{\blue{0.677}}^{\,***}$ & $\textit{\blue{0.737}}^{\,***}$ & $\textit{\blue{0.768}}^{\,***}$ & $0.721^{\,***}$ \\ 
  Factor & $0.751^{\,***}$ & $0.803^{\,***}$ & $0.792^{\,***}$ & $0.805^{\,***}$ & $0.838^{\,***}$ & $\textit{\blue{0.810}}^{\,***}$ & $0.684^{\,***}$ & $0.759^{\,***}$ & $0.780^{\,***}$ & $0.725^{\,***}$ \\ 
  FarmPredict & $0.757^{\,***}$ & $0.795^{\,***}$ & $0.793^{\,***}$ & $0.810^{\,***}$ & $0.839^{\,***}$ & $0.815^{\,***}$ & $0.687^{\,***}$ & $0.809^{\,***}$ & $0.784^{\,***}$ & $0.728^{\,***}$ \\ 
  Target Factor & $0.786^{\,***}$ & $0.811^{\,***}$ & $0.857^{\,***}$ & 0.954 & $0.858^{\,***}$ & $0.836^{\,***}$ & $0.812^{\,***}$ & $\textbf{\blue{0.731}}^{\,***}$ & $0.802^{\,***}$ & $0.808^{\,***}$ \\ 
  CSR & $0.779^{\,***}$ & $0.796^{\,***}$ & $0.785^{\,***}$ & $0.850^{\,***}$ & $\textbf{\blue{0.801}}^{\,***}$ & $0.819^{\,***}$ & $0.683^{\,***}$ & $0.810^{\,***}$ & $0.791^{\,***}$ & $0.730^{\,***}$ \\ 
  Random Forest & $\textbf{\blue{0.706}}^{\,***}$ & $\textit{\blue{0.772}}^{\,***}$ & $\textbf{\blue{0.770}}^{\,***}$ & $0.858^{\,***}$ & $0.849^{\,***}$ & $\textbf{\blue{0.757}}^{\,***}$ & $\textbf{\blue{0.650}}^{\,***}$ & $0.759^{\,***}$ & $\textbf{\blue{0.734}}^{\,***}$ & $\textit{\blue{0.694}}^{\,***}$ \\ 
   \cmidrule(lr){1-11}
\multicolumn{1}{c}{Estimator/Model \;\;\;} & inf.sg11$\;$ & inf.sg12$\;$ & inf.sg13$\;$ & inf.sg14$\;$ & inf.sg15$\;$ & inf.sg16$\;$ & inf.sg17$\;$ & inf.sg18$\;$ & inf.sg19$\;$ &  \\ 
   \cmidrule(lr){1-10}
AR & 1.000 & 1.000 & 1.000 & 1.000 & 1.000 & 1.000 & 1.000 & 1.000 & 1.000 &  \\ 
  Augmented AR & 1.147 & 1.037 & $0.889^{\,***}$ & 1.010 & 1.023 & 1.117 & 1.061 & $0.436^{\,***}$ & 1.022 &  \\ 
  Ridge & $0.736^{\,***}$ & $0.874^{\,***}$ & $0.941^{\,**}$ & 0.987 & $0.899^{\,***}$ & $0.786^{\,***}$ & $0.742^{\,***}$ & $0.415^{\,***}$ & $0.613^{\,***}$ &  \\ 
  adaLASSO & $\textit{\blue{0.706}}^{\,***}$ & $0.861^{\,***}$ & $0.827^{\,***}$ & $0.979^{\,*}$ & $0.898^{\,***}$ & $0.713^{\,***}$ & $\textit{\blue{0.737}}^{\,***}$ & $0.389^{\,***}$ & $\textbf{\blue{0.599}}^{\,***}$ &  \\ 
  Factor & $0.736^{\,***}$ & $0.863^{\,***}$ & $0.892^{\,***}$ & $0.970^{\,**}$ & $0.899^{\,***}$ & $0.719^{\,***}$ & $\textbf{\blue{0.734}}^{\,***}$ & $\textbf{\blue{0.383}}^{\,***}$ & $0.605^{\,***}$ &  \\ 
  FarmPredict & $0.739^{\,***}$ & $\textit{\blue{0.861}}^{\,***}$ & $0.877^{\,***}$ & $0.978^{\,*}$ & $\textbf{\blue{0.896}}^{\,***}$ & $0.730^{\,***}$ & $0.738^{\,***}$ & $\textit{\blue{0.389}}^{\,***}$ & $\textit{\blue{0.604}}^{\,***}$ &  \\ 
  Target Factor & $0.774^{\,***}$ & $0.923^{\,***}$ & $\textbf{\blue{0.729}}^{\,***}$ & $\textbf{\blue{0.958}}^{\,**}$ & $0.924^{\,***}$ & $0.807^{\,***}$ & $0.831^{\,***}$ & $0.400^{\,***}$ & $0.829^{\,***}$ &  \\ 
  CSR & $0.722^{\,***}$ & $0.867^{\,***}$ & $0.848^{\,***}$ & $\textit{\blue{0.967}}^{\,**}$ & $\textit{\blue{0.897}}^{\,***}$ & $\textit{\blue{0.662}}^{\,***}$ & $0.776^{\,***}$ & $0.604^{\,***}$ & $0.653^{\,***}$ &  \\ 
  Random Forest & $\textbf{\blue{0.673}}^{\,***}$ & $\textbf{\blue{0.860}}^{\,***}$ & $\textit{\blue{0.798}}^{\,***}$ & $0.969^{\,**}$ & $0.913^{\,***}$ & $\textbf{\blue{0.643}}^{\,***}$ & $0.761^{\,***}$ & $0.480^{\,***}$ & $0.639^{\,***}$ &  \\ 
   \bottomrule
\end{tabular}
}
\\
\footnotesize
\justifying
\singlespacing
\noindent \textit{Notes:} see Table \ref{tab:rmse_disaggreg_bcb}. For a definition of the subgroups and their respective weights in the IPCA, see Table \ref{tab:groups_subgroups} in Appendix \ref{append:groups_subgroups}.
\end{table}

Some subgroups are equivalent to groups. It is the case of transportation (\texttt{inf.sg12} and \texttt{inf.g5}), courses, reading, and stationary (\texttt{inf.sg18}) that is equivalent to education (\texttt{inf.g8}), and communication (\texttt{inf.sg19} and \texttt{inf.g9}). Interestingly, the predictive accuracy obtained from disaggregation into subgroups is superior to that based on groups, both stacking the horizons and considering them individually. In particular, the improvements are remarkable for transportation and communication. The difference between both approaches is that, while one includes lags from other subgroups and not groups, the other includes lags from different groups and not subgroups. Note that, stacking the forecast horizons, while CSR reduces the RMSE of the AR model for \textit{group} transportation (\texttt{inf.g5}) by 5\%, the RF obtains an average reduction of 14\% for \textit{subgroup} transportation (\texttt{inf.sg12}). For transportation, while the ML methods using disaggregation into groups do not statistically outperform the AR model for horizons 1 to 4, 10, and 11 months ahead, some of these methods employing disaggregation into subgroups generate statistically significant improvements at the most demanding level ( i.e., 1\%) for all these horizons. For communication, disaggregation into subgroups can improve the already good performance of methods when employing disaggregation into groups.

%This result suggest that ``more granular information'', achieved through a higher disaggregation, can led to better disaggregate predictions.

%%%%%
\vspace{-0.1cm}
\subsubsection{Further remarks}

We find that ML models considering several predictors beat the AR model by a wide and statistically significant margin for all disaggregates considered. Comparing the use of group and subgroup disaggregations, the predictive benefits associated with a higher level of disaggregation appear to align with the results of \citet{duarte2007}, for Portugal, \citet{ibarra2012}, for Mexico, and \citet {bermingham2014}, for the US and Euro Area. However, upon revisiting the results of the aggregation of disaggregated forecasts (Tables \ref{tab:rmse_allsample} to \ref{tab:rmse_sample2}), we note that the use of disaggregation into subgroups does not generate more accurate forecasts than the direct approach or aggregation from less profound disaggregations. A potential explanation for this result is that we do not consider the aggregation of disaggregated forecasts generated by different models in this paper. As we have seen, a single model does not dominate the forecasts of all disaggregates and often does not even exhibit dominance over time for the same disaggregate (see Figures from \ref{fig:sel_model_bcb_horizon} to \ref{fig:sel_model_subgroups} in Appendix \ref{append:selected_models}). Additionally, there is a possibility of inaccurate disaggregated forecasts occurring at some point in time, which can lead to a deterioration of the predictive accuracy of the aggregation of disaggregated forecasts. Importantly, this deterioration cannot be attributed to the use of the most recent available weights for each item in the consumption basket. When we conduct the forecasting exercises again using the actual weights, the results demonstrate no significant changes.

%%%%%%%%%%%%%%%%%%%%%%%%%%%%%%%%%%%%%%%%%%

\section{Conclusion}
\label{sec:conclusion}

In this paper, we investigate the use of inflation disaggregations to forecast the aggregate via aggregation of disaggregated forecasts -- what became known as the bottom-up approach. We innovate by considering multi-horizon forecasts of several inflation disaggregates in a data-rich environment (i.e., considering many predictors), which is only possible by employing machine learning methods. Analyzing Brazilian inflation and exploring different levels of disaggregation for inflation, we conduct forecasting exercises from both direct and bottom-up forecast approaches. We highlight the relevance of considering the combination of disaggregated analysis and machine learning methods in the econometrician's toolbox. 

For many forecast horizons, the aggregation of disaggregated forecasts performs as well as survey-based expectations and models that generate forecasts directly from the aggregate. Our results reinforce the benefits of using models in a data-rich environment for inflation forecasting, including aggregating disaggregated forecasts generated from machine learning techniques, mainly during volatile periods. During the COVID-19 pandemic, model-based forecasts, including those based on disaggregated data, tend to provide more accurate forecasts than survey-based expectations. For example, the random forest model based on both aggregate and disaggregated inflation delivers great results for intermediate and longer horizons. The selection of predictors obtained by the adaLASSO and FarmPredict indicates the importance of considering a broad and diversified set of variables when forecasting inflation. Regarding the prediction of individual disaggregates, we find that ML models considering several predictors beat the AR model by a wide and statistically significant margin for all disaggregates and the vast majority of horizons.

This paper can be extended in many ways in future research. Firstly, it is possible to replicate the procedures analyzed here for other developed and emerging countries. Secondly, an important possibility is the formulation and implementation of a methodology that combines different models predicting different disaggregates. As we have seen, there is no dominant technique. Thus, to fully exploit the potential of disaggregated forecasting, it is necessary to combine different forecasts to obtain the aggregate forecast. Beyond combining different models in the dimension of disaggregates, one can also combine different models over horizons to improve the forecast of time-accumulated inflation (e.g., 12-month cumulative inflation). Thirdly, combining different levels of disaggregation could be valuable since increasing the level in some branches may be advantageous while for others, it does not. For example, in the Brazilian price index employed in this paper, some groups may be worth keeping in the final combination, while others may benefit from further breakdown into subgroups or deeper disaggregation. Finally, a fourth possibility not explored in this paper is to use breakeven inflation as a predictor and benchmark in inflation forecasting.

%%%%%%%%%%%%%%%%%%%%%%%%%%%%%%%%%%%%%%%%%%

\normalsize \onehalfspacing
%\newpage
\appendix
\vspace{1cm}
%\begin{Large}
%\noindent \textbf{Appendices}
%\end{Large}

%\setcounter{subsection}{}
%\renewcommand{\thesubsection}{A\arabic{subsection}}

%%%%%%%%%%%%%%%%%%%%%%%%%%%%%%%%%%%%%%%%%%

\section{Groups and subgroups of the IPCA}
\label{append:groups_subgroups}

\setcounter{table}{0}
\renewcommand{\thetable}{A\arabic{table}}

\setcounter{figure}{0}
\renewcommand{\thefigure}{A\arabic{figure}}

% \usepackage{booktabs}
\begin{table}[!ht]
\centering
\small
\caption{Description of groups e subgroups of the IPCA}
\label{tab:groups_subgroups}
\begin{tabular}{llr}
\toprule
\multicolumn{2}{c}{\textbf{Groups / subgroups}}          & \multicolumn{1}{c}{\textbf{Weight}} \\ \cmidrule(lr){1-3}
\\ \vspace{-0.8cm} \\ \multicolumn{2}{l}{Foods and beverages}                & 23.7\vspace{0.1cm}                                \\
\textit{}\;\; & \textit{Food at home}                        & \textit{15.8}                       \\
\textit{} & \textit{Food away from home}                 & \textit{7.9}                        \\
\\ \vspace{-0.8cm} \\ \multicolumn{2}{l}{Housing}                              & 15.5\vspace{0.1cm}                                \\
\textit{} & \textit{Utilities and maintenance}           & \textit{10.3}                       \\
\textit{} & \textit{Domestic fuels and energy}                    & \textit{5.2}                        \\
\\ \vspace{-0.8cm} \\ \multicolumn{2}{l}{Household   goods}                    & 4.1\vspace{0.1cm}                                 \\
\textit{} & \textit{Furniture and fixtures}              & \textit{2.0}                        \\
\textit{} & \textit{Appliances}                          & \textit{1.7}                        \\
\textit{} & \textit{Household operations}                & \textit{0.3}                        \\
\\ \vspace{-0.8cm} \\ \multicolumn{2}{l}{Apparel}                              & 5.5\vspace{0.1cm}                                 \\
\textit{} & \textit{Clothes}                             & \textit{3.6}                        \\
\textit{} & \textit{Footwear and accessories}            & \textit{1.6}                        \\
\textit{} & \textit{Jewelry}                             & \textit{0.3}                        \\
\textit{} & \textit{Fabrics}                             & \textit{0.1}                        \\
\\ \vspace{-0.8cm} \\ \multicolumn{2}{l}{Transportation}                       & 19.0\vspace{0.1cm}                                \\
\textit{} & \textit{Transportation}                      & \textit{19.0}                       \\
\\ \vspace{-0.8cm} \\ \multicolumn{2}{l}{Medical   and personal care}          & 12.1\vspace{0.1cm}                                \\
\textit{} & \textit{Pharmaceutical and optical products \qquad \qquad \qquad} & \textit{3.6}                        \\
\textit{} & \textit{Medical services}                    & \textit{5.5}                        \\
\textit{} & \textit{Personal care}                       & \textit{3.0}                        \\
\\ \vspace{-0.8cm} \\ \multicolumn{2}{l}{Personal   expenses}                  & 10.6\vspace{0.1cm}                                \\
\textit{} & \textit{Personal services}                   & \textit{6.6}                        \\
\textit{} & \textit{Recreation and tobbaco}              & \textit{4.0}                        \\
\\ \vspace{-0.8cm} \\ \multicolumn{2}{l}{Education}                            & 5.1\vspace{0.1cm}                                 \\
\textit{} & \textit{Courses, reading, and stationery}    & \textit{5.1}                        \\
\\ \vspace{-0.8cm} \\ \multicolumn{2}{l}{Communication}                        & 4.3\vspace{0.1cm}                                 \\
\textit{} & \textit{Communication}                       & \textit{4.3}                        \\ \bottomrule
\end{tabular}
\\
\vspace{0.4cm}
\begin{minipage}{10.7cm}
    \footnotesize
    \justifying
    \onehalfspacing
    \noindent \textit{Notes:} The column ``Weight'' shows the average weight of each group and subgroup in the IPCA from January 2014 to June 2022. The text indicates each group and subgroup by \texttt{inf.g\#} and \texttt{inf.sg\#}, respectively, where the \# indicates the order in which each appears in this table.
\end{minipage}
\end{table}

%%%%%%%%%%%%%%%%%%%%%%%%%%%%%%%%%%%%%%%%%%

\newpage

\section{Predictive variables}
\label{append:variables}

\setcounter{table}{0}
\renewcommand{\thetable}{B\arabic{table}}

\setcounter{figure}{0}
\renewcommand{\thefigure}{B\arabic{figure}}

% \usepackage[normalem]{ulem}
%{\footnotesize
\FloatBarrier
\begin{table}[!ht]
%\begin{longtable}{lllcccc}%[!ht]
\centering
\caption{Description of predictive variables}
\label{tab:variables}
\vspace{-0.1cm}
\resizebox{\linewidth}{!}{
\begin{tabular}{rllcccc}
\toprule
%\hline
\multicolumn{1}{c}{\textbf{\#}} & \multicolumn{1}{c}{\textbf{Abbreviation}} & \multicolumn{1}{c}{\textbf{Description}}                          & \textbf{Unit}     & \textbf{Source} & \textbf{Lag} & \textbf{Transformation} \\ 
  \cmidrule(lr){1-7}
\vspace{-0.4cm} \\
\multicolumn{7}{c}{{\ul \textbf{A. Prices and Money}}} \vspace{0.2cm}                                                                                                                                                                         \\
1                               & inf, \, inf.bcb\#,                                     & IPCA and disaggregations                     & index      & IBGE            & 1            & \% change                       \\
  & \hspace{0.1cm} inf.g\#, \, inf.sg\# &  &  &  &  & \\
2                               & expec                                      & Focus-based inflation expectation (available)                     & \hspace{0.1cm} \% per month \hspace{0.1cm}     & BCB            & 0            & --                      \\
3                               & inpc                                       & INPC                                                             & index      & IBGE             & 1            & \% change                       \\
4                               & ipca15                                    & IPCA-15                                                           & index      & IBGE            & 0            & \% change                       \\
5                               & ipc                                       & IPC-Br                                                            & index      & FGV             & 1            & \% change                       \\
6                               & igpm                                      & IGP-M                                                             & index      & FGV             & 1            & \% change                       \\
7                               & igpdi                                     & IGP-DI                                                            & index      & FGV             & 1            & \% change                       \\
8                               & igp10                                     & IGP-10                                                            & index      & FGV             & 1            & \% change                       \\
9                               & ipc\_fipe                                 & IPC-Fipe                                                          & index      & FGV             & 1            & \% change                       \\
10                              & ipa                                       & IPA                                                               & index      & FGV             & 1            & \% change                       \\
11                              & ipa\_ind                                  & IPA                                                               & index      & FGV             & 1            & \% change                       \\
12                              & ipa\_agr                                  & IPA                                                               & index      & FGV             & 1            & \% change                       \\
13                              & incc                                      & INCC                                                              & index      & FGV             & 1            & \% change                       \\
14                              & bm\_broad                                 & Broad Monetary Base -- end-of-period balance                       & index             & BCB             & 2            & \% change               \\
15                              & bm                                        & Monetary Base -- working day balance average                       & index             & BCB             & 2            & \% change               \\
16                              & m1                                        & Money supply M1 -- working day balance average                     & index             & BCB             & 2            & \% change               \\
17                              & m2                                        & Money supply M2 -- end-of-period balance                           & index             & BCB             & 2            & \% change               \\
18                              & m3                                        & Money supply M3 -- end-of-period balance                           & index             & BCB             & 2            & \% change               \\
19                              & m4                                        & Money supply M4 -- end-of-period balance                           & index             & BCB             & 2            & \% change               \\ 
  \vspace{-0.3cm} \\
\multicolumn{7}{c}{{\ul \textbf{B. Commodities   Prices}}} \vspace{0.2cm}                                                                                                                                                                      \\
20                              & icbr                                      & Brazilian Commodity index (all)                                   & index             & BCB             & 1            & \% change               \\
21                              & icbr\_agr                                 & Brazilian Commodity index -- agriculture                           & index             & BCB             & 1            & \% change               \\
22                              & icbr\_metal                               & Brazilian Commodity index -- metal                                 & index             & BCB             & 1            & \% change               \\
23                              & icbr\_energy                              & Brazilian Commodity index -- energy                                & index             & BCB             & 1            & \% change               \\ 
  \vspace{-0.3cm} \\
\multicolumn{7}{c}{{\ul \textbf{C. Economic Activity}}} \vspace{0.2cm}                                              \\
24                              & ibcbr                                     & Brazilian IBC-Br Economic Activity index                          & index             & BCB             & 3            & \% change               \\
25          & month\_gdp                            & GDP monthly -- current prices                                     & R\$ million      & BCB             & 1            & \% change \\
26                              & tcu                                       & Total capacity utilization -- manufacturing industry \hspace{2cm}                & \%                & FGV             & 1            & first difference        \\ 
27	& pimpf			& Industry Production -- general			& index	& IBGE	& 2	& \% change	\\
28  & pmc\_total    & Retail sales volume -- total    & index & IBGE            & 2            & \% change               \\
29  & pmc\_fuel                              & Retail sales volume -- fuels and oils                              & index & IBGE            & 2            & \% change               \\
30  & pmc\_supermarket                       & Retail sales volume -- supermarkets and food products              & index & IBGE            & 2            & \% change               \\
31  & pmc\_clothing                          & Retail sales volume -- fabrics, clothing and shoes                 & index & IBGE            & 2            & \% change               \\
32  & pmc\_house                             & Retail sales volume -- furniture and appliances                    & index & IBGE            & 2            & \% change               \\
33  & pmc\_drugstore                         & Retail sales volume -- pharmaceutical and cosmetic articles    & index             & IBGE            & 2            & \% change               \\
34  & pmc\_paper                             & Retail sales volume -- books, newspapers and stationery            & index & IBGE            & 2            & \% change               \\
35  & pmc\_office                            & Retail sales volume -- office and eletronical equipments           & index & IBGE            & 2            & \% change               \\
36  & pmc\_others                            & Retail sales volume -- others                                      & index & IBGE            & 2            & \% change               \\
37  & pmc\_building                           & Retail sales volume -- building material                           & index & IBGE            & 2            & \% change               \\
38  & pmc\_auto                               & Retail sales volume -- automotive and parts                        & index             & IBGE            & 2            & \% change   \\
39   & steel  & Steel production   & index   & BCB   & 1 & --  \\
40                              & prod\_vehicles                            & Vehicle production -- total                                        & units             & Anfavea         & 1            & \% change               \\
41                              & prod\_agr\_mach                           & Production of agricultural machinery -- total                      & units             & Anfavea         & 1            & \% change               \\
42                              & vehicle\_sales                            & Vehicle sales by dealerships -- total                              & units             & Fenabrave    & 1            & \% change               \\
\vspace{-0.3cm} \\
\multicolumn{7}{c}{{\ul \textbf{D. Employment}}} \vspace{0.2cm}                                                                                                                                                                                \\
43                              & unem                                      & Unemployment (combination of PME and PNADC)                       & \%                & IBGE            & 2            & first difference        \\
44                              & employment                                      & Registered employess by economic activity - Total      & units             & IBGE            & 1            & \% change        \\
45                              & aggreg\_wage                              & Overall Earnings (broad wage income)                              & R\$ (million)     & BCB            & 1            & \% change               \\
46                              & min\_wage                                 & Federal Minimum Wage                                              & R\$               & MTb             & 0            & \% change               \\ 
47                              & income                                 & Households gross disposable national income                                              & R\$ (million)               & BCB             & 2            & \% change               \\ 
\vspace{-0.3cm} \\
\multicolumn{7}{c}{{\ul \textbf{E. Electricity}}} \vspace{0.2cm}                                                                                                                                                                               \\
48                              & elec                                      & Electricity consumption -- total                                   & GWh         
      & Eletrobrás      & 2            & \% change               \\
49                              & elec\_res                                 & Electricity consumption -- residential                             & GWh               & Eletrobrás      & 2            & \% change               \\
50                              & elec\_com                                 & Electricity consumption -- commercial                              & GWh               & Eletrobrás      & 2            & \% change               \\
51                              & elec\_ind                                 & Electricity consumption -- industry                                & GWh               & Eletrobrás      & 2            & \% change               \\ 
  \vspace{-0.3cm} \\
\multicolumn{7}{c}{{\ul \textbf{F. Confidence}}} \vspace{0.2cm}    \\
52                              & cons\_confidence                          & Consumer Confidence index                                         & index             & Fecomercio      & 1            & \% change               \\
53                              & future\_expec                             & Future expectations index                                         & index             & Fecomercio      & 1            & \% change               \\ 
54                              & conditions                             & 	Current economic conditions index                                   & index             & Fecomercio      & 1            & \% change               \\ 
\bottomrule %\\ \vspace{-0.7cm} \\
%\multicolumn{7}{r}{(continued on next page)}
\end{tabular}
}
\\
\vspace{-0.1cm}
\footnotesize
\flushright
(continued on next page)
%\end{longtable}
\end{table}
\FloatBarrier

%%%%%%%%%%%%%%%%%%%%%%%%%%%%%%%%%%%%%%%%%%%%%%%%%%%%%%
%%%%%%%%%%%%%%%%%%%%%%%%%%%%%%%%%%%%%%%%%%%%%%%%%%%%%%

\newpage

\FloatBarrier
\begin{table}[!ht]
%\begin{longtable}{lllcccc}%[!ht]
\centering
\caption*{Table B1: Description of predictive variables (cont.)}
\vspace{-0.1cm}
\resizebox{\linewidth}{!}{
\begin{tabular}{rllcccc}
\toprule
%\hline
\multicolumn{1}{c}{\textbf{\#}} & \multicolumn{1}{c}{\textbf{Abbreviation}} & \multicolumn{1}{c}{\textbf{Description}}                          & \textbf{Unit}     & \textbf{Source} & \textbf{Lag} & \textbf{Transformation} \\ 
  \cmidrule(lr){1-7}
  \vspace{-0.3cm} \\
\multicolumn{7}{c}{{\ul \textbf{G. Finance}}} \vspace{0.2cm}                                                                                                                                                                                   \\
55                              & ibovespa                                  & Ibovespa index                                                    & \% per month      & B3    & 0            & -                       \\
56                              & irf\_m                                    & Anbima Market Index of the prefixed federal bonds                 & index             & Anbima          & 1            & \% change               \\
57                              & ima\_s                                    & Anbima Market Index of the federal bonds tied to the Selic   & index             & Anbima          & 1            & \% change               \\
58                              & ima\_b                                    & Anbima Market Index of the federal bonds tied to the IPCA & index             & Anbima          & 1            & \% change               \\
59                              & ima                                       & General Anbima Market index                                       & index             & Anbima          & 1            & \% change               \\
60                              & saving\_deposits                          & Saving deposits -- end-of-period balance                          & R\$ (mil)         & BCB             & 2            & \% change               \\
61                              & selic                                     & Selic Basic Interest rate                                         & \% per month      & BCB             & 0            & -- \\ %first difference                       \\
62                              & cdi                                       & Cetip DI Interbank Deposits rate                                  & \% per month      & Cetip           & 0            & -- \\ %first difference                       \\
63                              & tjlp                                      & TJLP Long-term Interest rate                                      & \% per year       & BCB             & 1            & -- \\ %first difference                       \\
64                              & ntnb                                      & 3-Year Treasury (real) Rate indexed to the IPCA (NTN-B)                                       & \% per year       & Anbima             & 0            & -- \\
65          & embi                                  & Emerging Markets Bond Index Plus -- Brazil                        & b.p. acc. month  & JP Morgan       & 0            & first difference        \\
66 & vix                    & CBOE Volatility Index (VIX)                                   & index            & CBOE      & 0 &                  \\
\vspace{-0.3cm} \\
\multicolumn{7}{c}{{\ul \textbf{H. Credit}}} \vspace{0.2cm}                                                                                                                                                                                    \\
67                              & cred\_total                               & Credit outstanding -- total                                        & R\$ (million)     & BCB             & 2            & \% change               \\
68                              & cred\_dgp                                 & Credit outstanding as a percentage of GDP                         & \% of GDP         & BCB             & 2            & first difference        \\
69                              & indebt\_house1                             & Household debt to income -- total                                          & \% of 12m income  & BCB             & 2            & first difference        \\ 
70                              & indebt\_house2                             & Household debt without mortgage loans                                          & \% of 12m income  & BCB             & 2            & first difference        \\ 
  \vspace{-0.3cm} \\
\multicolumn{7}{c}{{\ul \textbf{I. Government}}} \vspace{0.2cm}                                                                                                                                                 \\
71                              & net\_debt\_gdp                            & Net public debt (\% GDP) -Consolidated public sector              & \% of GDP          & BCB             & 2            & first difference        \\
72                              & net\_debt                                 & Net public debt -- Total -- Consolidated public sector              & R\$ (million)    & BCB             & 2            & \% change               \\
73                              & net\_debt\_fedgov\_bcb                    & Net public debt -- Federal Government and Central Bank             & R\$ (million)     & BCB             & 2            & \% change               \\
74                              & net\_debt\_states                         & Net public debt -- State governments                               & R\$ (million)     & BCB             & 2            & \% change               \\
75                              & net\_debt\_cities                         & Net public debt -- Municipal governments                           & R\$ (million)     & BCB             & 2            & \% change               \\
76                              & primary\_result                           & Primary result -- Consolidated public sector                       & R\$ (million)     & BCB             & 2            & \% change               \\
77                              & debt\_fedgov\_old                         & Gross general government debt -- Method used until 2007              & R\$ (million)     & BCB             & 2            & \% change               \\
78                              & debt\_fedgov\_new                         & Gross general government debt -- Method used since 2008            & R\$ (million)     & BCB             & 2            & \% change               \\
79                              & treasury\_emit                            & National Treasuary domestic securities -- Total issued             & R\$ (million)     & BCB             & 2            & \% change               \\
80                              & treasury\_mkt                             & National Treasuary domestic securities -- Total on market          & R\$ (million)     & BCB             & 2            & \% change               \\
81                              & treasury\_term                            & National Treasury securities debt -- medium term                   & months            & BCB             & 2            & first difference        \\
82                              & treasury\_dur                             & National Treasury securities debt -- medium duration               & months            & BCB             & 2            & first difference        \\ 
  \vspace{-0.3cm} \\
\multicolumn{7}{c}{{\ul \textbf{J. Exchange and   International Transactions}}} \vspace{0.2cm}                                                                                                                                                 \\
83                              & reer                                      & Real Effective Exchange Rate                                      & R\$/other         & BIS             & 1            & \% change               \\
84                              & usd\_brl\_end                             & USD-BRL rate -- end-of-period                                      & R\$/US\$            & BCB             & 0            & \% change               \\
85                              & usd\_brl\_av                              & USD-BRL rate -- monthly average                                    & R\$/US\$            & BCB             & 1            & \% change               \\
86                              & eur\_brl\_end                             & EUR-BRL rate -- end-of-period                                      & R\$/€            & Bloomberg             & 0            & \% change               \\
87                              & eur\_brl\_av                              & EUR-BRL rate -- monthly average                                    & R\$/€            & Bloomberg             & 1            & \% change               \\
88                              & current\_account                          & Current account -- net                                             & US\$ (million)    & BCB             & 2            & \% change               \\
89                              & trade\_balance                            & Balance on goods and services -- net (Brazilian trade balance)     & US\$ (million)    & BCB             & 2            & \% change               \\
90                              & exports                                   & Exports                                                           & US\$ (million)    & BCB             & 2            & \% change               \\
91                              & imports                                   & Imports                                                           & US\$ (million)    & BCB             & 2            & \% change               \\
\bottomrule %\hline
\end{tabular}
}
%\end{longtable}
\\
\footnotesize
\justifying
\singlespacing
\noindent \textit{Notes:} ``Lag'' column indicates the delay for each variable to become available and ``Transformations'' column indicates transformation implemented to guarantee the stationarity of the series.
\end{table}
\FloatBarrier

\begin{comment}

\newpage

\FloatBarrier
\begin{figure}[!t]
    \centering
    \caption{Explanatory variables correlation}
    \label{fig:variables_corr}
    \vspace{-0.2cm}
    \includegraphics[width=\linewidth]{Figures/cor_plot.pdf}
    \\
    \vspace{0.3cm}
    \footnotesize
    \justifying
    \noindent \textit{Notes:} Correlations were calculated considering the period from March 2007 to January 2021 according to data availability for all variables.
\end{figure}
\FloatBarrier

\end{comment}

%%%%%%%%%%%%%%%%%%%%%%%%%%%%%%%%%%%%%%%%%%

\newpage

\section{Forecast performance for disaggregates by horizon}
\label{append:disaggreg_by_horizon}

\setcounter{table}{0}
\renewcommand{\thetable}{C\arabic{table}}

\setcounter{figure}{0}
\renewcommand{\thefigure}{C\arabic{figure}}

\subsection{BCB disaggregates before and after the COVID-19 pandemic}

% latex table generated in R 4.1.3 by xtable 1.8-4 package
% version adapted by Gilberto Boaretto
% Mon Apr 03 11:29:14 2023
\begin{table}[ht]
\centering
\caption{\centering Out-of-sample RMSE for BCB disaggregate (in terms of RMSE of the AR model): Jan/2014 to Feb/2020, by disaggregate and horizon} 
\label{tab:rmse_disaggreg_bcb_sample1}
\resizebox{1\linewidth}{!}{
\begin{tabular}{llllllllllllll}
  \toprule
\multicolumn{1}{c}{Estimator/Model \;\;\;} & $h = 0\quad\,$ & $h = 1\quad\,$ & $h = 2\quad\,$ & $h = 3\quad\,$ & $h = 4\quad\,$ & $h = 5\quad\,$ & $h = 6\quad\,$ & $h = 7\quad\,$ & $h = 8\quad\,$ & $h = 9\quad\,$ & $h = 10\;\;$ & $h = 11\;\;$ & all $h$\;\;\;\; \\ 
  \cmidrule(lr){1-14}
\vspace{-0.3cm} \\ \multicolumn{13}{c}{\textbf{\underline{A. Monitored Prices}}} \\ \vspace{-0.3cm} &  &  &  &  &  &  &  &  &  &  &  &  &  \\ 
  AR & 1.000 & 1.000 & 1.000 & 1.000 & 1.000 & 1.000 & 1.000 & 1.000 & \textit{\blue{1.000}} & 1.000 & 1.000 & 1.000 & 1.000 \\ 
  Augmented AR & $\textbf{\blue{0.740}}^{\,***}$ & $\textbf{\blue{0.860}}^{\,***}$ & \textbf{\blue{0.971}} & 0.985 & 1.005 & 1.010 & 1.030 & 1.040 & 1.040 & 1.015 & 1.012 & 1.024 & $0.983^{\,*}$ \\ 
  Ridge & $\textit{\blue{0.744}}^{\,***}$ & 0.952 & 0.990 & \textbf{\blue{0.979}} & \textbf{\blue{0.987}} & \textit{\blue{0.975}} & \textit{\blue{0.994}} & \textit{\blue{0.991}} & \textbf{\blue{0.998}} & \textbf{\blue{0.977}} & $\textbf{\blue{0.942}}^{\,**}$ & $\textbf{\blue{0.929}}^{\,*}$ & $\textbf{\blue{0.958}}^{\,***}$ \\ 
  adaLASSO & $0.795^{\,***}$ & $0.929^{\,*}$ & \textit{\blue{0.982}} & \textit{\blue{0.985}} & \textit{\blue{0.991}} & \textbf{\blue{0.972}} & 0.999 & 0.998 & 1.012 & 1.010 & 0.967 & 0.963 & $\textit{\blue{0.970}}^{\,***}$ \\ 
  Factor & $0.777^{\,***}$ & $0.937^{\,*}$ & 1.018 & 1.011 & 1.013 & 0.995 & 1.019 & 1.028 & 1.036 & \textit{\blue{0.994}} & 0.976 & \textit{\blue{0.961}} & $0.983^{\,*}$ \\ 
  FarmPredict & $0.815^{\,***}$ & 0.957 & 1.008 & 1.004 & 1.009 & 1.011 & \textbf{\blue{0.988}} & \textbf{\blue{0.983}} & 1.013 & 0.998 & 0.972 & 0.967 & $0.979^{\,**}$ \\ 
  Target Factor & $0.796^{\,***}$ & \textit{\blue{0.921}} & 1.004 & 1.011 & 1.069 & 1.076 & 1.056 & 1.202 & 1.151 & 1.091 & 1.052 & 1.012 & 1.043 \\ 
  CSR & $0.950^{\,*}$ & $0.961^{\,*}$ & 0.987 & 0.996 & 1.013 & 1.030 & 1.028 & 1.041 & 1.030 & 1.008 & 0.997 & 1.012 & 1.005 \\ 
  Random Forest & $0.913^{\,*}$ & 0.991 & 1.032 & 1.026 & 1.060 & 1.076 & 1.085 & 1.050 & 1.047 & 1.014 & $\textit{\blue{0.967}}^{\,*}$ & 0.977 & 1.021 \\ 
  \vspace{-0.3cm} \\ \multicolumn{13}{c}{\textbf{\underline{B. Non-Tradables}}} \\ \vspace{-0.3cm} &  &  &  &  &  &  &  &  &  &  &  &  &  \\ 
  AR & 1.000 & 1.000 & 1.000 & 1.000 & 1.000 & 1.000 & 1.000 & 1.000 & 1.000 & \textit{\blue{1.000}} & \textit{\blue{1.000}} & \textbf{\blue{1.000}} & 1.000 \\ 
  Augmented AR & $\textit{\blue{0.784}}^{\,***}$ & $0.801^{\,**}$ & $0.830^{\,***}$ & $0.795^{\,***}$ & $0.781^{\,***}$ & $0.735^{\,***}$ & $0.819^{\,***}$ & $0.831^{\,**}$ & $0.914^{\,*}$ & 1.063 & 1.112 & 1.056 & $0.867^{\,***}$ \\ 
  Ridge & $\textbf{\blue{0.763}}^{\,***}$ & $\textit{\blue{0.791}}^{\,***}$ & $\textit{\blue{0.821}}^{\,***}$ & $\textit{\blue{0.792}}^{\,***}$ & $0.775^{\,***}$ & $0.735^{\,***}$ & $0.808^{\,***}$ & $0.820^{\,***}$ & $\textit{\blue{0.903}}^{\,**}$ & 1.024 & 1.059 & \textit{\blue{1.015}} & $\textit{\blue{0.850}}^{\,***}$ \\ 
  adaLASSO & $0.850^{\,**}$ & $\textbf{\blue{0.790}}^{\,**}$ & $\textbf{\blue{0.783}}^{\,***}$ & $\textbf{\blue{0.773}}^{\,***}$ & $\textbf{\blue{0.746}}^{\,***}$ & $\textbf{\blue{0.709}}^{\,***}$ & $\textit{\blue{0.786}}^{\,***}$ & $\textit{\blue{0.778}}^{\,***}$ & $0.922^{\,*}$ & 1.039 & 1.089 & 1.087 & $0.851^{\,***}$ \\ 
  Factor & $0.875^{\,*}$ & $0.837^{\,**}$ & $0.843^{\,**}$ & $0.828^{\,***}$ & $0.791^{\,***}$ & $0.758^{\,***}$ & $0.814^{\,**}$ & $0.813^{\,**}$ & $0.911^{\,**}$ & 1.043 & 1.053 & 1.026 & $0.872^{\,***}$ \\ 
  FarmPredict & $0.878^{\,*}$ & $0.827^{\,**}$ & $0.835^{\,***}$ & $0.799^{\,***}$ & $\textit{\blue{0.770}}^{\,***}$ & $0.741^{\,***}$ & $0.814^{\,***}$ & $0.784^{\,***}$ & $0.923^{\,*}$ & 1.051 & 1.070 & 1.038 & $0.866^{\,***}$ \\ 
  Target Factor & 1.054 & $0.805^{\,**}$ & $0.895^{\,*}$ & 1.027 & 0.965 & $0.750^{\,***}$ & $0.832^{\,**}$ & 0.907 & 1.004 & 1.085 & 1.058 & 1.105 & $0.949^{\,**}$ \\ 
  CSR & $0.882^{\,***}$ & $0.863^{\,***}$ & $0.887^{\,***}$ & $0.861^{\,***}$ & $0.834^{\,***}$ & $0.799^{\,***}$ & $0.843^{\,***}$ & $0.864^{\,**}$ & 0.979 & 1.076 & 1.074 & 1.035 & $0.906^{\,***}$ \\ 
  Random Forest & $0.820^{\,***}$ & $0.803^{\,***}$ & $0.828^{\,***}$ & $0.797^{\,***}$ & $0.775^{\,***}$ & $\textit{\blue{0.721}}^{\,***}$ & $\textbf{\blue{0.765}}^{\,***}$ & $\textbf{\blue{0.777}}^{\,***}$ & $\textbf{\blue{0.841}}^{\,***}$ & \textbf{\blue{0.942}} & \textbf{\blue{0.997}} & 1.038 & $\textbf{\blue{0.832}}^{\,***}$ \\ 
  \vspace{-0.3cm} \\ \multicolumn{13}{c}{\textbf{\underline{C. Tradables}}} \\ \vspace{-0.3cm} &  &  &  &  &  &  &  &  &  &  &  &  &  \\ 
  AR & 1.000 & 1.000 & 1.000 & 1.000 & 1.000 & 1.000 & 1.000 & 1.000 & 1.000 & 1.000 & 1.000 & 1.000 & 1.000 \\ 
  Augmented AR & $0.786^{\,**}$ & $\textit{\blue{0.869}}^{\,***}$ & 0.966 & 1.047 & 1.074 & 1.049 & 1.000 & 0.972 & 0.979 & 1.011 & 1.061 & 1.039 & 0.996 \\ 
  Ridge & $0.763^{\,**}$ & $0.884^{\,**}$ & $\textbf{\blue{0.900}}^{\,**}$ & 0.951 & $\textit{\blue{0.948}}^{\,*}$ & 1.060 & $0.857^{\,**}$ & $0.846^{\,**}$ & $0.869^{\,**}$ & $\textbf{\blue{0.855}}^{\,***}$ & 1.047 & $\textbf{\blue{0.842}}^{\,***}$ & $0.907^{\,***}$ \\ 
  adaLASSO & $\textit{\blue{0.727}}^{\,***}$ & $0.911^{\,**}$ & $0.930^{\,*}$ & $\textit{\blue{0.948}}^{\,*}$ & $0.956^{\,*}$ & $\textbf{\blue{0.856}}^{\,**}$ & $\textbf{\blue{0.831}}^{\,***}$ & $0.875^{\,**}$ & 0.938 & $\textit{\blue{0.871}}^{\,**}$ & 0.932 & $\textit{\blue{0.853}}^{\,***}$ & $0.889^{\,***}$ \\ 
  Factor & $0.741^{\,***}$ & $\textbf{\blue{0.866}}^{\,***}$ & 0.952 & 1.002 & 0.993 & 0.909 & $0.858^{\,**}$ & $\textit{\blue{0.844}}^{\,**}$ & $\textbf{\blue{0.861}}^{\,**}$ & $0.892^{\,**}$ & $\textit{\blue{0.885}}^{\,**}$ & $0.888^{\,**}$ & $0.894^{\,***}$ \\ 
  FarmPredict & $0.747^{\,***}$ & $0.894^{\,**}$ & $0.914^{\,**}$ & 0.997 & 0.985 & 0.910 & $\textit{\blue{0.853}}^{\,**}$ & $\textbf{\blue{0.843}}^{\,**}$ & $\textit{\blue{0.868}}^{\,**}$ & $0.887^{\,***}$ & $\textbf{\blue{0.863}}^{\,**}$ & $0.858^{\,***}$ & $\textit{\blue{0.887}}^{\,***}$ \\ 
  Target Factor & $\textbf{\blue{0.722}}^{\,**}$ & 0.936 & 0.932 & 1.004 & 1.119 & 1.010 & $0.903^{\,*}$ & $0.863^{\,*}$ & $0.892^{\,*}$ & 1.001 & 1.000 & 0.964 & $0.953^{\,**}$ \\ 
  CSR & $0.844^{\,***}$ & 1.046 & 1.176 & 1.065 & 0.979 & 0.928 & 0.965 & 0.949 & $0.919^{\,*}$ & $0.927^{\,*}$ & 0.971 & 0.973 & 0.981 \\ 
  Random Forest & $0.768^{\,***}$ & $0.880^{\,**}$ & $\textit{\blue{0.913}}^{\,**}$ & $\textbf{\blue{0.923}}^{\,**}$ & $\textbf{\blue{0.888}}^{\,**}$ & $\textit{\blue{0.880}}^{\,*}$ & $0.868^{\,**}$ & $0.879^{\,*}$ & $0.905^{\,*}$ & $0.894^{\,**}$ & $0.896^{\,**}$ & $0.883^{\,**}$ & $\textbf{\blue{0.884}}^{\,***}$ \\ 
   \bottomrule
\end{tabular}
}
\\
\footnotesize
\justifying
\singlespacing
\onehalfspacing
\noindent \textit{Notes:} $^{***}$, $^{**}$, and $^{*}$ indicate that for a specific disaggregate and forecast horizon, a model $m$ performed statistically better than an AR model at 1, 5, and 10\% significance levels in a one-tailed Diebold-Mariano test with $\hip_0: \text{MSE}\big(\widehat{\pi}_{i,\,t+h\, |\,t}^{m}\big) = \text{MSE}\big(\pi_{i,\,t+h\,|\,t}^{\text{AR}}\big) \; \textit{ versus } \hip_1: \text{MSE}\big(\widehat{\pi}_{i,\,t+h\,|\,t}^{m}\big) < \text{MSE}\big(\pi_{i,\,t+h\,|\,t}^{\text{AR}}\big)$. The value highlighted in bold blue indicates the best model for each horizon in terms of RMSE ratio with respect to the AR model, and the values in blue italics indicate the second and third best models.
\end{table}

\newpage

% latex table generated in R 4.1.3 by xtable 1.8-4 package
% version adapted by Gilberto Boaretto
% Mon Apr 03 11:31:20 2023
\begin{table}[ht]
\centering
\caption{\centering Out-of-sample RMSE for BCB disaggregate (in terms of RMSE of the AR model): Mar/2020 to Jun/2022, by disaggregate and horizon} 
\label{tab:rmse_disaggreg_bcb_sample2}
\resizebox{1\linewidth}{!}{
\begin{tabular}{llllllllllllll}
  \toprule
\multicolumn{1}{c}{Estimator/Model \;\;\;} & $h = 0\quad\,$ & $h = 1\quad\,$ & $h = 2\quad\,$ & $h = 3\quad\,$ & $h = 4\quad\,$ & $h = 5\quad\,$ & $h = 6\quad\,$ & $h = 7\quad\,$ & $h = 8\quad\,$ & $h = 9\quad\,$ & $h = 10\;\;$ & $h = 11\;\;$ & all $h$\;\;\;\; \\ 
  \cmidrule(lr){1-14}
\vspace{-0.3cm} \\ \multicolumn{13}{c}{\textbf{\underline{A. Monitored Prices}}} \\ \vspace{-0.3cm} &  &  &  &  &  &  &  &  &  &  &  &  &  \\ 
  AR & 1.000 & 1.000 & 1.000 & 1.000 & 1.000 & \textbf{\blue{1.000}} & \textbf{\blue{1.000}} & 1.000 & 1.000 & 1.000 & 1.000 & 1.000 & 1.000 \\ 
  Augmented AR & $\textbf{\blue{0.634}}^{\,***}$ & \textbf{\blue{0.955}} & 1.066 & 1.066 & 1.049 & 1.040 & 1.022 & 1.029 & 1.036 & 1.064 & 1.034 & 1.046 & 1.012 \\ 
  Ridge & $\textit{\blue{0.665}}^{\,***}$ & \textit{\blue{0.955}} & \textbf{\blue{0.984}} & \textbf{\blue{0.991}} & \textbf{\blue{0.981}} & \textit{\blue{1.006}} & \textit{\blue{1.011}} & \textbf{\blue{0.978}} & 0.975 & \textbf{\blue{0.997}} & \textit{\blue{0.977}} & \textit{\blue{0.962}} & $\textbf{\blue{0.962}}^{\,***}$ \\ 
  adaLASSO & $0.706^{\,***}$ & 0.999 & \textit{\blue{0.991}} & \textbf{\blue{0.991}} & \textbf{\blue{0.981}} & \textit{\blue{1.006}} & \textit{\blue{1.011}} & \textit{\blue{0.978}} & \textit{\blue{0.975}} & 0.997 & \textit{\blue{0.977}} & \textit{\blue{0.962}} & $\textit{\blue{0.968}}^{\,**}$ \\ 
  Factor & $0.679^{\,***}$ & 1.010 & 0.997 & \textbf{\blue{0.991}} & \textbf{\blue{0.981}} & \textit{\blue{1.006}} & 1.012 & 0.985 & \textbf{\blue{0.971}} & \textbf{\blue{0.997}} & \textit{\blue{0.977}} & \textit{\blue{0.962}} & $0.969^{\,**}$ \\ 
  FarmPredict & $0.722^{\,***}$ & 1.007 & 0.996 & \textbf{\blue{0.991}} & \textbf{\blue{0.981}} & \textit{\blue{1.006}} & \textit{\blue{1.011}} & 0.980 & 0.975 & \textbf{\blue{0.997}} & \textit{\blue{0.977}} & \textit{\blue{0.962}} & $0.971^{\,**}$ \\ 
  Target Factor & $0.692^{\,***}$ & 1.160 & 1.092 & 1.165 & 1.056 & 1.083 & 1.121 & 1.083 & 1.134 & 1.103 & 1.101 & 1.163 & 1.088 \\ 
  CSR & 1.001 & 1.053 & 1.063 & 1.080 & 1.117 & 1.081 & 1.042 & 1.052 & 1.106 & 1.116 & 1.052 & 1.047 & 1.068 \\ 
  Random Forest & $0.821^{\,**}$ & 1.056 & 1.048 & 1.101 & 1.055 & 1.076 & 1.115 & 1.089 & 1.077 & 1.058 & \textbf{\blue{0.967}} & \textbf{\blue{0.928}} & 1.036 \\ 
  \vspace{-0.3cm} \\ \multicolumn{13}{c}{\textbf{\underline{B. Non-Tradables}}} \\ \vspace{-0.3cm} &  &  &  &  &  &  &  &  &  &  &  &  &  \\ 
  AR & 1.000 & 1.000 & 1.000 & 1.000 & 1.000 & 1.000 & 1.000 & 1.000 & 1.000 & 1.000 & 1.000 & 1.000 & 1.000 \\ 
  Augmented AR & 0.973 & 0.889 & 0.863 & 0.860 & 0.855 & 0.878 & 0.849 & 0.907 & 0.893 & $0.818^{\,**}$ & $0.844^{\,**}$ & $0.891^{\,*}$ & $0.872^{\,***}$ \\ 
  Ridge & 0.957 & 0.869 & 0.841 & $0.833^{\,*}$ & 0.828 & 0.858 & 0.853 & 0.910 & 0.894 & $0.830^{\,**}$ & $0.848^{\,**}$ & $0.884^{\,**}$ & $0.862^{\,***}$ \\ 
  adaLASSO & 1.001 & $0.773^{\,**}$ & $0.771^{\,*}$ & $0.784^{\,**}$ & $0.752^{\,**}$ & $0.775^{\,*}$ & $0.862^{\,*}$ & 1.005 & 0.989 & $0.891^{\,*}$ & $\textit{\blue{0.805}}^{\,***}$ & $\textit{\blue{0.825}}^{\,***}$ & $0.841^{\,***}$ \\ 
  Factor & \textit{\blue{0.921}} & $\textit{\blue{0.750}}^{\,**}$ & $\textit{\blue{0.754}}^{\,*}$ & $\textit{\blue{0.763}}^{\,**}$ & $\textit{\blue{0.720}}^{\,**}$ & $0.803^{\,*}$ & $\textit{\blue{0.841}}^{\,*}$ & 0.987 & 1.031 & 0.957 & $0.878^{\,**}$ & $0.872^{\,**}$ & $0.851^{\,***}$ \\ 
  FarmPredict & 0.938 & $\textbf{\blue{0.736}}^{\,**}$ & $\textbf{\blue{0.707}}^{\,**}$ & $\textbf{\blue{0.715}}^{\,**}$ & $\textbf{\blue{0.699}}^{\,**}$ & $\textbf{\blue{0.737}}^{\,**}$ & 0.896 & 1.009 & 1.018 & 0.932 & $0.857^{\,***}$ & $0.833^{\,**}$ & $\textit{\blue{0.832}}^{\,***}$ \\ 
  Target Factor & 1.003 & 0.925 & 0.898 & 1.014 & 0.884 & 0.883 & 0.983 & 1.286 & 1.070 & 0.919 & $0.818^{\,*}$ & 0.985 & 0.963 \\ 
  CSR & \textbf{\blue{0.916}} & $0.866^{\,**}$ & $0.886^{\,*}$ & 0.928 & 0.920 & $0.848^{\,*}$ & $0.882^{\,*}$ & \textit{\blue{0.903}} & $\textbf{\blue{0.821}}^{\,**}$ & $\textbf{\blue{0.775}}^{\,***}$ & $0.853^{\,***}$ & 0.991 & $0.887^{\,***}$ \\ 
  Random Forest & 0.975 & 0.951 & $0.799^{\,**}$ & $0.814^{\,**}$ & $0.784^{\,**}$ & $\textit{\blue{0.743}}^{\,***}$ & $\textbf{\blue{0.840}}^{\,**}$ & \textbf{\blue{0.869}} & \textit{\blue{0.880}} & $\textit{\blue{0.810}}^{\,*}$ & $\textbf{\blue{0.780}}^{\,***}$ & $\textbf{\blue{0.788}}^{\,***}$ & $\textbf{\blue{0.824}}^{\,***}$ \\ 
  \vspace{-0.3cm} \\ \multicolumn{13}{c}{\textbf{\underline{C. Tradables}}} \\ \vspace{-0.3cm} &  &  &  &  &  &  &  &  &  &  &  &  &  \\ 
  AR & 1.000 & 1.000 & 1.000 & 1.000 & 1.000 & 1.000 & 1.000 & 1.000 & 1.000 & 1.000 & 1.000 & 1.000 & 1.000 \\ 
  Augmented AR & 0.932 & 0.978 & 1.033 & 1.100 & 1.092 & 1.053 & 0.994 & $0.915^{\,*}$ & $0.913^{\,*}$ & 0.955 & 1.019 & 1.076 & 1.012 \\ 
  Ridge & \textit{\blue{0.912}} & 0.976 & 1.024 & 1.020 & 0.974 & 1.021 & $0.953^{\,*}$ & $0.909^{\,***}$ & 0.977 & 1.046 & 1.006 & $0.946^{\,*}$ & $0.983^{\,*}$ \\ 
  adaLASSO & $\textbf{\blue{0.842}}^{\,*}$ & 0.989 & 0.959 & $0.930^{\,**}$ & $0.849^{\,***}$ & $0.830^{\,***}$ & $0.871^{\,***}$ & $0.842^{\,***}$ & $0.891^{\,**}$ & $0.908^{\,*}$ & $0.883^{\,**}$ & $\textit{\blue{0.810}}^{\,***}$ & $0.881^{\,***}$ \\ 
  Factor & 0.964 & 1.024 & 1.000 & $\textit{\blue{0.895}}^{\,***}$ & $0.841^{\,***}$ & $\textit{\blue{0.798}}^{\,***}$ & $0.825^{\,***}$ & $0.929^{\,**}$ & 1.018 & 1.070 & 1.006 & $0.928^{\,**}$ & $0.938^{\,***}$ \\ 
  FarmPredict & 0.944 & 1.079 & 1.063 & $0.928^{\,**}$ & $0.857^{\,***}$ & $0.825^{\,***}$ & $0.864^{\,***}$ & $0.962^{\,*}$ & 1.027 & 1.053 & 1.031 & 0.949 & $0.963^{\,***}$ \\ 
  Target Factor & 0.938 & $\textbf{\blue{0.796}}^{\,**}$ & $\textbf{\blue{0.872}}^{\,*}$ & 0.946 & $0.838^{\,**}$ & $0.838^{\,***}$ & $\textbf{\blue{0.694}}^{\,***}$ & $\textbf{\blue{0.627}}^{\,***}$ & $\textit{\blue{0.741}}^{\,**}$ & $\textbf{\blue{0.706}}^{\,***}$ & 0.975 & $0.875^{\,*}$ & $\textit{\blue{0.820}}^{\,***}$ \\ 
  CSR & 0.972 & 1.044 & $0.926^{\,*}$ & $0.918^{\,*}$ & $\textit{\blue{0.838}}^{\,**}$ & $0.834^{\,***}$ & $0.826^{\,***}$ & $0.792^{\,***}$ & $0.789^{\,***}$ & $0.784^{\,***}$ & $\textit{\blue{0.858}}^{\,***}$ & $0.877^{\,***}$ & $0.863^{\,***}$ \\ 
  Random Forest & 0.933 & \textit{\blue{0.941}} & $\textit{\blue{0.897}}^{\,**}$ & $\textbf{\blue{0.875}}^{\,***}$ & $\textbf{\blue{0.808}}^{\,***}$ & $\textbf{\blue{0.753}}^{\,***}$ & $\textit{\blue{0.745}}^{\,***}$ & $\textit{\blue{0.710}}^{\,***}$ & $\textbf{\blue{0.734}}^{\,***}$ & $\textit{\blue{0.773}}^{\,***}$ & $\textbf{\blue{0.757}}^{\,***}$ & $\textbf{\blue{0.703}}^{\,***}$ & $\textbf{\blue{0.792}}^{\,***}$ \\ 
   \bottomrule
\end{tabular}
}
\\
\footnotesize
\justifying
\singlespacing
\noindent \textit{Notes:} see Table \ref{tab:rmse_disaggreg_bcb_sample1}.
\end{table}

\newpage

%%%%%
\subsection{Groups}
%%%%%

% latex table generated in R 4.1.3 by xtable 1.8-4 package
% version adapted by Gilberto Boaretto
% Thu Mar 30 17:33:02 2023
\begin{table}[!ht]
\centering
\caption{\centering Out-of-sample RMSE for IBGE groups (in terms of RMSE of the AR model): Jan/2014 to Jun/2022, by disaggregate and horizon} 
\label{tab:rmse_disaggreg_groups_append}
\resizebox{1\linewidth}{!}{
\begin{tabular}{llllllllllllll}
  \toprule
\multicolumn{1}{c}{Estimator/Model \;\;\;} & $h = 0\quad\,$ & $h = 1\quad\,$ & $h = 2\quad\,$ & $h = 3\quad\,$ & $h = 4\quad\,$ & $h = 5\quad\,$ & $h = 6\quad\,$ & $h = 7\quad\,$ & $h = 8\quad\,$ & $h = 9\quad\,$ & $h = 10\;\;$ & $h = 11\;\;$ & all $h$\;\;\;\; \\ 
  \cmidrule(lr){1-14}
\vspace{-0.3cm} \\ \multicolumn{14}{c}{\underline{\textbf{A. Foods and beverages} (\texttt{inf.g1})}} \\ \vspace{-0.3cm} &  &  &  &  &  &  &  &  &  &  &  &  &  \\ 
  AR & 1.000 & 1.000 & 1.000 & 1.000 & 1.000 & 1.000 & 1.000 & 1.000 & 1.000 & 1.000 & 1.000 & 1.000 & 1.000 \\ 
  Augmented AR & 0.959 & 0.979 & 1.036 & 1.034 & 1.045 & 1.030 & 1.006 & 0.994 & 0.938 & $0.929^{\,**}$ & 1.025 & 1.069 & 1.007 \\ 
  Ridge & $0.898^{\,*}$ & 0.916 & $0.884^{\,*}$ & $0.903^{\,**}$ & $0.899^{\,**}$ & 0.951 & 0.964 & 0.965 & $0.913^{\,**}$ & $\textit{\blue{0.831}}^{\,***}$ & $\textit{\blue{0.811}}^{\,***}$ & $\textit{\blue{0.776}}^{\,***}$ & $\textit{\blue{0.886}}^{\,***}$ \\ 
  adaLASSO & $\textit{\blue{0.805}}^{\,**}$ & $\textit{\blue{0.839}}^{\,***}$ & $\textit{\blue{0.877}}^{\,**}$ & $\textit{\blue{0.877}}^{\,**}$ & $\textit{\blue{0.896}}^{\,**}$ & $0.937^{\,*}$ & 1.012 & 1.018 & 0.991 & $0.896^{\,**}$ & $0.884^{\,*}$ & $0.788^{\,***}$ & $0.900^{\,***}$ \\ 
  Factor & $0.841^{\,**}$ & $0.889^{\,**}$ & 0.962 & 0.992 & 0.964 & 0.995 & 1.014 & 0.966 & $0.910^{\,**}$ & $0.833^{\,***}$ & $0.820^{\,***}$ & $0.797^{\,***}$ & $0.911^{\,***}$ \\ 
  FarmPredict & $0.847^{\,**}$ & $0.888^{\,*}$ & 0.907 & 0.962 & 0.933 & 0.984 & 1.024 & 0.966 & $0.913^{\,**}$ & $0.872^{\,***}$ & $0.856^{\,**}$ & $0.799^{\,***}$ & $0.909^{\,***}$ \\ 
  Target Factor & 0.925 & $0.908^{\,*}$ & 0.945 & 1.010 & 0.952 & $\textit{\blue{0.915}}^{\,*}$ & $\textbf{\blue{0.894}}^{\,**}$ & \textbf{\blue{0.924}} & $\textbf{\blue{0.862}}^{\,**}$ & 0.966 & 1.013 & $0.847^{\,***}$ & $0.932^{\,***}$ \\ 
  CSR & $0.886^{\,**}$ & 0.955 & 1.067 & 1.058 & 0.983 & 0.944 & $0.935^{\,*}$ & 0.938 & 1.023 & $0.886^{\,***}$ & 0.920 & $0.856^{\,**}$ & $0.954^{\,**}$ \\ 
  Random Forest & $\textbf{\blue{0.789}}^{\,***}$ & $\textbf{\blue{0.831}}^{\,***}$ & $\textbf{\blue{0.836}}^{\,***}$ & $\textbf{\blue{0.859}}^{\,***}$ & $\textbf{\blue{0.848}}^{\,***}$ & $\textbf{\blue{0.890}}^{\,**}$ & $\textit{\blue{0.926}}^{\,*}$ & $\textit{\blue{0.927}}^{\,*}$ & $\textit{\blue{0.890}}^{\,**}$ & $\textbf{\blue{0.817}}^{\,***}$ & $\textbf{\blue{0.792}}^{\,***}$ & $\textbf{\blue{0.761}}^{\,***}$ & $\textbf{\blue{0.844}}^{\,***}$ \\ 
  \vspace{-0.3cm} \\ \multicolumn{14}{c}{\underline{\textbf{B. Housing} (\texttt{inf.g2})}} \\ \vspace{-0.3cm} &  &  &  &  &  &  &  &  &  &  &  &  &  \\ 
  AR & 1.000 & 1.000 & 1.000 & 1.000 & 1.000 & 1.000 & 1.000 & 1.000 & 1.000 & 1.000 & 1.000 & 1.000 & 1.000 \\ 
  Augmented AR & $\textbf{\blue{0.780}}^{\,***}$ & $\textbf{\blue{0.886}}^{\,**}$ & 0.981 & 1.000 & 0.992 & 0.967 & 1.006 & 1.012 & 1.032 & 0.999 & 1.017 & 1.017 & $0.977^{\,**}$ \\ 
  Ridge & $0.873^{\,**}$ & $0.914^{\,**}$ & \textbf{\blue{0.952}} & $\textbf{\blue{0.948}}^{\,**}$ & $\textbf{\blue{0.928}}^{\,**}$ & $\textit{\blue{0.883}}^{\,***}$ & $\textbf{\blue{0.895}}^{\,***}$ & $\textit{\blue{0.936}}^{\,*}$ & $\textit{\blue{0.900}}^{\,**}$ & $0.902^{\,**}$ & $\textit{\blue{0.861}}^{\,**}$ & $0.914^{\,*}$ & $\textbf{\blue{0.908}}^{\,***}$ \\ 
  adaLASSO & $0.862^{\,***}$ & $0.946^{\,*}$ & \textbf{\blue{0.952}} & $0.949^{\,**}$ & $0.948^{\,*}$ & $0.884^{\,***}$ & $0.899^{\,**}$ & $\textbf{\blue{0.935}}^{\,*}$ & $0.907^{\,**}$ & $0.907^{\,**}$ & $\textbf{\blue{0.860}}^{\,**}$ & $0.914^{\,*}$ & $\textit{\blue{0.913}}^{\,***}$ \\ 
  Factor & $0.889^{\,**}$ & 0.967 & 0.952 & $\textbf{\blue{0.948}}^{\,**}$ & $\textit{\blue{0.948}}^{\,*}$ & $0.884^{\,***}$ & $\textit{\blue{0.896}}^{\,**}$ & $0.936^{\,**}$ & $\textbf{\blue{0.897}}^{\,**}$ & $\textit{\blue{0.901}}^{\,**}$ & $0.861^{\,**}$ & $\textit{\blue{0.913}}^{\,*}$ & $0.915^{\,***}$ \\ 
  FarmPredict & $0.875^{\,**}$ & 0.949 & 0.957 & 0.979 & 0.965 & $\textbf{\blue{0.879}}^{\,***}$ & $0.901^{\,**}$ & $0.944^{\,*}$ & $0.917^{\,*}$ & $\textbf{\blue{0.898}}^{\,**}$ & $0.862^{\,**}$ & $\textbf{\blue{0.913}}^{\,*}$ & $0.919^{\,***}$ \\ 
  Target Factor & $\textit{\blue{0.807}}^{\,***}$ & $\textit{\blue{0.901}}^{\,**}$ & 1.184 & 1.051 & 1.044 & $0.930^{\,*}$ & 0.981 & 1.068 & 0.976 & 1.033 & 0.943 & 0.971 & 0.992 \\ 
  CSR & $0.916^{\,***}$ & $0.934^{\,***}$ & 0.986 & 1.027 & 0.977 & $0.943^{\,*}$ & $0.945^{\,**}$ & 0.984 & $0.925^{\,**}$ & $0.929^{\,**}$ & $0.907^{\,**}$ & $0.924^{\,*}$ & $0.949^{\,***}$ \\ 
  Random Forest & $0.898^{\,**}$ & 0.948 & 1.001 & 0.980 & 0.966 & $0.944^{\,*}$ & 0.987 & 1.002 & 0.972 & 0.974 & $0.903^{\,*}$ & 0.960 & $0.961^{\,***}$ \\ 
  \vspace{-0.3cm} \\ \multicolumn{14}{c}{\underline{\textbf{C. Household goods} (\texttt{inf.g3})}} \\ \vspace{-0.3cm} &  &  &  &  &  &  &  &  &  &  &  &  &  \\ 
  AR & \textit{\blue{1.000}} & 1.000 & 1.000 & 1.000 & 1.000 & 1.000 & 1.000 & 1.000 & 1.000 & 1.000 & 1.000 & 1.000 & 1.000 \\ 
  Augmented AR & 1.053 & 1.046 & 1.025 & 0.984 & 0.999 & 1.055 & 1.064 & 1.041 & 1.078 & 1.091 & 1.125 & 1.109 & 1.062 \\ 
  Ridge & 1.140 & 1.077 & 1.015 & 0.976 & 1.040 & 0.955 & $0.926^{\,**}$ & $0.878^{\,***}$ & $0.929^{\,**}$ & $0.908^{\,**}$ & $0.919^{\,**}$ & $0.888^{\,***}$ & $0.956^{\,***}$ \\ 
  adaLASSO & 1.034 & 1.031 & 1.030 & 0.996 & 0.981 & $0.904^{\,**}$ & $0.855^{\,***}$ & $0.800^{\,***}$ & $0.856^{\,***}$ & $0.858^{\,***}$ & $0.854^{\,***}$ & $0.849^{\,***}$ & $0.905^{\,***}$ \\ 
  Factor & 1.015 & $0.913^{\,*}$ & \textit{\blue{0.940}} & $\textbf{\blue{0.905}}^{\,**}$ & 0.966 & $0.869^{\,***}$ & $0.853^{\,***}$ & $0.885^{\,***}$ & $0.937^{\,*}$ & $0.883^{\,**}$ & $0.885^{\,**}$ & $0.844^{\,***}$ & $0.900^{\,***}$ \\ 
  FarmPredict & \textbf{\blue{0.995}} & $\textit{\blue{0.912}}^{\,*}$ & 0.951 & $0.928^{\,*}$ & 0.972 & $0.883^{\,**}$ & $0.872^{\,***}$ & $0.890^{\,**}$ & $0.942^{\,*}$ & $0.900^{\,**}$ & $0.906^{\,**}$ & $0.870^{\,***}$ & $0.912^{\,***}$ \\ 
  Target Factor & 1.030 & 1.000 & 0.954 & $\textit{\blue{0.914}}^{\,*}$ & 1.152 & $0.905^{\,**}$ & $0.864^{\,***}$ & $0.849^{\,***}$ & $0.879^{\,***}$ & $0.839^{\,***}$ & $0.829^{\,***}$ & $0.800^{\,***}$ & $0.905^{\,***}$ \\ 
  CSR & 1.125 & $\textbf{\blue{0.893}}^{\,*}$ & $\textbf{\blue{0.928}}^{\,*}$ & 0.945 & $\textbf{\blue{0.885}}^{\,**}$ & $\textbf{\blue{0.834}}^{\,***}$ & $\textbf{\blue{0.800}}^{\,***}$ & $\textbf{\blue{0.757}}^{\,***}$ & $\textit{\blue{0.817}}^{\,***}$ & $\textit{\blue{0.807}}^{\,***}$ & $\textit{\blue{0.812}}^{\,***}$ & $\textit{\blue{0.797}}^{\,***}$ & $\textit{\blue{0.849}}^{\,***}$ \\ 
  Random Forest & 1.053 & 0.914 & 0.961 & $0.919^{\,*}$ & $\textit{\blue{0.916}}^{\,*}$ & $\textit{\blue{0.839}}^{\,***}$ & $\textit{\blue{0.809}}^{\,***}$ & $\textit{\blue{0.764}}^{\,***}$ & $\textbf{\blue{0.800}}^{\,***}$ & $\textbf{\blue{0.790}}^{\,***}$ & $\textbf{\blue{0.803}}^{\,***}$ & $\textbf{\blue{0.777}}^{\,***}$ & $\textbf{\blue{0.845}}^{\,***}$ \\ 
  \vspace{-0.3cm} \\ \multicolumn{14}{c}{\underline{\textbf{D. Apparel} (\texttt{inf.g4})}} \\ \vspace{-0.3cm} &  &  &  &  &  &  &  &  &  &  &  &  &  \\ 
  AR & 1.000 & 1.000 & 1.000 & 1.000 & 1.000 & 1.000 & 1.000 & 1.000 & 1.000 & 1.000 & 1.000 & 1.000 & 1.000 \\ 
  Augmented AR & $\textit{\blue{0.718}}^{\,***}$ & $\textbf{\blue{0.710}}^{\,***}$ & $\textit{\blue{0.731}}^{\,***}$ & $0.896^{\,**}$ & 0.971 & $0.876^{\,**}$ & $0.808^{\,***}$ & $0.801^{\,***}$ & $0.912^{\,*}$ & 0.975 & 1.082 & 1.026 & $0.874^{\,***}$ \\ 
  Ridge & $\textbf{\blue{0.714}}^{\,***}$ & $\textit{\blue{0.710}}^{\,***}$ & $\textbf{\blue{0.709}}^{\,***}$ & $0.835^{\,***}$ & $0.893^{\,**}$ & $\textit{\blue{0.834}}^{\,***}$ & $0.797^{\,***}$ & $0.797^{\,***}$ & $0.858^{\,***}$ & $0.941^{\,*}$ & 0.987 & 0.966 & $\textbf{\blue{0.836}}^{\,***}$ \\ 
  adaLASSO & $0.752^{\,***}$ & $0.796^{\,***}$ & $0.797^{\,***}$ & $\textit{\blue{0.798}}^{\,***}$ & $\textit{\blue{0.880}}^{\,**}$ & $\textbf{\blue{0.805}}^{\,***}$ & $0.827^{\,***}$ & $0.789^{\,***}$ & $\textbf{\blue{0.806}}^{\,***}$ & 0.974 & 1.001 & $0.931^{\,*}$ & $\textit{\blue{0.844}}^{\,***}$ \\ 
  Factor & $0.739^{\,***}$ & $0.816^{\,***}$ & $0.793^{\,***}$ & $\textbf{\blue{0.775}}^{\,***}$ & $\textbf{\blue{0.878}}^{\,**}$ & $0.848^{\,***}$ & $0.823^{\,***}$ & $0.849^{\,***}$ & 0.976 & $0.929^{\,*}$ & 0.984 & $\textit{\blue{0.924}}^{\,**}$ & $0.863^{\,***}$ \\ 
  FarmPredict & $0.907^{\,*}$ & $0.912^{\,*}$ & $0.867^{\,**}$ & $0.824^{\,***}$ & 0.930 & $0.865^{\,**}$ & $0.905^{\,**}$ & $0.908^{\,**}$ & 1.001 & 0.950 & 1.000 & 0.948 & $0.920^{\,***}$ \\ 
  Target Factor & $0.755^{\,***}$ & $0.801^{\,***}$ & $0.800^{\,***}$ & $0.909^{\,*}$ & 0.959 & $0.862^{\,*}$ & $\textbf{\blue{0.742}}^{\,***}$ & $\textbf{\blue{0.775}}^{\,***}$ & $0.856^{\,***}$ & 1.030 & 1.152 & 1.014 & $0.884^{\,***}$ \\ 
  CSR & 0.922 & 0.996 & $0.938^{\,*}$ & $0.889^{\,**}$ & $0.918^{\,*}$ & $0.861^{\,**}$ & $\textit{\blue{0.789}}^{\,***}$ & $0.829^{\,***}$ & $0.879^{\,***}$ & $\textit{\blue{0.903}}^{\,*}$ & \textit{\blue{0.933}} & $0.944^{\,*}$ & $0.898^{\,***}$ \\ 
  Random Forest & $0.831^{\,***}$ & $0.846^{\,***}$ & $0.902^{\,**}$ & $0.897^{\,**}$ & 0.969 & $0.869^{\,**}$ & $0.810^{\,***}$ & $\textit{\blue{0.784}}^{\,***}$ & $\textit{\blue{0.823}}^{\,***}$ & $\textbf{\blue{0.902}}^{\,**}$ & $\textbf{\blue{0.921}}^{\,*}$ & $\textbf{\blue{0.906}}^{\,**}$ & $0.866^{\,***}$ \\ 
  \vspace{-0.3cm} \\ \multicolumn{14}{c}{\underline{\textbf{E. Transportation} (\texttt{inf.g5})}} \\ \vspace{-0.3cm} &  &  &  &  &  &  &  &  &  &  &  &  &  \\ 
  AR & 1.000 & 1.000 & \textit{\blue{1.000}} & \textit{\blue{1.000}} & 1.000 & 1.000 & 1.000 & 1.000 & 1.000 & 1.000 & 1.000 & 1.000 & 1.000 \\ 
  Augmented AR & $0.854^{\,***}$ & 1.002 & 1.061 & 1.017 & \textit{\blue{0.982}} & \textit{\blue{0.964}} & 0.977 & 1.003 & 0.962 & 0.994 & 1.062 & 1.121 & 1.001 \\ 
  Ridge & $0.859^{\,***}$ & 1.139 & 1.116 & 1.057 & 0.994 & 0.970 & $\textbf{\blue{0.907}}^{\,**}$ & $\textbf{\blue{0.868}}^{\,***}$ & $\textbf{\blue{0.877}}^{\,***}$ & $\textbf{\blue{0.898}}^{\,**}$ & 0.997 & 1.004 & $0.966^{\,**}$ \\ 
  adaLASSO & $\textit{\blue{0.786}}^{\,***}$ & \textit{\blue{0.998}} & 1.108 & 1.070 & 1.002 & 0.983 & $0.927^{\,**}$ & $0.889^{\,***}$ & $0.893^{\,***}$ & $0.917^{\,**}$ & 0.996 & 1.001 & $0.961^{\,***}$ \\ 
  Factor & $\textbf{\blue{0.778}}^{\,***}$ & \textbf{\blue{0.966}} & 1.112 & 1.070 & 0.997 & 0.992 & $0.932^{\,**}$ & $0.889^{\,***}$ & $\textit{\blue{0.891}}^{\,**}$ & $0.927^{\,**}$ & \textit{\blue{0.985}} & \textit{\blue{0.988}} & $\textit{\blue{0.959}}^{\,***}$ \\ 
  FarmPredict & $0.790^{\,***}$ & 1.014 & 1.097 & 1.065 & 0.999 & 0.989 & $0.933^{\,**}$ & $0.910^{\,***}$ & $0.898^{\,**}$ & $0.932^{\,*}$ & 0.986 & 1.004 & $0.966^{\,***}$ \\ 
  Target Factor & $0.870^{\,**}$ & 1.126 & 1.213 & 1.188 & 1.037 & 1.023 & 0.951 & $0.902^{\,**}$ & 0.937 & 0.943 & 1.041 & 1.155 & 1.026 \\ 
  CSR & $0.891^{\,**}$ & 1.001 & \textbf{\blue{0.990}} & \textbf{\blue{0.994}} & \textbf{\blue{0.947}} & $\textbf{\blue{0.942}}^{\,**}$ & $0.925^{\,**}$ & $\textit{\blue{0.876}}^{\,***}$ & $0.932^{\,*}$ & $0.942^{\,*}$ & 0.995 & 1.015 & $\textbf{\blue{0.951}}^{\,***}$ \\ 
  Random Forest & $0.896^{\,**}$ & 1.035 & 1.122 & 1.075 & 1.012 & 0.985 & $\textit{\blue{0.921}}^{\,**}$ & $0.891^{\,***}$ & $0.909^{\,***}$ & $\textit{\blue{0.911}}^{\,**}$ & \textbf{\blue{0.972}} & \textbf{\blue{0.959}} & $0.968^{\,***}$ \\ 
  \vspace{-0.3cm} \\ \multicolumn{14}{c}{\underline{\textbf{F. Medical and personal care} (\texttt{inf.g6})}} \\ \vspace{-0.3cm} &  &  &  &  &  &  &  &  &  &  &  &  &  \\ 
  AR & 1.000 & 1.000 & 1.000 & 1.000 & 1.000 & 1.000 & 1.000 & 1.000 & 1.000 & 1.000 & 1.000 & 1.000 & 1.000 \\ 
  Augmented AR & 0.991 & 0.987 & $\textbf{\blue{0.861}}^{\,***}$ & $\textit{\blue{0.792}}^{\,***}$ & $\textbf{\blue{0.781}}^{\,***}$ & $\textbf{\blue{0.803}}^{\,***}$ & $\textbf{\blue{0.833}}^{\,***}$ & $\textbf{\blue{0.872}}^{\,***}$ & $\textit{\blue{0.914}}^{\,**}$ & $\textit{\blue{0.898}}^{\,**}$ & $\textbf{\blue{0.900}}^{\,***}$ & 0.996 & $\textbf{\blue{0.881}}^{\,***}$ \\ 
  Ridge & \textit{\blue{0.972}} & $\textit{\blue{0.977}}^{\,**}$ & $\textit{\blue{0.896}}^{\,***}$ & $0.822^{\,***}$ & $\textit{\blue{0.798}}^{\,***}$ & $\textit{\blue{0.811}}^{\,***}$ & $0.846^{\,***}$ & $\textit{\blue{0.883}}^{\,***}$ & $\textbf{\blue{0.913}}^{\,***}$ & $0.921^{\,***}$ & $0.916^{\,***}$ & $\textit{\blue{0.959}}^{\,**}$ & $\textit{\blue{0.889}}^{\,***}$ \\ 
  adaLASSO & 0.974 & 1.016 & $0.897^{\,**}$ & $0.905^{\,*}$ & 0.931 & 1.218 & 1.030 & 1.051 & 0.984 & 1.013 & 0.944 & 0.961 & 0.995 \\ 
  Factor & \textbf{\blue{0.970}} & 0.996 & 0.977 & 0.998 & $0.923^{\,*}$ & 0.992 & 0.993 & 1.008 & 1.081 & 1.023 & 0.981 & 0.994 & 0.994 \\ 
  FarmPredict & 0.974 & 1.041 & 1.045 & 1.013 & 0.988 & 1.023 & 1.020 & 1.025 & 1.043 & 1.004 & 0.980 & 1.023 & 1.015 \\ 
  Target Factor & 0.990 & \textbf{\blue{0.958}} & 0.948 & $\textbf{\blue{0.764}}^{\,***}$ & $0.826^{\,**}$ & 1.026 & $\textit{\blue{0.834}}^{\,***}$ & 0.926 & $0.919^{\,**}$ & $\textbf{\blue{0.877}}^{\,**}$ & $0.914^{\,*}$ & 0.964 & $0.909^{\,***}$ \\ 
  CSR & 1.058 & 1.050 & $0.918^{\,**}$ & $0.857^{\,***}$ & 0.965 & 1.062 & 1.037 & 1.020 & 0.957 & $0.915^{\,*}$ & 0.919 & \textbf{\blue{0.955}} & $0.974^{\,*}$ \\ 
  Random Forest & 0.985 & 1.002 & $0.899^{\,*}$ & $0.908^{\,*}$ & 0.940 & 1.056 & 0.954 & $0.907^{\,**}$ & $0.929^{\,*}$ & $0.908^{\,**}$ & $\textit{\blue{0.904}}^{\,**}$ & 0.969 & $0.945^{\,***}$ \\ 
  \bottomrule
\end{tabular}
}
\\
\vspace{-0.1cm}
\footnotesize
\flushright
(continued on next page)
\end{table}

\newpage

% latex table generated in R 4.1.3 by xtable 1.8-4 package
% version adapted by Gilberto Boaretto
% Thu Mar 30 17:33:02 2023
\begin{table}[!ht]
\centering
\caption*{\centering Table C3: Out-of-sample RMSE for IBGE groups (in terms of RMSE of the AR model): Jan/2014 to Jun/2022, by disaggregate and horizon (cont.)}
\resizebox{1\linewidth}{!}{
\begin{tabular}{llllllllllllll}
  \toprule
\multicolumn{1}{c}{Estimator/Model \;\;\;} & $h = 0\quad\,$ & $h = 1\quad\,$ & $h = 2\quad\,$ & $h = 3\quad\,$ & $h = 4\quad\,$ & $h = 5\quad\,$ & $h = 6\quad\,$ & $h = 7\quad\,$ & $h = 8\quad\,$ & $h = 9\quad\,$ & $h = 10\;\;$ & $h = 11\;\;$ & all $h$\;\;\;\; \\ 
  \cmidrule(lr){1-14}
  \vspace{-0.3cm} \\ \multicolumn{14}{c}{\underline{\textbf{G. Personal expenses} (\texttt{inf.g7})}} \\ \vspace{-0.3cm} &  &  &  &  &  &  &  &  &  &  &  &  &  \\ 
  AR & 1.000 & 1.000 & 1.000 & 1.000 & 1.000 & 1.000 & 1.000 & 1.000 & 1.000 & 1.000 & 1.000 & \textit{\blue{1.000}} & 1.000 \\ 
  Augmented AR & 0.953 & 0.962 & 1.006 & 0.981 & 0.997 & 1.023 & 1.040 & 1.039 & 1.058 & 1.015 & 0.977 & 1.034 & 1.006 \\ 
  Ridge & 0.936 & $0.884^{\,**}$ & $0.884^{\,***}$ & $0.806^{\,***}$ & $0.878^{\,***}$ & 0.978 & 0.967 & $0.875^{\,**}$ & 0.949 & 0.937 & 0.943 & 1.021 & $0.917^{\,***}$ \\ 
  adaLASSO & $0.871^{\,**}$ & $\textbf{\blue{0.840}}^{\,***}$ & $\textbf{\blue{0.845}}^{\,***}$ & $\textbf{\blue{0.774}}^{\,***}$ & $0.840^{\,***}$ & 0.969 & 0.949 & $\textbf{\blue{0.863}}^{\,**}$ & $\textbf{\blue{0.927}}^{\,*}$ & $\textit{\blue{0.897}}^{\,**}$ & \textit{\blue{0.907}} & 1.005 & $0.886^{\,***}$ \\ 
  Factor & $\textit{\blue{0.864}}^{\,**}$ & $0.856^{\,***}$ & $0.858^{\,***}$ & $0.783^{\,***}$ & $\textbf{\blue{0.789}}^{\,***}$ & $\textbf{\blue{0.895}}^{\,**}$ & $\textbf{\blue{0.905}}^{\,*}$ & $\textit{\blue{0.863}}^{\,**}$ & 0.946 & $0.899^{\,**}$ & 0.919 & \textbf{\blue{0.982}} & $\textbf{\blue{0.876}}^{\,***}$ \\ 
  FarmPredict & $0.900^{\,*}$ & $\textit{\blue{0.850}}^{\,***}$ & $0.882^{\,**}$ & $\textit{\blue{0.781}}^{\,***}$ & $\textit{\blue{0.812}}^{\,***}$ & $\textit{\blue{0.909}}^{\,*}$ & \textit{\blue{0.941}} & $0.876^{\,**}$ & \textit{\blue{0.931}} & $\textbf{\blue{0.871}}^{\,**}$ & $\textbf{\blue{0.874}}^{\,*}$ & 1.001 & $\textit{\blue{0.882}}^{\,***}$ \\ 
  Target Factor & 0.998 & 0.973 & 1.061 & 0.914 & 0.971 & 1.026 & 0.984 & 0.931 & 1.098 & 0.936 & 1.070 & 1.178 & 1.008 \\ 
  CSR & $\textbf{\blue{0.839}}^{\,***}$ & 0.946 & $0.888^{\,**}$ & $0.822^{\,***}$ & 0.936 & 0.957 & 0.949 & $0.881^{\,**}$ & 1.066 & 0.922 & 0.936 & 1.018 & $0.927^{\,***}$ \\ 
  Random Forest & $0.881^{\,**}$ & $0.851^{\,***}$ & $\textit{\blue{0.852}}^{\,***}$ & $0.838^{\,***}$ & 0.943 & 1.040 & 1.024 & $0.903^{\,*}$ & 0.955 & 0.933 & 0.949 & 1.048 & $0.931^{\,***}$ \\ 
  \vspace{-0.3cm} \\ \multicolumn{14}{c}{\underline{\textbf{H. Education} (\texttt{inf.g8})}} \\ \vspace{-0.3cm} &  &  &  &  &  &  &  &  &  &  &  &  &  \\ 
  AR & 1.000 & 1.000 & 1.000 & 1.000 & 1.000 & 1.000 & 1.000 & 1.000 & 1.000 & 1.000 & 1.000 & 1.000 & 1.000 \\ 
  Augmented AR & $0.443^{\,***}$ & $\textit{\blue{0.396}}^{\,***}$ & $0.395^{\,***}$ & $\textbf{\blue{0.367}}^{\,***}$ & $\textit{\blue{0.351}}^{\,***}$ & $\textbf{\blue{0.358}}^{\,***}$ & $0.422^{\,***}$ & $\textit{\blue{0.411}}^{\,***}$ & $0.507^{\,***}$ & 0.885 & \textit{\blue{0.869}} & \textit{\blue{0.879}} & $0.451^{\,***}$ \\ 
  Ridge & $\textbf{\blue{0.429}}^{\,***}$ & $\textbf{\blue{0.386}}^{\,***}$ & $\textbf{\blue{0.377}}^{\,***}$ & $0.383^{\,***}$ & $0.364^{\,***}$ & $0.380^{\,***}$ & $0.419^{\,***}$ & $0.418^{\,***}$ & $0.495^{\,***}$ & $\textbf{\blue{0.823}}^{\,*}$ & \textbf{\blue{0.853}} & \textbf{\blue{0.835}} & $\textbf{\blue{0.446}}^{\,***}$ \\ 
  adaLASSO & $\textit{\blue{0.432}}^{\,***}$ & $0.400^{\,***}$ & $0.388^{\,***}$ & $0.385^{\,***}$ & $0.361^{\,***}$ & $0.384^{\,***}$ & $0.420^{\,***}$ & $0.419^{\,***}$ & $0.488^{\,***}$ & \textit{\blue{0.874}} & 0.933 & 0.897 & $0.457^{\,***}$ \\ 
  Factor & $0.432^{\,***}$ & $0.408^{\,***}$ & $0.381^{\,***}$ & $\textit{\blue{0.374}}^{\,***}$ & $\textbf{\blue{0.349}}^{\,***}$ & $\textit{\blue{0.370}}^{\,***}$ & $\textbf{\blue{0.409}}^{\,***}$ & $\textbf{\blue{0.404}}^{\,***}$ & $\textbf{\blue{0.478}}^{\,***}$ & $0.878^{\,*}$ & $0.923^{\,*}$ & 0.881 & $\textit{\blue{0.450}}^{\,***}$ \\ 
  FarmPredict & $\textit{\blue{0.432}}^{\,***}$ & $0.400^{\,***}$ & $0.387^{\,***}$ & $0.385^{\,***}$ & $0.362^{\,***}$ & $0.384^{\,***}$ & $\textit{\blue{0.418}}^{\,***}$ & $0.419^{\,***}$ & $\textit{\blue{0.486}}^{\,***}$ & 0.882 & 0.930 & 0.897 & $0.457^{\,***}$ \\ 
  Target Factor & $0.536^{\,***}$ & $0.425^{\,***}$ & $\textit{\blue{0.380}}^{\,***}$ & $0.390^{\,***}$ & $0.377^{\,***}$ & $0.442^{\,***}$ & $0.455^{\,***}$ & $0.457^{\,***}$ & $0.504^{\,***}$ & 0.894 & 0.915 & 0.943 & $0.487^{\,***}$ \\ 
  CSR & $0.636^{\,***}$ & 0.823 & $0.740^{\,***}$ & $0.663^{\,***}$ & $0.630^{\,***}$ & $0.668^{\,***}$ & $0.654^{\,***}$ & $0.528^{\,***}$ & $0.629^{\,***}$ & 0.924 & 1.031 & 1.020 & $0.698^{\,***}$ \\ 
  Random Forest & $0.569^{\,***}$ & $0.508^{\,***}$ & $0.538^{\,***}$ & $0.542^{\,***}$ & $0.490^{\,***}$ & $0.543^{\,***}$ & $0.520^{\,***}$ & $0.547^{\,***}$ & $0.607^{\,***}$ & 0.927 & 0.915 & 0.916 & $0.573^{\,***}$ \\ 
  \vspace{-0.3cm} \\ \multicolumn{14}{c}{\underline{\textbf{I. Communication} (\texttt{inf.g9})}} \\ \vspace{-0.3cm} &  &  &  &  &  &  &  &  &  &  &  &  &  \\ 
  AR & 1.000 & 1.000 & 1.000 & 1.000 & 1.000 & 1.000 & 1.000 & 1.000 & 1.000 & 1.000 & 1.000 & 1.000 & 1.000 \\ 
  Augmented AR & 1.095 & 1.040 & 1.002 & 1.029 & 1.067 & 1.060 & 1.086 & 1.041 & 1.077 & 1.074 & 1.183 & 1.030 & 1.066 \\ 
  Ridge & $0.748^{\,***}$ & $\textbf{\blue{0.806}}^{\,***}$ & $\textit{\blue{0.769}}^{\,***}$ & $0.836^{\,**}$ & $\textit{\blue{0.822}}^{\,*}$ & $0.859^{\,**}$ & $0.864^{\,**}$ & $0.740^{\,**}$ & $0.771^{\,***}$ & $0.773^{\,***}$ & $0.842^{\,**}$ & $0.874^{\,*}$ & $0.805^{\,***}$ \\ 
  adaLASSO & $\textbf{\blue{0.733}}^{\,***}$ & $\textit{\blue{0.810}}^{\,***}$ & $0.773^{\,***}$ & $\textit{\blue{0.828}}^{\,***}$ & $0.823^{\,*}$ & $\textbf{\blue{0.848}}^{\,**}$ & $\textbf{\blue{0.855}}^{\,**}$ & $\textbf{\blue{0.732}}^{\,**}$ & $\textit{\blue{0.767}}^{\,***}$ & $\textbf{\blue{0.712}}^{\,***}$ & $0.819^{\,*}$ & $\textbf{\blue{0.824}}^{\,*}$ & $\textbf{\blue{0.790}}^{\,***}$ \\ 
  Factor & $\textit{\blue{0.747}}^{\,***}$ & $0.820^{\,***}$ & $\textbf{\blue{0.764}}^{\,***}$ & $\textbf{\blue{0.820}}^{\,***}$ & $0.830^{\,**}$ & $0.868^{\,**}$ & $0.868^{\,**}$ & $\textit{\blue{0.733}}^{\,**}$ & $0.777^{\,***}$ & $0.733^{\,***}$ & $\textbf{\blue{0.787}}^{\,**}$ & $\textit{\blue{0.835}}^{\,*}$ & $\textit{\blue{0.795}}^{\,***}$ \\ 
  FarmPredict & $0.748^{\,***}$ & $0.832^{\,***}$ & $0.770^{\,***}$ & $0.832^{\,**}$ & $\textbf{\blue{0.818}}^{\,*}$ & $\textit{\blue{0.853}}^{\,**}$ & $\textit{\blue{0.859}}^{\,**}$ & $0.735^{\,**}$ & $\textbf{\blue{0.764}}^{\,***}$ & $\textit{\blue{0.733}}^{\,***}$ & $\textit{\blue{0.817}}^{\,**}$ & $0.858^{\,*}$ & $0.798^{\,***}$ \\ 
  Target Factor & 1.031 & 1.426 & 1.096 & 1.210 & 1.221 & 1.149 & 1.552 & 0.922 & 1.104 & 0.916 & 1.230 & 1.103 & 1.164 \\ 
  CSR & $0.776^{\,***}$ & $0.865^{\,**}$ & $0.842^{\,**}$ & $0.915^{\,*}$ & 0.930 & $0.894^{\,**}$ & 0.925 & $0.781^{\,***}$ & $0.851^{\,**}$ & $0.889^{\,**}$ & $0.832^{\,*}$ & $0.887^{\,*}$ & $0.863^{\,***}$ \\ 
  Random Forest & $0.768^{\,***}$ & $0.822^{\,***}$ & $0.831^{\,**}$ & $0.880^{\,*}$ & 0.873 & $0.900^{\,*}$ & $0.893^{\,*}$ & $0.762^{\,**}$ & $0.796^{\,**}$ & $0.798^{\,***}$ & 0.921 & $0.852^{\,*}$ & $0.838^{\,***}$ \\ 
   \bottomrule
\end{tabular}
}
\\
\footnotesize
\justifying
\singlespacing
\noindent \textit{Notes:} see Table \ref{tab:rmse_disaggreg_bcb_sample1}.
\\
%\footnotesize
%\justifying
%\singlespacing
%\onehalfspacing
%\noindent \textit{Notes:} $^{***}$, $^{**}$, and $^{*}$ indicate that for a specific disaggregate and forecast horizon, a model $m$ performed statistically better than an AR model in a one-tailed Diebold-Mariano test with $\hip_0: \text{MSE}\big(\widehat{\pi}_{i,\,t+h\, |\,t}^{m}\big) = \text{MSE}\big(\pi_{i,\,t+h\,|\,t}^{\text{AR}}\big) \; \textit{ versus } \hip_1: \text{MSE}\big(\widehat{\pi}_{i,\,t+h\,|\,t}^{m}\big) < \text{MSE}\big(\pi_{i,\,t+h\,|\,t}^{\text{AR}}\big)$. The highlighted value in blue bold indicates the best model for each horizon in terms of RMSE ratio with respect to the AR model, and the values in blue italics indicate the second and third best models.
\end{table}

\newpage

%%%%%
\subsection{Subgroups}
%%%%%

% latex table generated in R 4.1.3 by xtable 1.8-4 package
% version adapted by Gilberto Boaretto
% Thu Mar 30 22:45:40 2023
\begin{table}[!ht]
\centering
\caption{\centering Out-of-sample RMSE for IBGE subgroups (in terms of RMSE of the AR model): Jan/2014 to Jun/2022, by disaggregate and horizon} 
\label{tab:rmse_disaggreg_subgroups_append}
\resizebox{1\linewidth}{!}{
\begin{tabular}{llllllllllllll}
  \toprule
\multicolumn{1}{c}{Estimator/Model \;\;\;} & $h = 0\quad\,$ & $h = 1\quad\,$ & $h = 2\quad\,$ & $h = 3\quad\,$ & $h = 4\quad\,$ & $h = 5\quad\,$ & $h = 6\quad\,$ & $h = 7\quad\,$ & $h = 8\quad\,$ & $h = 9\quad\,$ & $h = 10\;\;$ & $h = 11\;\;$ & all $h$\;\;\;\; \\ 
  \cmidrule(lr){1-14}
\vspace{-0.3cm} \\ \multicolumn{14}{c}{\underline{\textbf{A. Food at home} (\texttt{inf.sg1})}} \\ \vspace{-0.3cm} &  &  &  &  &  &  &  &  &  &  &  &  &  \\ 
  AR & 1.000 & 1.000 & 1.000 & 1.000 & 1.000 & 1.000 & 1.000 & 1.000 & 1.000 & 1.000 & 1.000 & 1.000 & 1.000 \\ 
  Augmented AR & 0.991 & 1.105 & 1.123 & 1.026 & 1.053 & 1.058 & 1.019 & 1.064 & 1.016 & 0.975 & 1.152 & 1.106 & 1.062 \\ 
  Ridge & 0.940 & $0.832^{\,***}$ & $\textit{\blue{0.790}}^{\,***}$ & $\textit{\blue{0.765}}^{\,***}$ & $\textit{\blue{0.766}}^{\,***}$ & $0.788^{\,***}$ & $0.745^{\,***}$ & $0.730^{\,***}$ & $0.754^{\,***}$ & $\textit{\blue{0.705}}^{\,***}$ & $\textit{\blue{0.719}}^{\,***}$ & $0.562^{\,***}$ & $\textit{\blue{0.742}}^{\,***}$ \\ 
  adaLASSO & $\textit{\blue{0.718}}^{\,***}$ & $\textit{\blue{0.831}}^{\,***}$ & 0.916 & $0.790^{\,***}$ & $0.777^{\,***}$ & $0.797^{\,***}$ & $0.759^{\,***}$ & $0.790^{\,***}$ & $0.815^{\,***}$ & $0.759^{\,***}$ & $0.788^{\,***}$ & $\textit{\blue{0.561}}^{\,***}$ & $0.766^{\,***}$ \\ 
  Factor & $0.799^{\,**}$ & $0.857^{\,***}$ & $0.834^{\,***}$ & $0.797^{\,***}$ & $0.804^{\,***}$ & $0.814^{\,***}$ & $0.777^{\,***}$ & $0.730^{\,***}$ & $0.754^{\,***}$ & $\textit{\blue{0.705}}^{\,***}$ & $\textit{\blue{0.719}}^{\,***}$ & $0.574^{\,***}$ & $0.751^{\,***}$ \\ 
  FarmPredict & $0.815^{\,**}$ & $0.843^{\,***}$ & $0.819^{\,***}$ & $0.792^{\,***}$ & $0.785^{\,***}$ & $0.804^{\,***}$ & $0.776^{\,***}$ & $0.736^{\,***}$ & $\textit{\blue{0.753}}^{\,***}$ & $0.758^{\,***}$ & $0.768^{\,***}$ & $0.579^{\,***}$ & $0.757^{\,***}$ \\ 
  Target Factor & $0.762^{\,**}$ & $0.864^{\,***}$ & 0.948 & $0.870^{\,**}$ & $0.829^{\,***}$ & $\textit{\blue{0.785}}^{\,***}$ & $\textbf{\blue{0.703}}^{\,***}$ & $0.718^{\,***}$ & $0.773^{\,***}$ & $0.816^{\,***}$ & $0.831^{\,**}$ & $0.623^{\,***}$ & $0.786^{\,***}$ \\ 
  CSR & $0.783^{\,***}$ & $0.878^{\,**}$ & 0.970 & $0.885^{\,*}$ & $0.825^{\,***}$ & $0.807^{\,***}$ & $0.730^{\,***}$ & $\textbf{\blue{0.707}}^{\,***}$ & $0.753^{\,***}$ & $0.714^{\,***}$ & $0.804^{\,***}$ & $0.610^{\,***}$ & $0.779^{\,***}$ \\ 
  Random Forest & $\textbf{\blue{0.709}}^{\,***}$ & $\textbf{\blue{0.766}}^{\,***}$ & $\textbf{\blue{0.761}}^{\,***}$ & $\textbf{\blue{0.738}}^{\,***}$ & $\textbf{\blue{0.731}}^{\,***}$ & $\textbf{\blue{0.747}}^{\,***}$ & $\textit{\blue{0.713}}^{\,***}$ & $\textit{\blue{0.708}}^{\,***}$ & $\textbf{\blue{0.744}}^{\,***}$ & $\textbf{\blue{0.702}}^{\,***}$ & $\textbf{\blue{0.703}}^{\,***}$ & $\textbf{\blue{0.554}}^{\,***}$ & $\textbf{\blue{0.706}}^{\,***}$ \\ 
  \vspace{-0.3cm} \\ \multicolumn{14}{c}{\underline{\textbf{B. Food away from home} (\texttt{inf.sg2})}} \\ \vspace{-0.3cm} &  &  &  &  &  &  &  &  &  &  &  &  &  \\ 
  AR & 1.000 & 1.000 & 1.000 & 1.000 & 1.000 & 1.000 & 1.000 & 1.000 & 1.000 & 1.000 & 1.000 & 1.000 & 1.000 \\ 
  Augmented AR & 1.058 & 0.982 & 1.106 & 1.206 & 1.088 & 1.051 & 0.987 & 1.038 & 0.938 & 1.030 & 1.074 & 1.097 & 1.053 \\ 
  Ridge & $0.883^{\,*}$ & $0.899^{\,*}$ & $0.863^{\,**}$ & $0.903^{\,**}$ & $0.869^{\,**}$ & $0.835^{\,***}$ & $0.768^{\,***}$ & $0.769^{\,***}$ & $0.763^{\,**}$ & $\textit{\blue{0.742}}^{\,***}$ & $\textit{\blue{0.672}}^{\,***}$ & $\textit{\blue{0.749}}^{\,***}$ & $0.798^{\,***}$ \\ 
  adaLASSO & $\textit{\blue{0.796}}^{\,***}$ & $\textit{\blue{0.803}}^{\,***}$ & $0.821^{\,***}$ & $0.883^{\,**}$ & $\textbf{\blue{0.795}}^{\,***}$ & $\textit{\blue{0.783}}^{\,***}$ & $\textbf{\blue{0.741}}^{\,***}$ & $0.766^{\,***}$ & $\textbf{\blue{0.747}}^{\,***}$ & $\textbf{\blue{0.730}}^{\,***}$ & $\textbf{\blue{0.628}}^{\,***}$ & $\textbf{\blue{0.720}}^{\,***}$ & $\textbf{\blue{0.759}}^{\,***}$ \\ 
  Factor & $0.822^{\,**}$ & $0.822^{\,***}$ & $0.827^{\,***}$ & $0.902^{\,**}$ & $0.893^{\,**}$ & $0.825^{\,***}$ & $0.796^{\,***}$ & $\textbf{\blue{0.738}}^{\,***}$ & $\textit{\blue{0.756}}^{\,***}$ & $0.797^{\,***}$ & $0.714^{\,***}$ & $0.823^{\,***}$ & $0.803^{\,***}$ \\ 
  FarmPredict & $0.826^{\,**}$ & $0.846^{\,**}$ & $\textit{\blue{0.802}}^{\,***}$ & $0.890^{\,**}$ & $0.868^{\,**}$ & $0.827^{\,***}$ & $0.768^{\,***}$ & $\textit{\blue{0.760}}^{\,***}$ & $0.764^{\,**}$ & $0.783^{\,***}$ & $0.712^{\,***}$ & $0.775^{\,***}$ & $0.795^{\,***}$ \\ 
  Target Factor & $0.881^{\,**}$ & $0.890^{\,*}$ & $0.879^{\,**}$ & $0.901^{\,**}$ & $0.826^{\,***}$ & $\textbf{\blue{0.777}}^{\,***}$ & $0.787^{\,***}$ & $0.764^{\,***}$ & $0.781^{\,**}$ & $0.791^{\,***}$ & $0.728^{\,***}$ & $0.820^{\,***}$ & $0.811^{\,***}$ \\ 
  CSR & $0.813^{\,***}$ & $0.822^{\,***}$ & $0.808^{\,***}$ & $\textbf{\blue{0.839}}^{\,***}$ & $\textit{\blue{0.804}}^{\,***}$ & $0.796^{\,***}$ & $\textit{\blue{0.754}}^{\,***}$ & $0.791^{\,***}$ & $0.844^{\,*}$ & $0.750^{\,***}$ & $0.792^{\,**}$ & $0.770^{\,***}$ & $0.796^{\,***}$ \\ 
  Random Forest & $\textbf{\blue{0.784}}^{\,***}$ & $\textbf{\blue{0.799}}^{\,***}$ & $\textbf{\blue{0.788}}^{\,***}$ & $\textit{\blue{0.860}}^{\,***}$ & $0.831^{\,***}$ & $0.811^{\,***}$ & $0.763^{\,***}$ & $0.768^{\,***}$ & $0.766^{\,**}$ & $0.746^{\,***}$ & $0.674^{\,***}$ & $0.754^{\,***}$ & $\textit{\blue{0.772}}^{\,***}$ \\ 
  \vspace{-0.3cm} \\ \multicolumn{14}{c}{\underline{\textbf{C. Utilities and maintenance} (\texttt{inf.sg3})}} \\ \vspace{-0.3cm} &  &  &  &  &  &  &  &  &  &  &  &  &  \\ 
  AR & 1.000 & 1.000 & 1.000 & 1.000 & 1.000 & 1.000 & 1.000 & 1.000 & 1.000 & 1.000 & 1.000 & 1.000 & 1.000 \\ 
  Augmented AR & 1.031 & 1.061 & 1.088 & 1.090 & 1.007 & 1.064 & 1.019 & 1.074 & 1.058 & 1.143 & 1.157 & 1.140 & 1.080 \\ 
  Ridge & 0.908 & $0.853^{\,**}$ & $0.801^{\,**}$ & $0.857^{\,**}$ & $0.797^{\,***}$ & $0.831^{\,***}$ & $0.780^{\,***}$ & $0.814^{\,***}$ & $\textit{\blue{0.721}}^{\,***}$ & $0.854^{\,***}$ & $0.805^{\,***}$ & $0.711^{\,***}$ & $0.804^{\,***}$ \\ 
  adaLASSO & $0.763^{\,***}$ & $0.788^{\,***}$ & $0.779^{\,***}$ & $0.816^{\,***}$ & $0.798^{\,***}$ & $0.824^{\,***}$ & $0.785^{\,***}$ & $\textit{\blue{0.811}}^{\,***}$ & $0.727^{\,***}$ & $\textbf{\blue{0.824}}^{\,***}$ & $\textit{\blue{0.798}}^{\,***}$ & $\textit{\blue{0.690}}^{\,***}$ & $\textit{\blue{0.781}}^{\,***}$ \\ 
  Factor & $\textit{\blue{0.762}}^{\,***}$ & $0.806^{\,***}$ & $0.774^{\,***}$ & $0.843^{\,**}$ & $0.785^{\,***}$ & $0.831^{\,***}$ & $\textbf{\blue{0.774}}^{\,***}$ & $0.819^{\,***}$ & $0.730^{\,***}$ & $0.869^{\,***}$ & $0.825^{\,***}$ & $0.722^{\,***}$ & $0.792^{\,***}$ \\ 
  FarmPredict & $\textbf{\blue{0.760}}^{\,***}$ & $0.809^{\,***}$ & $0.780^{\,***}$ & $0.839^{\,***}$ & $0.785^{\,***}$ & $0.820^{\,***}$ & $0.783^{\,***}$ & $0.820^{\,***}$ & $0.729^{\,***}$ & $0.868^{\,**}$ & $0.831^{\,**}$ & $0.730^{\,***}$ & $0.793^{\,***}$ \\ 
  Target Factor & 0.923 & $\textbf{\blue{0.766}}^{\,***}$ & $\textit{\blue{0.763}}^{\,***}$ & 0.958 & $0.848^{\,***}$ & 1.026 & $0.841^{\,***}$ & $0.832^{\,***}$ & $0.752^{\,***}$ & $0.884^{\,**}$ & $0.844^{\,**}$ & $0.863^{\,**}$ & $0.857^{\,***}$ \\ 
  CSR & 0.836 & $0.849^{\,***}$ & $0.781^{\,***}$ & $\textit{\blue{0.781}}^{\,***}$ & $\textbf{\blue{0.747}}^{\,***}$ & $\textbf{\blue{0.784}}^{\,***}$ & $\textit{\blue{0.776}}^{\,***}$ & $\textbf{\blue{0.793}}^{\,***}$ & $\textbf{\blue{0.712}}^{\,***}$ & $\textit{\blue{0.831}}^{\,***}$ & $0.822^{\,**}$ & $0.754^{\,***}$ & $0.785^{\,***}$ \\ 
  Random Forest & $0.805^{\,**}$ & $\textit{\blue{0.770}}^{\,***}$ & $\textbf{\blue{0.711}}^{\,***}$ & $\textbf{\blue{0.771}}^{\,***}$ & $\textit{\blue{0.772}}^{\,***}$ & $\textit{\blue{0.806}}^{\,***}$ & $0.783^{\,***}$ & $0.825^{\,***}$ & $0.726^{\,***}$ & $0.847^{\,***}$ & $\textbf{\blue{0.792}}^{\,***}$ & $\textbf{\blue{0.675}}^{\,***}$ & $\textbf{\blue{0.770}}^{\,***}$ \\ 
  \vspace{-0.3cm} \\ \multicolumn{14}{c}{\underline{\textbf{D. Domestic fuels and energy} (\texttt{inf.sg4})}} \\ \vspace{-0.3cm} &  &  &  &  &  &  &  &  &  &  &  &  &  \\ 
  AR & 1.000 & 1.000 & 1.000 & 1.000 & 1.000 & 1.000 & 1.000 & 1.000 & 1.000 & 1.000 & 1.000 & 1.000 & 1.000 \\ 
  Augmented AR & $0.802^{\,***}$ & $0.912^{\,**}$ & 1.033 & 1.050 & 0.976 & $0.965^{\,*}$ & 1.003 & 1.025 & 1.074 & 1.035 & 1.056 & 1.055 & 1.000 \\ 
  Ridge & $0.776^{\,***}$ & $\textbf{\blue{0.760}}^{\,***}$ & $\textbf{\blue{0.849}}^{\,***}$ & $\textit{\blue{0.838}}^{\,***}$ & $\textbf{\blue{0.832}}^{\,***}$ & $\textbf{\blue{0.780}}^{\,***}$ & $\textbf{\blue{0.763}}^{\,***}$ & $\textbf{\blue{0.837}}^{\,***}$ & $\textbf{\blue{0.812}}^{\,***}$ & $\textit{\blue{0.772}}^{\,***}$ & $\textbf{\blue{0.723}}^{\,***}$ & $\textbf{\blue{0.815}}^{\,***}$ & $\textbf{\blue{0.794}}^{\,***}$ \\ 
  adaLASSO & $\textit{\blue{0.756}}^{\,***}$ & $0.782^{\,***}$ & $0.887^{\,**}$ & $\textit{\blue{0.838}}^{\,***}$ & $\textit{\blue{0.856}}^{\,***}$ & $\textbf{\blue{0.780}}^{\,***}$ & $\textbf{\blue{0.763}}^{\,***}$ & $0.843^{\,***}$ & $0.816^{\,***}$ & $\textbf{\blue{0.769}}^{\,***}$ & $0.743^{\,***}$ & $\textit{\blue{0.832}}^{\,***}$ & $\textit{\blue{0.803}}^{\,***}$ \\ 
  Factor & $0.773^{\,***}$ & $0.786^{\,***}$ & $0.887^{\,**}$ & $\textbf{\blue{0.836}}^{\,***}$ & $0.857^{\,***}$ & $\textbf{\blue{0.780}}^{\,***}$ & $0.763^{\,***}$ & $0.843^{\,***}$ & $0.817^{\,***}$ & $0.773^{\,***}$ & $\textit{\blue{0.743}}^{\,***}$ & $0.834^{\,***}$ & $0.805^{\,***}$ \\ 
  FarmPredict & $0.771^{\,***}$ & $0.784^{\,***}$ & $\textit{\blue{0.874}}^{\,**}$ & $0.848^{\,***}$ & $0.889^{\,**}$ & $0.788^{\,***}$ & $0.784^{\,***}$ & $\textit{\blue{0.840}}^{\,***}$ & $\textit{\blue{0.816}}^{\,***}$ & $0.775^{\,***}$ & $0.753^{\,***}$ & $0.834^{\,***}$ & $0.810^{\,***}$ \\ 
  Target Factor & $\textbf{\blue{0.720}}^{\,***}$ & $\textit{\blue{0.772}}^{\,***}$ & 1.545 & 0.984 & $0.907^{\,**}$ & $0.898^{\,**}$ & $0.901^{\,**}$ & 0.964 & 0.985 & 0.931 & $0.833^{\,**}$ & $0.911^{\,*}$ & 0.954 \\ 
  CSR & $0.788^{\,***}$ & $0.831^{\,***}$ & $0.905^{\,**}$ & $0.892^{\,**}$ & $0.894^{\,**}$ & $0.860^{\,**}$ & $0.835^{\,***}$ & $0.930^{\,*}$ & $0.849^{\,***}$ & $0.815^{\,***}$ & $0.780^{\,***}$ & $0.847^{\,***}$ & $0.850^{\,***}$ \\ 
  Random Forest & $0.789^{\,***}$ & $0.816^{\,***}$ & $0.925^{\,*}$ & $0.878^{\,***}$ & $0.895^{\,**}$ & $0.866^{\,***}$ & $0.875^{\,***}$ & $0.920^{\,**}$ & $0.888^{\,**}$ & $0.841^{\,**}$ & $0.769^{\,***}$ & $0.866^{\,**}$ & $0.858^{\,***}$ \\ 
  \vspace{-0.3cm} \\ \multicolumn{14}{c}{\underline{\textbf{E. Furniture and fixtures} (\texttt{inf.sg5})}} \\ \vspace{-0.3cm} &  &  &  &  &  &  &  &  &  &  &  &  &  \\ 
  AR & 1.000 & 1.000 & 1.000 & 1.000 & 1.000 & 1.000 & 1.000 & 1.000 & 1.000 & 1.000 & 1.000 & 1.000 & 1.000 \\ 
  Augmented AR & 1.071 & 1.015 & 1.049 & 1.129 & 1.013 & 1.003 & 1.025 & 1.056 & 1.039 & 1.138 & 1.159 & 1.081 & 1.063 \\ 
  Ridge & 1.152 & 0.976 & $0.888^{\,**}$ & $0.867^{\,**}$ & $0.832^{\,***}$ & $0.734^{\,***}$ & $0.775^{\,***}$ & $0.842^{\,***}$ & $0.795^{\,***}$ & $0.894^{\,**}$ & $0.871^{\,***}$ & $0.816^{\,***}$ & $0.854^{\,***}$ \\ 
  adaLASSO & $0.888^{\,*}$ & $0.910^{\,*}$ & $0.888^{\,**}$ & $\textit{\blue{0.853}}^{\,***}$ & $\textbf{\blue{0.797}}^{\,***}$ & $\textbf{\blue{0.724}}^{\,***}$ & $\textit{\blue{0.770}}^{\,***}$ & $0.851^{\,***}$ & $0.784^{\,***}$ & $0.869^{\,***}$ & $0.837^{\,***}$ & $0.799^{\,***}$ & $\textit{\blue{0.823}}^{\,***}$ \\ 
  Factor & $\textbf{\blue{0.878}}^{\,*}$ & $\textit{\blue{0.876}}^{\,**}$ & $0.832^{\,***}$ & $0.905^{\,**}$ & $0.868^{\,***}$ & $0.737^{\,***}$ & $0.780^{\,***}$ & $0.856^{\,***}$ & $0.812^{\,***}$ & $0.896^{\,**}$ & $0.862^{\,***}$ & $0.823^{\,***}$ & $0.838^{\,***}$ \\ 
  FarmPredict & $\textit{\blue{0.882}}^{\,*}$ & $0.895^{\,**}$ & $\textbf{\blue{0.819}}^{\,***}$ & $0.882^{\,**}$ & $0.844^{\,***}$ & $0.734^{\,***}$ & $0.784^{\,***}$ & $0.865^{\,***}$ & $0.821^{\,***}$ & $0.914^{\,**}$ & $0.877^{\,***}$ & $0.825^{\,***}$ & $0.839^{\,***}$ \\ 
  Target Factor & 0.927 & 0.927 & $0.844^{\,**}$ & $0.915^{\,*}$ & 0.989 & $0.816^{\,***}$ & $0.813^{\,***}$ & $\textit{\blue{0.822}}^{\,***}$ & $\textit{\blue{0.760}}^{\,***}$ & $0.858^{\,***}$ & $0.835^{\,***}$ & $0.847^{\,**}$ & $0.858^{\,***}$ \\ 
  CSR & $0.899^{\,*}$ & $\textbf{\blue{0.848}}^{\,***}$ & $\textit{\blue{0.824}}^{\,***}$ & $\textbf{\blue{0.826}}^{\,***}$ & $\textit{\blue{0.809}}^{\,***}$ & $\textit{\blue{0.732}}^{\,***}$ & $\textbf{\blue{0.767}}^{\,***}$ & $\textbf{\blue{0.815}}^{\,***}$ & $\textbf{\blue{0.758}}^{\,***}$ & $\textit{\blue{0.830}}^{\,***}$ & $\textit{\blue{0.811}}^{\,***}$ & $\textit{\blue{0.782}}^{\,***}$ & $\textbf{\blue{0.801}}^{\,***}$ \\ 
  Random Forest & 0.926 & $0.905^{\,**}$ & $0.917^{\,*}$ & 0.968 & 0.964 & $0.820^{\,***}$ & $0.843^{\,**}$ & $0.861^{\,***}$ & $0.767^{\,***}$ & $\textbf{\blue{0.791}}^{\,***}$ & $\textbf{\blue{0.753}}^{\,***}$ & $\textbf{\blue{0.712}}^{\,***}$ & $0.849^{\,***}$ \\ 
  \vspace{-0.3cm} \\ \multicolumn{14}{c}{\underline{\textbf{F. Appliances} (\texttt{inf.sg6})}} \\ \vspace{-0.3cm} &  &  &  &  &  &  &  &  &  &  &  &  &  \\ 
  AR & 1.000 & 1.000 & 1.000 & 1.000 & 1.000 & 1.000 & 1.000 & 1.000 & 1.000 & 1.000 & 1.000 & 1.000 & 1.000 \\ 
  Augmented AR & 1.052 & 1.099 & 1.007 & 1.004 & 1.031 & 1.084 & 1.111 & 1.073 & 1.153 & 1.106 & 1.083 & 1.165 & 1.085 \\ 
  Ridge & 0.954 & $0.875^{\,**}$ & $0.850^{\,***}$ & $0.883^{\,**}$ & $0.883^{\,**}$ & $0.824^{\,***}$ & $0.775^{\,***}$ & $0.735^{\,***}$ & $0.780^{\,***}$ & $0.803^{\,***}$ & $0.795^{\,***}$ & $0.839^{\,**}$ & $0.827^{\,***}$ \\ 
  adaLASSO & 0.911 & $0.865^{\,**}$ & $0.836^{\,***}$ & $0.876^{\,**}$ & $0.880^{\,**}$ & $0.814^{\,***}$ & $0.757^{\,***}$ & $\textit{\blue{0.714}}^{\,***}$ & $0.789^{\,**}$ & $0.831^{\,***}$ & $0.829^{\,**}$ & $0.868^{\,**}$ & $0.826^{\,***}$ \\ 
  Factor & $0.903^{\,*}$ & $\textit{\blue{0.817}}^{\,***}$ & $0.840^{\,***}$ & $0.850^{\,***}$ & $0.876^{\,**}$ & $\textit{\blue{0.794}}^{\,***}$ & $0.768^{\,***}$ & $0.732^{\,***}$ & $0.771^{\,***}$ & $0.804^{\,***}$ & $0.798^{\,***}$ & $\textit{\blue{0.830}}^{\,**}$ & $\textit{\blue{0.810}}^{\,***}$ \\ 
  FarmPredict & $0.903^{\,*}$ & $0.820^{\,***}$ & $0.858^{\,**}$ & $\textit{\blue{0.846}}^{\,***}$ & $0.879^{\,**}$ & $0.811^{\,***}$ & $0.787^{\,***}$ & $0.728^{\,***}$ & $0.780^{\,***}$ & $0.800^{\,***}$ & $0.797^{\,***}$ & $0.831^{\,**}$ & $0.815^{\,***}$ \\ 
  Target Factor & $\textit{\blue{0.874}}^{\,*}$ & $0.870^{\,**}$ & $0.848^{\,***}$ & $0.897^{\,**}$ & 0.932 & $0.800^{\,***}$ & $0.861^{\,**}$ & $0.786^{\,***}$ & $\textit{\blue{0.749}}^{\,***}$ & $0.781^{\,***}$ & $0.824^{\,***}$ & $0.852^{\,**}$ & $0.836^{\,***}$ \\ 
  CSR & 1.057 & $0.836^{\,**}$ & $\textit{\blue{0.815}}^{\,***}$ & $0.886^{\,**}$ & $\textit{\blue{0.852}}^{\,***}$ & $0.798^{\,***}$ & $\textit{\blue{0.750}}^{\,***}$ & $0.786^{\,***}$ & $0.753^{\,***}$ & $\textit{\blue{0.772}}^{\,***}$ & $\textit{\blue{0.766}}^{\,***}$ & $0.833^{\,**}$ & $0.819^{\,***}$ \\ 
  Random Forest & $\textbf{\blue{0.827}}^{\,**}$ & $\textbf{\blue{0.794}}^{\,***}$ & $\textbf{\blue{0.781}}^{\,***}$ & $\textbf{\blue{0.806}}^{\,***}$ & $\textbf{\blue{0.800}}^{\,***}$ & $\textbf{\blue{0.755}}^{\,***}$ & $\textbf{\blue{0.718}}^{\,***}$ & $\textbf{\blue{0.688}}^{\,***}$ & $\textbf{\blue{0.712}}^{\,***}$ & $\textbf{\blue{0.731}}^{\,***}$ & $\textbf{\blue{0.734}}^{\,***}$ & $\textbf{\blue{0.795}}^{\,***}$ & $\textbf{\blue{0.757}}^{\,***}$ \\ 
   \bottomrule
\end{tabular}
}
\\
\vspace{-0.1cm}
\footnotesize
\flushright
(continued on next page)
\end{table}

\newpage

% latex table generated in R 4.1.3 by xtable 1.8-4 package
% version adapted by Gilberto Boaretto
% Thu Mar 30 22:45:40 2023
\begin{table}[!ht]
\centering
\caption*{\centering Table C4: Out-of-sample RMSE for IBGE subgroups (in terms of RMSE of the AR model): Jan/2014 to Jun/2022, by disaggregate and horizon (cont.)}
\resizebox{1\linewidth}{!}{
\begin{tabular}{llllllllllllll}
  \toprule
\multicolumn{1}{c}{Estimator/Model \;\;\;} & $h = 0\quad\,$ & $h = 1\quad\,$ & $h = 2\quad\,$ & $h = 3\quad\,$ & $h = 4\quad\,$ & $h = 5\quad\,$ & $h = 6\quad\,$ & $h = 7\quad\,$ & $h = 8\quad\,$ & $h = 9\quad\,$ & $h = 10\;\;$ & $h = 11\;\;$ & all $h$\;\;\;\; \\ 
  \cmidrule(lr){1-14}
\vspace{-0.3cm} \\ \multicolumn{14}{c}{\underline{\textbf{G. Household operations} (\texttt{inf.sg7})}} \\ \vspace{-0.3cm} &  &  &  &  &  &  &  &  &  &  &  &  &  \\ 
  AR & 1.000 & 1.000 & 1.000 & 1.000 & 1.000 & 1.000 & 1.000 & 1.000 & 1.000 & 1.000 & 1.000 & 1.000 & 1.000 \\ 
  Augmented AR & 1.079 & 1.047 & 0.995 & 1.062 & 0.989 & 1.064 & 1.066 & 1.103 & 1.107 & 1.040 & 0.995 & 1.108 & 1.050 \\ 
  Ridge & $\textit{\blue{0.705}}^{\,***}$ & $\textit{\blue{0.700}}^{\,***}$ & $0.607^{\,***}$ & $0.681^{\,***}$ & $0.649^{\,***}$ & $0.733^{\,***}$ & $0.817^{\,***}$ & $0.730^{\,***}$ & $0.738^{\,***}$ & $0.630^{\,***}$ & $0.594^{\,***}$ & $0.685^{\,***}$ & $0.681^{\,***}$ \\ 
  adaLASSO & $0.720^{\,***}$ & $0.708^{\,***}$ & $\textit{\blue{0.606}}^{\,***}$ & $\textit{\blue{0.679}}^{\,***}$ & $0.653^{\,***}$ & $0.749^{\,***}$ & $0.798^{\,***}$ & $\textit{\blue{0.714}}^{\,***}$ & $\textit{\blue{0.716}}^{\,***}$ & $0.632^{\,***}$ & $0.592^{\,***}$ & $\textbf{\blue{0.651}}^{\,***}$ & $\textit{\blue{0.677}}^{\,***}$ \\ 
  Factor & $0.712^{\,***}$ & $0.711^{\,***}$ & $0.623^{\,***}$ & $0.696^{\,***}$ & $\textit{\blue{0.646}}^{\,***}$ & $0.733^{\,***}$ & $0.820^{\,***}$ & $0.731^{\,***}$ & $0.724^{\,***}$ & $0.632^{\,***}$ & $0.594^{\,***}$ & $0.682^{\,***}$ & $0.684^{\,***}$ \\ 
  FarmPredict & $0.709^{\,***}$ & $0.721^{\,***}$ & $0.613^{\,***}$ & $0.688^{\,***}$ & $0.652^{\,***}$ & $0.735^{\,***}$ & $0.817^{\,***}$ & $0.725^{\,***}$ & $0.738^{\,***}$ & $0.640^{\,***}$ & $0.607^{\,***}$ & $0.695^{\,***}$ & $0.687^{\,***}$ \\ 
  Target Factor & $0.839^{\,**}$ & $0.772^{\,***}$ & $0.698^{\,***}$ & $0.858^{\,*}$ & $0.783^{\,***}$ & $0.837^{\,***}$ & 0.926 & $0.822^{\,**}$ & $0.851^{\,**}$ & $0.811^{\,**}$ & $0.688^{\,**}$ & 0.938 & $0.812^{\,***}$ \\ 
  CSR & $0.726^{\,***}$ & $0.711^{\,***}$ & $0.657^{\,***}$ & $0.712^{\,***}$ & $0.670^{\,***}$ & $\textit{\blue{0.697}}^{\,***}$ & $\textit{\blue{0.766}}^{\,***}$ & $0.741^{\,***}$ & $0.733^{\,***}$ & $\textit{\blue{0.612}}^{\,***}$ & $\textit{\blue{0.579}}^{\,***}$ & $0.671^{\,***}$ & $0.683^{\,***}$ \\ 
  Random Forest & $\textbf{\blue{0.668}}^{\,***}$ & $\textbf{\blue{0.684}}^{\,***}$ & $\textbf{\blue{0.593}}^{\,***}$ & $\textbf{\blue{0.657}}^{\,***}$ & $\textbf{\blue{0.610}}^{\,***}$ & $\textbf{\blue{0.684}}^{\,***}$ & $\textbf{\blue{0.756}}^{\,***}$ & $\textbf{\blue{0.695}}^{\,***}$ & $\textbf{\blue{0.709}}^{\,***}$ & $\textbf{\blue{0.607}}^{\,***}$ & $\textbf{\blue{0.556}}^{\,***}$ & $\textit{\blue{0.663}}^{\,***}$ & $\textbf{\blue{0.650}}^{\,***}$ \\ 
\vspace{-0.3cm} \\ \multicolumn{14}{c}{\underline{\textbf{H. Clothes} (\texttt{inf.sg8})}} \\ \vspace{-0.3cm} &  &  &  &  &  &  &  &  &  &  &  &  &  \\ 
  AR & 1.000 & 1.000 & 1.000 & 1.000 & 1.000 & 1.000 & 1.000 & 1.000 & 1.000 & 1.000 & 1.000 & 1.000 & 1.000 \\ 
  Augmented AR & $0.824^{\,***}$ & $0.874^{\,*}$ & $0.804^{\,***}$ & $0.838^{\,***}$ & $0.891^{\,*}$ & $0.871^{\,*}$ & $0.883^{\,*}$ & $0.897^{\,*}$ & 0.924 & 1.024 & 1.014 & 0.940 & $0.902^{\,***}$ \\ 
  Ridge & $0.804^{\,***}$ & $0.887^{\,**}$ & $0.781^{\,***}$ & $0.861^{\,**}$ & $0.857^{\,*}$ & 0.966 & $0.902^{\,*}$ & $0.875^{\,**}$ & $0.880^{\,**}$ & 0.939 & 0.938 & 0.912 & $0.886^{\,***}$ \\ 
  adaLASSO & $0.745^{\,***}$ & $0.818^{\,***}$ & $\textit{\blue{0.704}}^{\,***}$ & $\textbf{\blue{0.629}}^{\,***}$ & $\textbf{\blue{0.684}}^{\,***}$ & $\textbf{\blue{0.701}}^{\,***}$ & $0.770^{\,***}$ & $\textit{\blue{0.760}}^{\,***}$ & $\textbf{\blue{0.687}}^{\,***}$ & $0.775^{\,**}$ & $0.796^{\,***}$ & $0.781^{\,***}$ & $\textit{\blue{0.737}}^{\,***}$ \\ 
  Factor & $\textbf{\blue{0.722}}^{\,***}$ & $\textit{\blue{0.793}}^{\,***}$ & $0.720^{\,***}$ & $\textit{\blue{0.631}}^{\,***}$ & $\textit{\blue{0.694}}^{\,***}$ & $\textit{\blue{0.711}}^{\,***}$ & $\textit{\blue{0.754}}^{\,***}$ & $0.845^{\,***}$ & $0.847^{\,**}$ & $0.779^{\,**}$ & $0.806^{\,***}$ & $\textit{\blue{0.772}}^{\,***}$ & $0.759^{\,***}$ \\ 
  FarmPredict & $0.873^{\,**}$ & $0.896^{\,*}$ & $0.789^{\,***}$ & $0.688^{\,***}$ & $0.735^{\,***}$ & $0.721^{\,***}$ & $0.819^{\,***}$ & $0.876^{\,**}$ & $0.890^{\,*}$ & $0.781^{\,**}$ & $0.834^{\,***}$ & $0.800^{\,***}$ & $0.809^{\,***}$ \\ 
  Target Factor & $\textit{\blue{0.723}}^{\,***}$ & $\textbf{\blue{0.715}}^{\,***}$ & $\textbf{\blue{0.659}}^{\,***}$ & $0.646^{\,***}$ & $0.704^{\,***}$ & $0.754^{\,***}$ & $\textbf{\blue{0.701}}^{\,***}$ & $\textbf{\blue{0.686}}^{\,***}$ & $0.779^{\,***}$ & $0.769^{\,**}$ & $0.829^{\,**}$ & $0.782^{\,***}$ & $\textbf{\blue{0.731}}^{\,***}$ \\ 
  CSR & 0.966 & 1.039 & $0.784^{\,***}$ & $0.745^{\,***}$ & $0.791^{\,**}$ & $0.780^{\,***}$ & $0.818^{\,***}$ & $0.802^{\,***}$ & $\textit{\blue{0.717}}^{\,***}$ & $\textit{\blue{0.755}}^{\,**}$ & $\textit{\blue{0.772}}^{\,***}$ & $0.773^{\,***}$ & $0.810^{\,***}$ \\ 
  Random Forest & $0.813^{\,***}$ & $0.834^{\,***}$ & $0.776^{\,***}$ & $0.698^{\,***}$ & $0.745^{\,***}$ & $0.731^{\,***}$ & $0.780^{\,***}$ & $0.765^{\,***}$ & $0.725^{\,***}$ & $\textbf{\blue{0.746}}^{\,**}$ & $\textbf{\blue{0.762}}^{\,***}$ & $\textbf{\blue{0.749}}^{\,***}$ & $0.759^{\,***}$ \\ 
  \vspace{-0.3cm} \\ \multicolumn{14}{c}{\underline{\textbf{I. Footwear and accessories} (\texttt{inf.sg9})}} \\ \vspace{-0.3cm} &  &  &  &  &  &  &  &  &  &  &  &  &  \\ 
  AR & 1.000 & 1.000 & 1.000 & 1.000 & 1.000 & 1.000 & 1.000 & 1.000 & 1.000 & 1.000 & 1.000 & 1.000 & 1.000 \\ 
  Augmented AR & 1.023 & 1.063 & 0.957 & 0.959 & 1.011 & 1.049 & 0.978 & $0.932^{\,*}$ & 1.000 & 1.145 & 1.218 & 1.062 & 1.040 \\ 
  Ridge & 0.938 & 0.925 & $0.874^{\,*}$ & $0.776^{\,**}$ & 0.884 & 0.951 & $0.838^{\,**}$ & $0.794^{\,***}$ & $0.800^{\,**}$ & $0.730^{\,*}$ & $0.791^{\,*}$ & $0.798^{\,***}$ & $0.832^{\,***}$ \\ 
  adaLASSO & $0.812^{\,***}$ & $\textit{\blue{0.782}}^{\,***}$ & $\textbf{\blue{0.780}}^{\,***}$ & $\textit{\blue{0.672}}^{\,***}$ & $\textit{\blue{0.769}}^{\,*}$ & $\textit{\blue{0.860}}^{\,**}$ & $0.876^{\,**}$ & $0.789^{\,***}$ & $0.794^{\,**}$ & $\textit{\blue{0.668}}^{\,*}$ & $0.734^{\,**}$ & $0.738^{\,***}$ & $\textit{\blue{0.768}}^{\,***}$ \\ 
  Factor & $0.830^{\,***}$ & $0.860^{\,**}$ & $0.792^{\,***}$ & $\textbf{\blue{0.670}}^{\,***}$ & $0.786^{\,*}$ & $0.914^{\,*}$ & $0.862^{\,**}$ & $\textit{\blue{0.787}}^{\,***}$ & $0.817^{\,*}$ & $0.670^{\,*}$ & $\textit{\blue{0.717}}^{\,**}$ & $\textit{\blue{0.731}}^{\,***}$ & $0.780^{\,***}$ \\ 
  FarmPredict & $0.884^{\,**}$ & 0.919 & $0.837^{\,**}$ & $0.676^{\,***}$ & $0.773^{\,*}$ & $0.885^{\,*}$ & $\textit{\blue{0.805}}^{\,***}$ & $0.792^{\,***}$ & $0.797^{\,**}$ & $0.676^{\,*}$ & $0.727^{\,**}$ & $0.739^{\,***}$ & $0.784^{\,***}$ \\ 
  Target Factor & $\textbf{\blue{0.741}}^{\,***}$ & $0.788^{\,***}$ & $\textit{\blue{0.783}}^{\,***}$ & $0.741^{\,***}$ & $0.818^{\,*}$ & $0.911^{\,*}$ & $0.831^{\,**}$ & $0.810^{\,**}$ & $0.841^{\,*}$ & 0.851 & $0.731^{\,**}$ & $0.765^{\,***}$ & $0.802^{\,***}$ \\ 
  CSR & 0.889 & 1.004 & $0.802^{\,***}$ & $0.685^{\,***}$ & $0.789^{\,*}$ & $0.910^{\,*}$ & $0.831^{\,***}$ & $0.824^{\,**}$ & $\textbf{\blue{0.719}}^{\,**}$ & $0.670^{\,*}$ & $0.737^{\,**}$ & $0.736^{\,***}$ & $0.791^{\,***}$ \\ 
  Random Forest & $\textit{\blue{0.769}}^{\,***}$ & $\textbf{\blue{0.779}}^{\,***}$ & $0.792^{\,***}$ & $0.681^{\,***}$ & $\textbf{\blue{0.763}}^{\,*}$ & $\textbf{\blue{0.840}}^{\,**}$ & $\textbf{\blue{0.752}}^{\,***}$ & $\textbf{\blue{0.715}}^{\,***}$ & $\textit{\blue{0.732}}^{\,**}$ & $\textbf{\blue{0.652}}^{\,**}$ & $\textbf{\blue{0.698}}^{\,**}$ & $\textbf{\blue{0.707}}^{\,***}$ & $\textbf{\blue{0.734}}^{\,***}$ \\ 
  \vspace{-0.3cm} \\ \multicolumn{14}{c}{\underline{\textbf{J. Jewelry} (\texttt{inf.sg10})}} \\ \vspace{-0.3cm} &  &  &  &  &  &  &  &  &  &  &  &  &  \\ 
  AR & 1.000 & 1.000 & 1.000 & 1.000 & 1.000 & 1.000 & 1.000 & 1.000 & 1.000 & 1.000 & 1.000 & 1.000 & 1.000 \\ 
  Augmented AR & 1.084 & 1.097 & 1.129 & 1.074 & 1.119 & 1.029 & 1.106 & 1.110 & 0.940 & 1.166 & 1.137 & 1.280 & 1.106 \\ 
  Ridge & 0.714 & $0.691^{\,***}$ & $\textit{\blue{0.704}}^{\,***}$ & $\textit{\blue{0.697}}^{\,***}$ & $\textit{\blue{0.681}}^{\,***}$ & $\textbf{\blue{0.677}}^{\,***}$ & $\textbf{\blue{0.662}}^{\,***}$ & $\textbf{\blue{0.717}}^{\,***}$ & $\textbf{\blue{0.678}}^{\,***}$ & $\textbf{\blue{0.697}}^{\,***}$ & $\textbf{\blue{0.666}}^{\,***}$ & $\textit{\blue{0.739}}^{\,***}$ & $\textbf{\blue{0.693}}^{\,***}$ \\ 
  adaLASSO & $\textbf{\blue{0.678}}^{\,***}$ & $\textbf{\blue{0.652}}^{\,***}$ & 0.716 & $0.732^{\,***}$ & $0.713^{\,***}$ & $0.727^{\,***}$ & $0.693^{\,***}$ & $0.762^{\,***}$ & $0.722^{\,***}$ & $0.739^{\,***}$ & $0.691^{\,***}$ & $0.827^{\,***}$ & $0.721^{\,***}$ \\ 
  Factor & $0.722^{\,**}$ & $0.705^{\,***}$ & $0.732^{\,***}$ & $0.751^{\,***}$ & $0.764^{\,***}$ & $0.723^{\,***}$ & $0.709^{\,***}$ & $\textit{\blue{0.745}}^{\,***}$ & $0.683^{\,***}$ & $\textbf{\blue{0.697}}^{\,***}$ & $\textit{\blue{0.671}}^{\,***}$ & $0.810^{\,***}$ & $0.725^{\,***}$ \\ 
  FarmPredict & $0.723^{\,**}$ & $0.704^{\,***}$ & $0.731^{\,***}$ & $0.740^{\,***}$ & $0.725^{\,***}$ & $0.733^{\,***}$ & $0.726^{\,***}$ & $0.822^{\,***}$ & $\textit{\blue{0.679}}^{\,***}$ & $0.705^{\,***}$ & $0.680^{\,***}$ & $0.774^{\,***}$ & $0.728^{\,***}$ \\ 
  Target Factor & $0.771^{\,**}$ & $0.777^{\,***}$ & 0.811 & $0.798^{\,**}$ & $0.765^{\,***}$ & $0.802^{\,***}$ & $0.867^{\,***}$ & $0.883^{\,***}$ & $0.807^{\,***}$ & $0.788^{\,***}$ & $0.801^{\,**}$ & $0.814^{\,***}$ & $0.808^{\,***}$ \\ 
  CSR & $0.705^{\,***}$ & $0.683^{\,**}$ & 0.710 & $0.699^{\,*}$ & $0.717^{\,***}$ & $0.712^{\,***}$ & $0.701^{\,***}$ & $0.776^{\,***}$ & $0.761^{\,***}$ & $0.746^{\,***}$ & $0.726^{\,***}$ & $0.822^{\,***}$ & $0.730^{\,***}$ \\ 
  Random Forest & $\textit{\blue{0.683}}^{\,***}$ & $\textit{\blue{0.664}}^{\,***}$ & $\textbf{\blue{0.674}}^{\,***}$ & $\textbf{\blue{0.682}}^{\,***}$ & $\textbf{\blue{0.679}}^{\,***}$ & $\textit{\blue{0.687}}^{\,***}$ & $\textit{\blue{0.682}}^{\,***}$ & $0.746^{\,***}$ & $0.706^{\,***}$ & $0.722^{\,***}$ & $0.675^{\,***}$ & $\textbf{\blue{0.730}}^{\,***}$ & $\textit{\blue{0.694}}^{\,***}$ \\ 
  \vspace{-0.3cm} \\ \multicolumn{14}{c}{\underline{\textbf{K. Fabrics} (\texttt{inf.sg11})}} \\ \vspace{-0.3cm} &  &  &  &  &  &  &  &  &  &  &  &  &  \\ 
  AR & 1.000 & 1.000 & 1.000 & 1.000 & 1.000 & 1.000 & 1.000 & 1.000 & 1.000 & 1.000 & 1.000 & 1.000 & 1.000 \\ 
  Augmented AR & 1.170 & 1.134 & 1.125 & 1.114 & 1.155 & 1.140 & 1.130 & 1.099 & 1.076 & 1.343 & 1.136 & 1.041 & 1.147 \\ 
  Ridge & 0.811 & $0.833^{\,***}$ & $0.759^{\,***}$ & $0.766^{\,***}$ & $0.724^{\,***}$ & $0.734^{\,***}$ & $0.820^{\,***}$ & $0.807^{\,***}$ & $0.789^{\,***}$ & $0.643^{\,***}$ & $0.595^{\,***}$ & $0.690^{\,***}$ & $0.736^{\,***}$ \\ 
  adaLASSO & $0.799^{\,***}$ & $0.833^{\,***}$ & 0.758 & $\textit{\blue{0.726}}^{\,***}$ & $0.730^{\,***}$ & $\textit{\blue{0.698}}^{\,***}$ & $\textit{\blue{0.766}}^{\,***}$ & $0.770^{\,***}$ & $0.755^{\,***}$ & $\textit{\blue{0.600}}^{\,***}$ & $\textbf{\blue{0.547}}^{\,***}$ & $\textit{\blue{0.639}}^{\,***}$ & $\textit{\blue{0.706}}^{\,***}$ \\ 
  Factor & $\textit{\blue{0.773}}^{\,**}$ & $0.831^{\,***}$ & $\textit{\blue{0.747}}^{\,***}$ & $0.759^{\,***}$ & $0.727^{\,***}$ & $0.738^{\,***}$ & $0.822^{\,***}$ & $0.816^{\,***}$ & $0.804^{\,***}$ & $0.648^{\,***}$ & $0.611^{\,***}$ & $0.685^{\,***}$ & $0.736^{\,***}$ \\ 
  FarmPredict & $0.790^{\,**}$ & $\textit{\blue{0.818}}^{\,***}$ & $0.747^{\,***}$ & $0.761^{\,***}$ & $0.724^{\,***}$ & $0.743^{\,***}$ & $0.853^{\,***}$ & $0.836^{\,***}$ & $0.794^{\,***}$ & $0.643^{\,***}$ & $0.605^{\,***}$ & $0.693^{\,***}$ & $0.739^{\,***}$ \\ 
  Target Factor & $0.825^{\,**}$ & $0.900^{\,***}$ & 0.796 & $0.750^{\,**}$ & $\textit{\blue{0.723}}^{\,***}$ & $0.762^{\,***}$ & $0.828^{\,***}$ & $0.972^{\,***}$ & $0.811^{\,***}$ & $0.719^{\,***}$ & $0.600^{\,**}$ & $0.740^{\,***}$ & $0.774^{\,***}$ \\ 
  CSR & $0.820^{\,***}$ & $0.838^{\,**}$ & 0.757 & $0.846^{\,*}$ & $0.736^{\,***}$ & $0.719^{\,***}$ & $0.769^{\,***}$ & $\textit{\blue{0.756}}^{\,***}$ & $\textbf{\blue{0.706}}^{\,***}$ & $\textbf{\blue{0.584}}^{\,***}$ & $0.592^{\,***}$ & $0.677^{\,***}$ & $0.722^{\,***}$ \\ 
  Random Forest & $\textbf{\blue{0.729}}^{\,***}$ & $\textbf{\blue{0.745}}^{\,***}$ & $\textbf{\blue{0.694}}^{\,***}$ & $\textbf{\blue{0.695}}^{\,***}$ & $\textbf{\blue{0.660}}^{\,***}$ & $\textbf{\blue{0.678}}^{\,***}$ & $\textbf{\blue{0.746}}^{\,***}$ & $\textbf{\blue{0.733}}^{\,***}$ & $\textit{\blue{0.727}}^{\,***}$ & $0.604^{\,***}$ & $\textit{\blue{0.550}}^{\,***}$ & $\textbf{\blue{0.638}}^{\,***}$ & $\textbf{\blue{0.673}}^{\,***}$ \\ 
 \vspace{-0.3cm} \\ \multicolumn{14}{c}{\underline{\textbf{L. Transportation} (\texttt{inf.sg12})}} \\ \vspace{-0.3cm} &  &  &  &  &  &  &  &  &  &  &  &  &  \\ 
  AR & 1.000 & 1.000 & 1.000 & 1.000 & 1.000 & 1.000 & 1.000 & 1.000 & 1.000 & 1.000 & 1.000 & 1.000 & 1.000 \\ 
  Augmented AR & 0.858 & 1.000 & 1.116 & 1.063 & 1.022 & 1.000 & 0.977 & 1.042 & 1.052 & 1.039 & 1.083 & 1.135 & 1.037 \\ 
  Ridge & $0.977^{\,*}$ & $0.915^{\,*}$ & $0.963^{\,**}$ & $\textit{\blue{0.925}}^{\,**}$ & $\textit{\blue{0.885}}^{\,**}$ & $\textit{\blue{0.880}}^{\,***}$ & $\textit{\blue{0.832}}^{\,***}$ & $\textbf{\blue{0.830}}^{\,***}$ & $\textit{\blue{0.839}}^{\,**}$ & $0.787^{\,***}$ & $\textit{\blue{0.877}}^{\,***}$ & $\textit{\blue{0.844}}^{\,***}$ & $0.874^{\,***}$ \\ 
  adaLASSO & $0.709^{\,***}$ & $0.830^{\,***}$ & $0.955^{\,***}$ & $0.955^{\,**}$ & $0.907^{\,***}$ & $\textbf{\blue{0.880}}^{\,***}$ & $\textbf{\blue{0.825}}^{\,***}$ & $\textit{\blue{0.830}}^{\,***}$ & $0.852^{\,***}$ & $0.804^{\,***}$ & $0.900^{\,***}$ & $0.878^{\,***}$ & $0.861^{\,***}$ \\ 
  Factor & $\textbf{\blue{0.683}}^{\,**}$ & $\textbf{\blue{0.821}}^{\,***}$ & $0.967^{\,***}$ & $0.946^{\,**}$ & $0.890^{\,**}$ & $0.900^{\,***}$ & $0.837^{\,***}$ & $0.839^{\,***}$ & $0.853^{\,***}$ & $0.818^{\,***}$ & $0.908^{\,***}$ & $0.866^{\,***}$ & $0.863^{\,***}$ \\ 
  FarmPredict & $\textit{\blue{0.705}}^{\,**}$ & $\textit{\blue{0.827}}^{\,**}$ & $\textit{\blue{0.951}}^{\,***}$ & $0.930^{\,**}$ & $0.887^{\,**}$ & $0.895^{\,***}$ & $0.836^{\,***}$ & $0.842^{\,***}$ & $0.858^{\,**}$ & $0.817^{\,***}$ & $0.899^{\,***}$ & $0.871^{\,***}$ & $\textit{\blue{0.861}}^{\,***}$ \\ 
  Target Factor & $0.783^{\,**}$ & $0.912^{\,*}$ & $1.071^{\,**}$ & $1.082^{\,**}$ & $0.938^{\,***}$ & $0.966^{\,***}$ & $0.900^{\,***}$ & $0.855^{\,***}$ & $0.877^{\,**}$ & $\textbf{\blue{0.774}}^{\,***}$ & $0.930^{\,***}$ & $0.999^{\,***}$ & $0.923^{\,***}$ \\ 
  CSR & $0.812^{\,***}$ & $0.890^{\,***}$ & $\textbf{\blue{0.903}}^{\,***}$ & $\textbf{\blue{0.901}}^{\,***}$ & $\textbf{\blue{0.881}}^{\,***}$ & $0.901^{\,***}$ & $0.859^{\,***}$ & $0.843^{\,***}$ & $\textbf{\blue{0.830}}^{\,*}$ & $0.847^{\,***}$ & $0.884^{\,**}$ & $0.859^{\,***}$ & $0.867^{\,***}$ \\ 
  Random Forest & $0.789^{\,***}$ & $0.861^{\,***}$ & $0.954^{\,***}$ & $0.945^{\,***}$ & $0.897^{\,***}$ & $0.907^{\,***}$ & $0.844^{\,***}$ & $0.852^{\,***}$ & $0.864^{\,**}$ & $\textit{\blue{0.780}}^{\,***}$ & $\textbf{\blue{0.840}}^{\,***}$ & $\textbf{\blue{0.806}}^{\,***}$ & $\textbf{\blue{0.860}}^{\,***}$ \\ 
  \vspace{-0.3cm} \\ \multicolumn{14}{c}{\underline{\textbf{M. Pharmaceutical and optical products} (\texttt{inf.sg13})}} \\ \vspace{-0.3cm} &  &  &  &  &  &  &  &  &  &  &  &  &  \\ 
  AR & 1.000 & 1.000 & 1.000 & 1.000 & 1.000 & 1.000 & 1.000 & 1.000 & 1.000 & 1.000 & 1.000 & 1.000 & 1.000 \\ 
  Augmented AR & 1.086 & 1.047 & 0.812 & \textit{\blue{0.797}} & \textit{\blue{0.800}} & \textit{\blue{0.814}} & 0.852 & 0.868 & 0.910 & 0.885 & 0.907 & 1.042 & 0.889 \\ 
  Ridge & 1.273 & $1.217^{\,**}$ & $0.937^{\,**}$ & $0.895^{\,**}$ & $0.903^{\,***}$ & $0.897^{\,***}$ & $0.847^{\,***}$ & $0.882^{\,***}$ & $0.990^{\,***}$ & $0.902^{\,***}$ & $0.853^{\,***}$ & $1.010^{\,***}$ & $0.941^{\,***}$ \\ 
  adaLASSO & $\textbf{\blue{0.941}}^{\,***}$ & $0.981^{\,***}$ & $0.814^{\,***}$ & $0.822^{\,***}$ & $0.925^{\,***}$ & $0.909^{\,***}$ & $0.874^{\,***}$ & $0.755^{\,***}$ & $0.755^{\,***}$ & $0.796^{\,***}$ & $0.718^{\,***}$ & $\textit{\blue{0.774}}^{\,***}$ & $0.827^{\,***}$ \\ 
  Factor & $0.965^{\,***}$ & $0.996^{\,***}$ & $1.020^{\,***}$ & $0.932^{\,**}$ & $0.943^{\,***}$ & $0.918^{\,***}$ & $0.871^{\,***}$ & $0.811^{\,***}$ & $0.947^{\,***}$ & $0.822^{\,***}$ & $0.798^{\,***}$ & $0.780^{\,***}$ & $0.892^{\,***}$ \\ 
  FarmPredict & $0.971^{\,***}$ & $0.984^{\,***}$ & $1.019^{\,***}$ & $0.924^{\,***}$ & $0.946^{\,***}$ & $0.914^{\,***}$ & $0.874^{\,***}$ & $0.767^{\,***}$ & $0.865^{\,***}$ & $0.805^{\,**}$ & $0.795^{\,**}$ & $0.780^{\,***}$ & $0.877^{\,***}$ \\ 
  Target Factor & \textit{\blue{0.951}} & $\textbf{\blue{0.937}}^{\,***}$ & $\textbf{\blue{0.740}}^{\,***}$ & \textbf{\blue{0.698}} & $\textbf{\blue{0.717}}^{\,***}$ & \textbf{\blue{0.671}} & $\textbf{\blue{0.669}}^{\,***}$ & $\textbf{\blue{0.681}}^{\,***}$ & $\textbf{\blue{0.697}}^{\,***}$ & $\textit{\blue{0.727}}^{\,**}$ & $0.714^{\,**}$ & $0.775^{\,**}$ & $\textbf{\blue{0.729}}^{\,***}$ \\ 
  CSR & 1.126 & $1.061^{\,***}$ & $0.805^{\,***}$ & $0.922^{\,***}$ & $0.943^{\,***}$ & $0.928^{\,***}$ & $0.875^{\,***}$ & $0.764^{\,***}$ & $0.755^{\,***}$ & $0.734^{\,***}$ & $\textit{\blue{0.708}}^{\,**}$ & $0.788^{\,***}$ & $0.848^{\,***}$ \\ 
  Random Forest & $1.043^{\,**}$ & $\textit{\blue{0.967}}^{\,***}$ & $\textit{\blue{0.743}}^{\,***}$ & $0.809^{\,***}$ & $0.955^{\,***}$ & $0.900^{\,***}$ & $\textit{\blue{0.802}}^{\,***}$ & $\textit{\blue{0.697}}^{\,***}$ & $\textit{\blue{0.737}}^{\,***}$ & $\textbf{\blue{0.694}}^{\,***}$ & $\textbf{\blue{0.673}}^{\,***}$ & $\textbf{\blue{0.756}}^{\,***}$ & $\textit{\blue{0.798}}^{\,***}$ \\ 
   \bottomrule
\end{tabular}
}
\\
\vspace{-0.1cm}
\footnotesize
\flushright
(continued on next page)
\end{table}

\newpage

% latex table generated in R 4.1.3 by xtable 1.8-4 package
% version adapted by Gilberto Boaretto
% Thu Mar 30 22:45:40 2023
\begin{table}[!ht]
\centering
\caption*{\centering Table C4: Out-of-sample RMSE for IBGE subgroups (in terms of RMSE of the AR model): Jan/2014 to Jun/2022, by disaggregate and horizon (cont.)} 
\resizebox{1\linewidth}{!}{
\begin{tabular}{llllllllllllll}
  \toprule
\multicolumn{1}{c}{Estimator/Model \;\;\;} & $h = 0\quad\,$ & $h = 1\quad\,$ & $h = 2\quad\,$ & $h = 3\quad\,$ & $h = 4\quad\,$ & $h = 5\quad\,$ & $h = 6\quad\,$ & $h = 7\quad\,$ & $h = 8\quad\,$ & $h = 9\quad\,$ & $h = 10\;\;$ & $h = 11\;\;$ & all $h$\;\;\;\; \\ 
  \cmidrule(lr){1-14}
\vspace{-0.3cm} \\ \multicolumn{14}{c}{\underline{\textbf{N. Medical services} (\texttt{inf.sg14})}} \\ \vspace{-0.3cm} &  &  &  &  &  &  &  &  &  &  &  &  &  \\ 
  AR & 1.000 & 1.000 & 1.000 & 1.000 & 1.000 & 1.000 & \textit{\blue{1.000}} & 1.000 & 1.000 & 1.000 & 1.000 & 1.000 & 1.000 \\ 
  Augmented AR & $1.009^{\,***}$ & $1.012^{\,**}$ & 1.018 & 0.989 & 1.013 & $1.035^{\,*}$ & 1.002 & 1.025 & 1.007 & 1.028 & 0.999 & 0.980 & 1.010 \\ 
  Ridge & $0.983^{\,***}$ & $0.984^{\,***}$ & $0.989^{\,***}$ & $0.987^{\,***}$ & $0.979^{\,***}$ & $0.971^{\,***}$ & $1.012^{\,***}$ & $0.984^{\,***}$ & $1.003^{\,***}$ & $1.006^{\,***}$ & $0.978^{\,***}$ & $\textbf{\blue{0.977}}^{\,***}$ & $0.987^{\,***}$ \\ 
  adaLASSO & $0.952^{\,***}$ & $0.903^{\,***}$ & $\textit{\blue{0.904}}^{\,**}$ & $1.009^{\,***}$ & $1.025^{\,***}$ & $0.978^{\,***}$ & $1.039^{\,***}$ & $0.964^{\,***}$ & $0.991^{\,***}$ & $1.032^{\,***}$ & $0.964^{\,***}$ & $0.998^{\,***}$ & $0.979^{\,***}$ \\ 
  Factor & $0.945^{\,***}$ & $\textit{\blue{0.901}}^{\,***}$ & $0.934^{\,**}$ & $1.011^{\,***}$ & $0.982^{\,***}$ & $0.980^{\,***}$ & $1.008^{\,***}$ & $\textit{\blue{0.962}}^{\,***}$ & $0.982^{\,***}$ & $1.012^{\,***}$ & $0.957^{\,***}$ & $\textit{\blue{0.979}}^{\,***}$ & $0.970^{\,***}$ \\ 
  FarmPredict & $\textbf{\blue{0.931}}^{\,***}$ & $0.906^{\,***}$ & $0.928^{\,**}$ & $1.043^{\,***}$ & $0.988^{\,**}$ & $0.995^{\,***}$ & $1.046^{\,***}$ & $0.967^{\,***}$ & $\textit{\blue{0.972}}^{\,***}$ & $1.035^{\,***}$ & $\textit{\blue{0.948}}^{\,***}$ & $0.980^{\,***}$ & $0.978^{\,***}$ \\ 
  Target Factor & $0.949^{\,***}$ & $1.019^{\,***}$ & 0.967 & \textbf{\blue{0.932}} & $0.979^{\,**}$ & $\textbf{\blue{0.919}}^{\,**}$ & $\textbf{\blue{0.989}}^{\,**}$ & \textbf{\blue{0.852}} & \textbf{\blue{0.938}} & \textit{\blue{0.984}} & $\textbf{\blue{0.945}}^{\,**}$ & $1.009^{\,*}$ & \textbf{\blue{0.958}} \\ 
  CSR & $\textit{\blue{0.938}}^{\,***}$ & $\textbf{\blue{0.899}}^{\,***}$ & $\textbf{\blue{0.902}}^{\,**}$ & $0.979^{\,**}$ & $\textit{\blue{0.960}}^{\,**}$ & $\textit{\blue{0.966}}^{\,**}$ & $1.020^{\,***}$ & $0.973^{\,*}$ & $0.976^{\,***}$ & $1.030^{\,***}$ & $0.987^{\,***}$ & $0.998^{\,***}$ & $\textit{\blue{0.967}}^{\,***}$ \\ 
  Random Forest & $0.947^{\,***}$ & $0.944^{\,***}$ & $0.916^{\,*}$ & $\textit{\blue{0.967}}^{\,***}$ & $\textbf{\blue{0.957}}^{\,**}$ & $0.967^{\,***}$ & $1.015^{\,***}$ & $0.988^{\,**}$ & $0.975^{\,**}$ & $\textbf{\blue{0.975}}^{\,**}$ & $0.974^{\,***}$ & $1.023^{\,**}$ & $0.969^{\,***}$ \\
\vspace{-0.3cm} \\ \multicolumn{14}{c}{\underline{\textbf{O. Personal care} (\texttt{inf.sg15})}} \\ \vspace{-0.3cm} &  &  &  &  &  &  &  &  &  &  &  &  &  \\ 
  AR & 1.000 & 1.000 & 1.000 & 1.000 & 1.000 & 1.000 & 1.000 & 1.000 & 1.000 & 1.000 & 1.000 & 1.000 & 1.000 \\ 
  Augmented AR & 0.973 & 0.995 & 1.034 & 1.059 & 1.042 & 1.019 & 1.001 & 0.984 & 1.071 & 1.054 & 1.005 & 1.021 & 1.023 \\ 
  Ridge & 0.905 & 0.901 & $\textit{\blue{0.880}}^{\,**}$ & $\textit{\blue{0.863}}^{\,**}$ & $\textit{\blue{0.858}}^{\,***}$ & $0.953^{\,***}$ & $0.982^{\,***}$ & $0.923^{\,***}$ & $0.901^{\,***}$ & $0.907^{\,**}$ & $0.914^{\,***}$ & $0.833^{\,***}$ & $0.899^{\,***}$ \\ 
  adaLASSO & $0.922^{\,*}$ & $\textit{\blue{0.893}}^{\,*}$ & $0.895^{\,**}$ & $0.881^{\,***}$ & $0.870^{\,***}$ & $0.954^{\,***}$ & $0.975^{\,***}$ & $0.918^{\,***}$ & $\textit{\blue{0.878}}^{\,***}$ & $\textit{\blue{0.889}}^{\,***}$ & $\textbf{\blue{0.892}}^{\,***}$ & $0.833^{\,***}$ & $0.898^{\,***}$ \\ 
  Factor & $\textbf{\blue{0.881}}^{\,*}$ & $0.905^{\,**}$ & $0.882^{\,***}$ & $0.867^{\,**}$ & $0.871^{\,***}$ & $0.950^{\,***}$ & $0.981^{\,***}$ & $0.917^{\,***}$ & $0.924^{\,***}$ & $0.896^{\,**}$ & $\textit{\blue{0.910}}^{\,***}$ & $0.834^{\,***}$ & $0.899^{\,***}$ \\ 
  FarmPredict & $\textit{\blue{0.885}}^{\,*}$ & $0.900^{\,**}$ & $\textbf{\blue{0.869}}^{\,***}$ & $\textbf{\blue{0.851}}^{\,**}$ & $\textbf{\blue{0.849}}^{\,***}$ & $\textit{\blue{0.950}}^{\,***}$ & $0.983^{\,***}$ & $0.925^{\,***}$ & $0.916^{\,***}$ & $0.906^{\,**}$ & $0.918^{\,***}$ & $\textit{\blue{0.832}}^{\,***}$ & $\textbf{\blue{0.896}}^{\,***}$ \\ 
  Target Factor & 0.950 & \textbf{\blue{0.881}} & $0.921^{\,**}$ & $0.883^{\,*}$ & 0.881 & $\textbf{\blue{0.898}}^{\,***}$ & $\textbf{\blue{0.923}}^{\,***}$ & $\textbf{\blue{0.852}}^{\,***}$ & $0.935^{\,***}$ & $0.966^{\,***}$ & $1.084^{\,***}$ & $0.903^{\,**}$ & $0.924^{\,***}$ \\ 
  CSR & $0.924^{\,*}$ & $0.903^{\,***}$ & $0.888^{\,***}$ & $0.878^{\,***}$ & $0.889^{\,***}$ & $0.952^{\,***}$ & $\textit{\blue{0.966}}^{\,***}$ & $\textit{\blue{0.898}}^{\,***}$ & $\textbf{\blue{0.863}}^{\,***}$ & $\textbf{\blue{0.869}}^{\,***}$ & $0.915^{\,***}$ & $0.840^{\,***}$ & $\textit{\blue{0.897}}^{\,***}$ \\ 
  Random Forest & 0.899 & $0.925^{\,**}$ & $0.924^{\,*}$ & 0.903 & 0.868 & $0.951^{\,***}$ & $0.988^{\,**}$ & $0.922^{\,***}$ & $0.913^{\,***}$ & $0.928^{\,***}$ & $0.926^{\,***}$ & $\textbf{\blue{0.831}}^{\,***}$ & $0.913^{\,***}$ \\ 
  \vspace{-0.3cm} \\ \multicolumn{14}{c}{\underline{\textbf{P. Personal services} (\texttt{inf.sg16})}} \\ \vspace{-0.3cm} &  &  &  &  &  &  &  &  &  &  &  &  &  \\ 
  AR & 1.000 & 1.000 & 1.000 & 1.000 & 1.000 & 1.000 & 1.000 & 1.000 & 1.000 & 1.000 & 1.000 & 1.000 & 1.000 \\ 
  Augmented AR & 1.097 & 1.138 & 1.027 & 1.068 & 1.184 & 1.138 & 1.066 & 0.987 & 1.225 & 1.260 & 1.098 & 1.151 & 1.117 \\ 
  Ridge & 0.864 & $0.778^{\,**}$ & $0.765^{\,***}$ & $0.786^{\,**}$ & $0.826^{\,**}$ & $0.733^{\,***}$ & $0.781^{\,***}$ & $0.727^{\,***}$ & $0.821^{\,***}$ & $0.972^{\,***}$ & $0.817^{\,***}$ & $0.675^{\,**}$ & $0.786^{\,***}$ \\ 
  adaLASSO & 0.698 & $0.698^{\,**}$ & $0.707^{\,***}$ & $0.767^{\,**}$ & $0.816^{\,**}$ & $0.712^{\,***}$ & $0.753^{\,***}$ & $0.689^{\,***}$ & $0.663^{\,**}$ & $\textbf{\blue{0.769}}^{\,***}$ & $\textbf{\blue{0.649}}^{\,**}$ & $0.663^{\,**}$ & $0.713^{\,***}$ \\ 
  Factor & $0.711^{\,*}$ & $0.719^{\,***}$ & $0.734^{\,***}$ & $0.755^{\,***}$ & $0.833^{\,**}$ & $0.725^{\,***}$ & $0.749^{\,***}$ & $0.700^{\,***}$ & $\textbf{\blue{0.651}}^{\,***}$ & $\textit{\blue{0.778}}^{\,***}$ & $\textit{\blue{0.657}}^{\,***}$ & $0.649^{\,**}$ & $0.719^{\,***}$ \\ 
  FarmPredict & $0.742^{\,*}$ & $0.758^{\,***}$ & $0.747^{\,**}$ & $0.793^{\,***}$ & $0.821^{\,**}$ & $0.707^{\,***}$ & $0.757^{\,***}$ & $0.711^{\,***}$ & $\textit{\blue{0.656}}^{\,***}$ & $0.783^{\,***}$ & $0.666^{\,***}$ & $0.660^{\,**}$ & $0.730^{\,***}$ \\ 
  Target Factor & $0.792^{\,*}$ & $0.742^{\,**}$ & $0.807^{\,***}$ & $0.878^{\,**}$ & 0.786 & $0.786^{\,***}$ & $0.977^{\,**}$ & $0.735^{\,***}$ & $0.781^{\,***}$ & $0.838^{\,***}$ & $0.742^{\,***}$ & $0.804^{\,**}$ & $0.807^{\,***}$ \\ 
  CSR & \textit{\blue{0.655}} & $\textit{\blue{0.571}}^{\,**}$ & $\textit{\blue{0.609}}^{\,***}$ & $\textit{\blue{0.636}}^{\,**}$ & $\textbf{\blue{0.642}}^{\,***}$ & $\textit{\blue{0.605}}^{\,***}$ & $\textit{\blue{0.630}}^{\,***}$ & $\textit{\blue{0.629}}^{\,***}$ & $0.723^{\,***}$ & $0.929^{\,***}$ & $0.728^{\,***}$ & $\textit{\blue{0.646}}^{\,**}$ & $\textit{\blue{0.662}}^{\,***}$ \\ 
  Random Forest & $\textbf{\blue{0.606}}^{\,**}$ & $\textbf{\blue{0.564}}^{\,***}$ & $\textbf{\blue{0.576}}^{\,***}$ & $\textbf{\blue{0.619}}^{\,***}$ & $\textit{\blue{0.659}}^{\,***}$ & $\textbf{\blue{0.559}}^{\,***}$ & $\textbf{\blue{0.629}}^{\,***}$ & $\textbf{\blue{0.605}}^{\,***}$ & $0.716^{\,***}$ & $0.884^{\,***}$ & $0.731^{\,***}$ & $\textbf{\blue{0.619}}^{\,***}$ & $\textbf{\blue{0.643}}^{\,***}$ \\ 
  \vspace{-0.3cm} \\ \multicolumn{14}{c}{\underline{\textbf{Q. Recreation and tobbaco} (\texttt{inf.sg17})}} \\ \vspace{-0.3cm} &  &  &  &  &  &  &  &  &  &  &  &  &  \\ 
  AR & 1.000 & 1.000 & 1.000 & 1.000 & 1.000 & 1.000 & 1.000 & 1.000 & 1.000 & 1.000 & 1.000 & 1.000 & 1.000 \\ 
  Augmented AR & 0.988 & 1.019 & 1.021 & 1.135 & 1.042 & 1.032 & 1.017 & 1.076 & 1.144 & 1.129 & 1.029 & 1.056 & 1.061 \\ 
  Ridge & $0.730^{\,***}$ & $0.786^{\,***}$ & $0.791^{\,***}$ & $\textit{\blue{0.740}}^{\,***}$ & $0.695^{\,***}$ & $0.773^{\,***}$ & $0.749^{\,***}$ & $0.756^{\,***}$ & $0.703^{\,***}$ & $0.684^{\,***}$ & $0.774^{\,***}$ & $\textit{\blue{0.747}}^{\,***}$ & $0.742^{\,***}$ \\ 
  adaLASSO & $\textit{\blue{0.718}}^{\,***}$ & $0.791^{\,***}$ & $0.798^{\,***}$ & $0.744^{\,***}$ & $0.676^{\,***}$ & $\textit{\blue{0.759}}^{\,***}$ & $\textit{\blue{0.740}}^{\,***}$ & $\textbf{\blue{0.752}}^{\,***}$ & $0.699^{\,***}$ & $\textit{\blue{0.678}}^{\,***}$ & $0.772^{\,***}$ & $0.750^{\,***}$ & $\textit{\blue{0.737}}^{\,***}$ \\ 
  Factor & $\textbf{\blue{0.705}}^{\,***}$ & $\textbf{\blue{0.781}}^{\,***}$ & $\textbf{\blue{0.776}}^{\,***}$ & $0.742^{\,***}$ & $\textbf{\blue{0.654}}^{\,***}$ & $\textbf{\blue{0.756}}^{\,***}$ & $\textbf{\blue{0.737}}^{\,***}$ & $0.756^{\,***}$ & $0.705^{\,***}$ & $0.690^{\,***}$ & $0.792^{\,***}$ & $\textbf{\blue{0.744}}^{\,***}$ & $\textbf{\blue{0.734}}^{\,***}$ \\ 
  FarmPredict & $0.719^{\,***}$ & $\textit{\blue{0.783}}^{\,***}$ & $0.787^{\,***}$ & $0.740^{\,***}$ & $\textit{\blue{0.666}}^{\,***}$ & $0.781^{\,***}$ & $0.767^{\,***}$ & $\textit{\blue{0.753}}^{\,***}$ & $\textit{\blue{0.698}}^{\,***}$ & $0.680^{\,***}$ & $\textit{\blue{0.772}}^{\,***}$ & $0.747^{\,***}$ & $0.738^{\,***}$ \\ 
  Target Factor & $0.781^{\,**}$ & $0.810^{\,***}$ & $0.896^{\,***}$ & $1.108^{\,*}$ & $0.809^{\,***}$ & $0.824^{\,***}$ & 0.839 & $0.811^{\,**}$ & $0.754^{\,**}$ & $0.699^{\,**}$ & $0.835^{\,**}$ & 0.773 & $0.831^{\,***}$ \\ 
  CSR & $0.723^{\,***}$ & $0.854^{\,***}$ & $0.795^{\,***}$ & $\textbf{\blue{0.732}}^{\,***}$ & $0.727^{\,***}$ & $0.774^{\,***}$ & $0.767^{\,***}$ & $0.862^{\,***}$ & $0.794^{\,***}$ & $0.729^{\,***}$ & $0.796^{\,***}$ & $0.778^{\,***}$ & $0.776^{\,***}$ \\ 
  Random Forest & $0.753^{\,***}$ & $0.796^{\,***}$ & $\textit{\blue{0.784}}^{\,***}$ & $0.776^{\,***}$ & $0.757^{\,***}$ & $0.850^{\,***}$ & $0.815^{\,***}$ & $0.769^{\,***}$ & $\textbf{\blue{0.693}}^{\,***}$ & $\textbf{\blue{0.652}}^{\,***}$ & $\textbf{\blue{0.763}}^{\,***}$ & $0.747^{\,***}$ & $0.761^{\,***}$ \\ 
  \vspace{-0.3cm} \\ \multicolumn{14}{c}{\underline{\textbf{R. Courses, reading, and stationery} (\texttt{inf.sg18})}} \\ \vspace{-0.3cm} &  &  &  &  &  &  &  &  &  &  &  &  &  \\ 
  AR & 1.000 & 1.000 & 1.000 & 1.000 & 1.000 & 1.000 & 1.000 & 1.000 & 1.000 & 1.000 & 1.000 & 1.000 & 1.000 \\ 
  Augmented AR & $0.419^{\,***}$ & $0.411^{\,*}$ & $0.387^{\,***}$ & $0.330^{\,***}$ & $0.336^{\,*}$ & $0.328^{\,*}$ & $0.397^{\,*}$ & $0.443^{\,*}$ & 0.543 & 0.830 & 0.869 & 0.928 & $0.436^{\,***}$ \\ 
  Ridge & $0.410^{\,***}$ & $0.364^{\,**}$ & $0.345^{\,***}$ & $0.330^{\,**}$ & $0.349^{\,*}$ & 0.342 & $0.389^{\,*}$ & $0.440^{\,**}$ & $0.484^{\,**}$ & \textbf{\blue{0.765}} & \textbf{\blue{0.794}} & 0.844 & $0.415^{\,***}$ \\ 
  adaLASSO & $\textbf{\blue{0.346}}^{\,***}$ & $\textbf{\blue{0.350}}^{\,***}$ & $\textbf{\blue{0.328}}^{\,***}$ & $0.303^{\,***}$ & $\textit{\blue{0.307}}^{\,***}$ & $0.311^{\,***}$ & $0.354^{\,***}$ & $0.400^{\,***}$ & $0.453^{\,***}$ & $0.811^{\,**}$ & $0.823^{\,***}$ & $\textit{\blue{0.819}}^{\,***}$ & $0.389^{\,***}$ \\ 
  Factor & $0.353^{\,***}$ & $0.358^{\,***}$ & $0.338^{\,***}$ & $\textit{\blue{0.295}}^{\,***}$ & $\textbf{\blue{0.304}}^{\,***}$ & $\textbf{\blue{0.301}}^{\,***}$ & $\textbf{\blue{0.345}}^{\,***}$ & $\textbf{\blue{0.372}}^{\,***}$ & $\textbf{\blue{0.416}}^{\,**}$ & $0.796^{\,**}$ & $0.810^{\,***}$ & $0.820^{\,***}$ & $\textbf{\blue{0.383}}^{\,***}$ \\ 
  FarmPredict & $\textit{\blue{0.346}}^{\,**}$ & $\textit{\blue{0.350}}^{\,*}$ & $\textit{\blue{0.328}}^{\,***}$ & $0.303^{\,***}$ & $0.308^{\,***}$ & $0.311^{\,***}$ & $\textit{\blue{0.354}}^{\,***}$ & $\textit{\blue{0.400}}^{\,**}$ & $\textit{\blue{0.452}}^{\,*}$ & $0.803^{\,**}$ & $0.821^{\,***}$ & $0.822^{\,***}$ & $\textit{\blue{0.389}}^{\,***}$ \\ 
  Target Factor & $0.384^{\,***}$ & $0.369^{\,***}$ & $0.335^{\,***}$ & $\textbf{\blue{0.295}}^{\,***}$ & $0.330^{\,***}$ & $\textit{\blue{0.304}}^{\,***}$ & $0.395^{\,***}$ & $0.426^{\,***}$ & $0.456^{\,***}$ & $\textit{\blue{0.787}}^{\,**}$ & $0.845^{\,**}$ & $\textbf{\blue{0.761}}^{\,***}$ & $0.400^{\,***}$ \\ 
  CSR & 0.466 & 0.801 & $0.603^{\,***}$ & $0.508^{\,***}$ & $0.550^{\,**}$ & $0.528^{\,***}$ & $0.568^{\,***}$ & $0.502^{\,***}$ & $0.693^{\,***}$ & $0.834^{\,**}$ & $0.917^{\,***}$ & $0.910^{\,***}$ & $0.604^{\,***}$ \\ 
  Random Forest & $0.454^{\,***}$ & $0.455^{\,***}$ & $0.445^{\,***}$ & $0.425^{\,***}$ & $0.414^{\,***}$ & $0.420^{\,***}$ & $0.428^{\,***}$ & $0.497^{\,***}$ & $0.556^{\,***}$ & $0.848^{\,**}$ & $\textit{\blue{0.802}}^{\,***}$ & $0.843^{\,***}$ & $0.480^{\,***}$ \\ 
  \vspace{-0.3cm} \\ \multicolumn{14}{c}{\underline{\textbf{S. Communication} (\texttt{inf.sg19})}} \\ \vspace{-0.3cm} &  &  &  &  &  &  &  &  &  &  &  &  &  \\ 
  AR & 1.000 & 1.000 & 1.000 & 1.000 & 1.000 & 1.000 & 1.000 & 1.000 & 1.000 & 1.000 & 1.000 & 1.000 & 1.000 \\ 
  Augmented AR & 1.141 & 1.108 & 1.062 & 1.064 & 1.058 & 1.091 & 0.972 & $0.961^{\,*}$ & 0.919 & 0.928 & 0.988 & 1.030 & 1.022 \\ 
  Ridge & 0.653 & 0.705 & $\textit{\blue{0.618}}^{\,*}$ & $\textit{\blue{0.642}}^{\,**}$ & 0.622 & 0.664 & $0.677^{\,**}$ & $0.616^{\,***}$ & $0.553^{\,**}$ & $0.574^{\,*}$ & $0.535^{\,*}$ & $0.573^{\,***}$ & $0.613^{\,***}$ \\ 
  adaLASSO & $\textbf{\blue{0.638}}^{\,***}$ & $\textbf{\blue{0.700}}^{\,***}$ & $\textbf{\blue{0.613}}^{\,***}$ & $\textbf{\blue{0.636}}^{\,***}$ & $\textbf{\blue{0.616}}^{\,*}$ & $\textbf{\blue{0.656}}^{\,**}$ & $\textit{\blue{0.673}}^{\,**}$ & $\textbf{\blue{0.609}}^{\,***}$ & $\textit{\blue{0.551}}^{\,**}$ & $\textbf{\blue{0.528}}^{\,*}$ & $\textit{\blue{0.506}}^{\,**}$ & $\textit{\blue{0.537}}^{\,***}$ & $\textbf{\blue{0.599}}^{\,***}$ \\ 
  Factor & $0.650^{\,***}$ & $\textit{\blue{0.702}}^{\,**}$ & $0.619^{\,***}$ & $0.647^{\,***}$ & $\textit{\blue{0.621}}^{\,*}$ & $\textit{\blue{0.657}}^{\,*}$ & $0.674^{\,**}$ & $\textit{\blue{0.609}}^{\,***}$ & $0.569^{\,*}$ & $\textit{\blue{0.539}}^{\,*}$ & $\textbf{\blue{0.502}}^{\,**}$ & $0.545^{\,***}$ & $0.605^{\,***}$ \\ 
  FarmPredict & $\textit{\blue{0.645}}^{\,**}$ & 0.720 & $0.621^{\,**}$ & $0.643^{\,***}$ & $0.622^{\,*}$ & $0.660^{\,*}$ & $\textbf{\blue{0.673}}^{\,***}$ & $0.611^{\,***}$ & $\textbf{\blue{0.549}}^{\,**}$ & $0.544^{\,*}$ & $0.514^{\,**}$ & $\textbf{\blue{0.532}}^{\,***}$ & $\textit{\blue{0.604}}^{\,***}$ \\ 
  Target Factor & $0.856^{\,***}$ & $1.154^{\,***}$ & $0.874^{\,***}$ & $0.986^{\,***}$ & $0.912^{\,*}$ & $0.864^{\,*}$ & $0.934^{\,**}$ & $0.732^{\,**}$ & $0.667^{\,*}$ & 0.695 & $0.708^{\,**}$ & $0.660^{\,***}$ & $0.829^{\,***}$ \\ 
  CSR & 0.672 & 0.761 & $0.646^{\,***}$ & $0.684^{\,***}$ & $0.680^{\,*}$ & $0.679^{\,*}$ & $0.723^{\,***}$ & $0.631^{\,**}$ & $0.599^{\,**}$ & $0.657^{\,*}$ & $0.511^{\,**}$ & $0.663^{\,***}$ & $0.653^{\,***}$ \\ 
  Random Forest & $0.656^{\,***}$ & $0.711^{\,***}$ & $0.652^{\,***}$ & $0.662^{\,***}$ & $0.668^{\,*}$ & $0.694^{\,**}$ & $0.734^{\,***}$ & $0.649^{\,***}$ & $0.580^{\,**}$ & $0.602^{\,**}$ & $0.576^{\,**}$ & $0.560^{\,***}$ & $0.639^{\,***}$ \\ 
   \bottomrule
\end{tabular}
}
\\
\footnotesize
\justifying
\singlespacing
\noindent \textit{Notes:} see Table \ref{tab:rmse_disaggreg_bcb_sample1}.
\end{table}

%%%%%%%%%%%%%%%%%%%%%%%%%%%%%%%%%%%%%%%%%%

\section{Frequency of models with least squared forecast error}
\label{append:selected_models}

\setcounter{table}{0}
\renewcommand{\thetable}{D\arabic{table}}

\setcounter{figure}{0}
\renewcommand{\thefigure}{D\arabic{figure}}

\newpage

\begin{landscape}
    \begin{figure}[!ht]
        \centering
        \caption{Frequency each model attained the least forecast squared error for BCB disaggregation, by horizon and disaggregate (\%)}
        \label{fig:sel_model_bcb_horizon}
        \vspace{-0.2cm}
        \includegraphics[width=\linewidth]{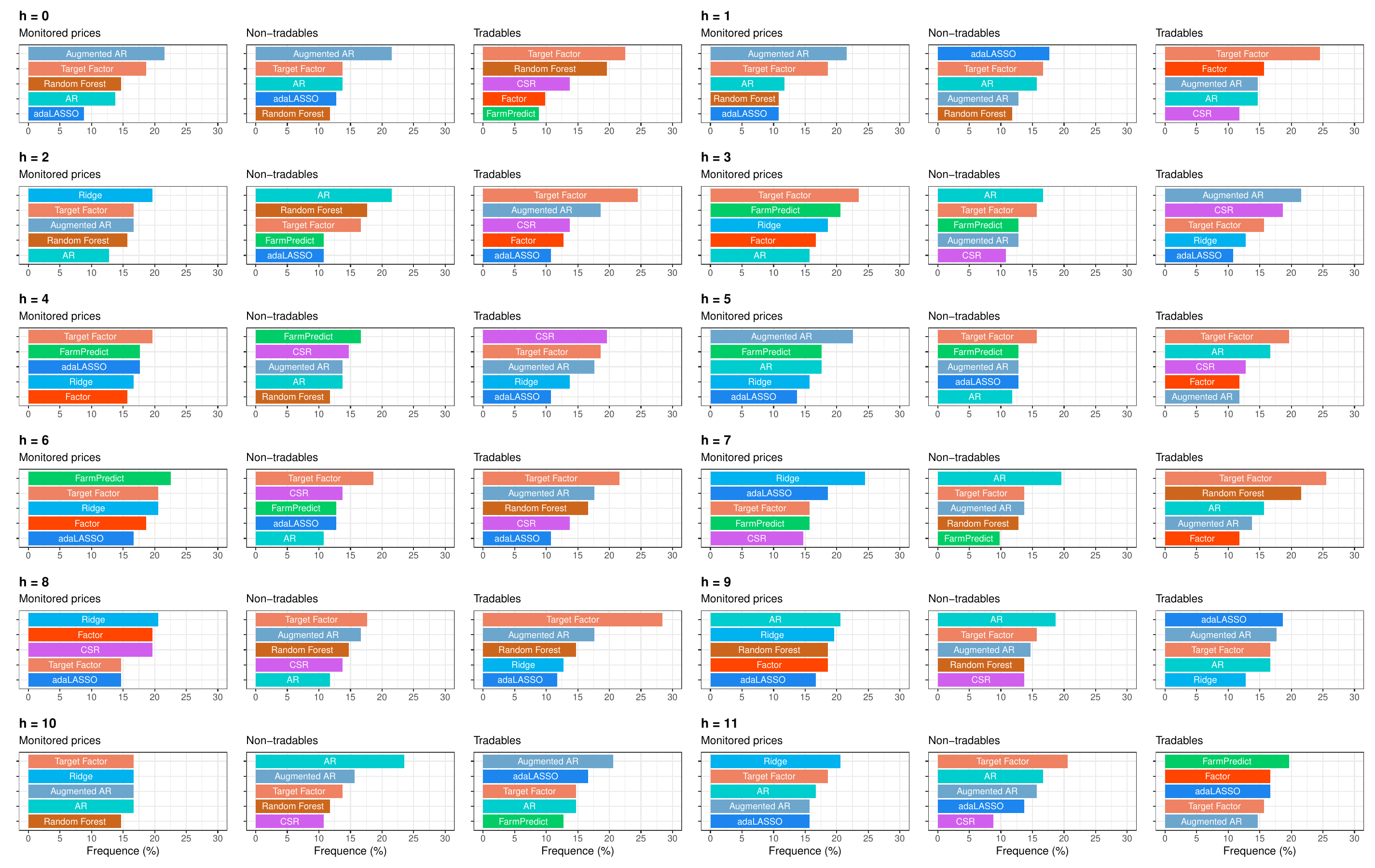}
    \end{figure}
\end{landscape}

\newpage

\begin{figure}[!ht]
    \centering
    \caption{Frequency each model attained the least forecast squared error for BCB disaggregation, by disaggregate and stacking the horizons (\%)}
    \label{fig:sel_model_bcb_stack}
    \vspace{-0.2cm}
    \includegraphics[width=\linewidth]{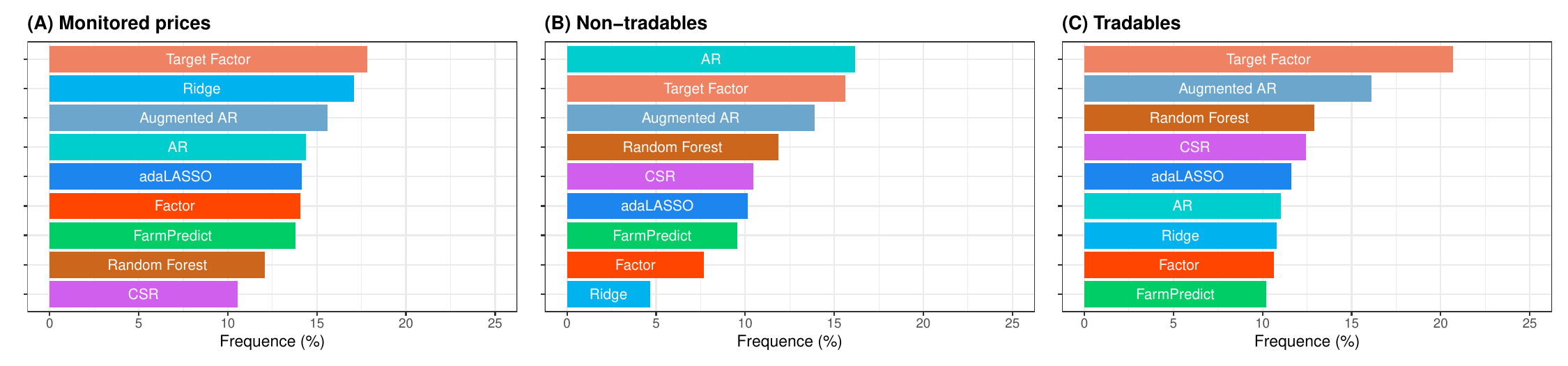}
\end{figure}

\vspace{1cm}

\begin{figure}[!ht]
    \centering
    \caption{Frequency each model attained the least forecast squared error for IBGE groups, by disaggregate and stacking the horizons (\%)}
    \label{fig:sel_model_groups}
    \vspace{-0.2cm}
    \includegraphics[width=\linewidth]{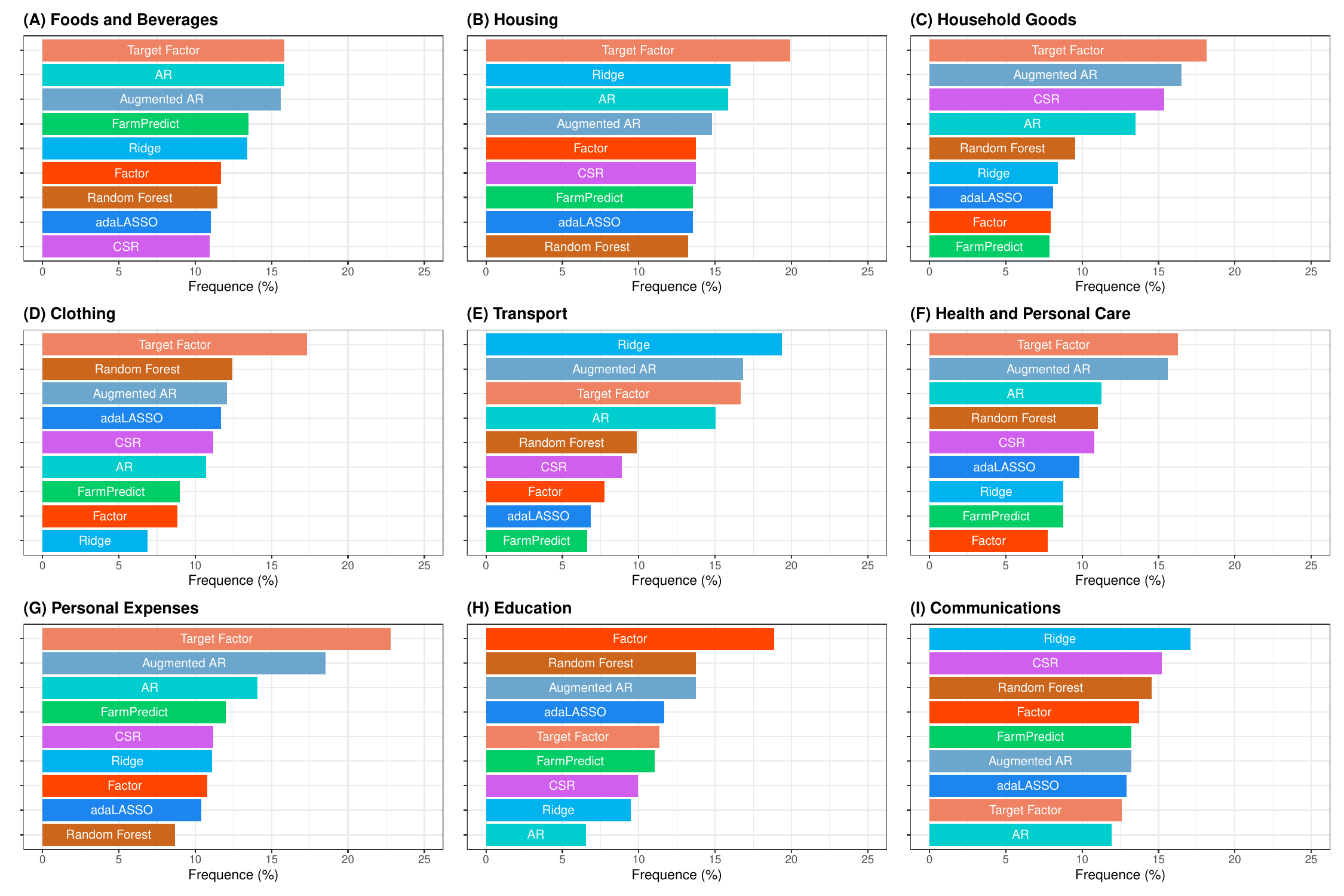}
\end{figure}

\newpage

\begin{figure}[H]
    \centering
    \caption{Frequency each model attained the least forecast squared error for IBGE subgroups, by disaggregate and stacking the horizons (\%)}
    \label{fig:sel_model_subgroups}
    \vspace{-0.2cm}
    \includegraphics[width=\linewidth]{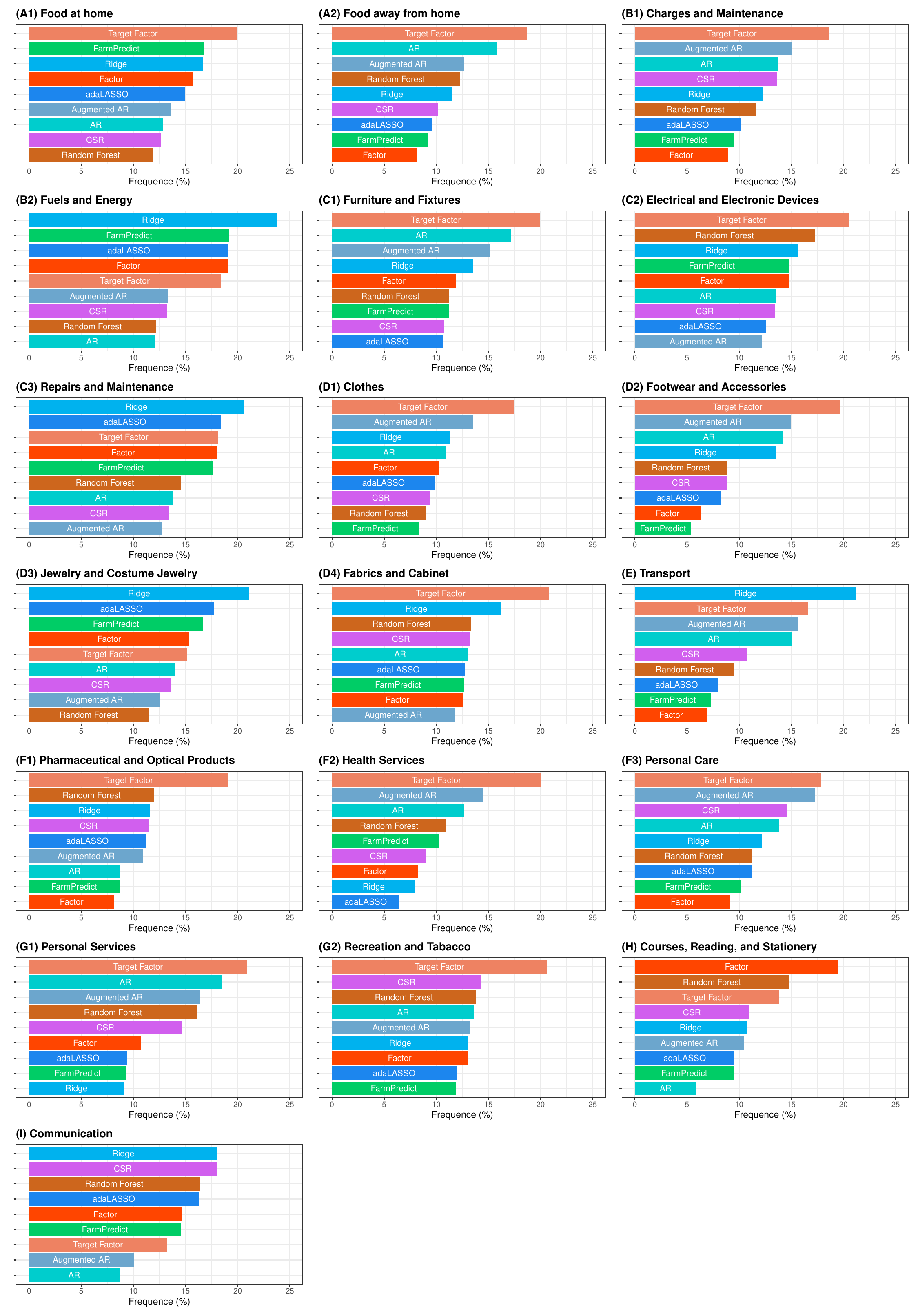}
\end{figure}

%%%%%%%%%%%%%%%%%%%%%%%%%%%%%%%%%%%%%%%%%%

%\mewpage

\small \singlespacing
\bibliographystyle{econometrica}
\bibliography{references}

%%%%%%%%%%%%%%%%%%%%%%%%%%%%%%%%%%%%%%%%%%

\end{document}